\documentstyle[12pt,axodraw,epsfig]{article}

\newcommand{\mysection}{\setcounter{equation}{0}\section}

\newcommand{\xo}{\mbox{$x_1^0$}}
\newcommand{\xt}{\mbox{$x_2^0$}}
\begin{document}
\begin{flushright} 
IPPP/04/70\\
DCPT/04/140\\
SINP/TNP/04-18\\
TIFR/TH/04-29\\
\hfill {\tt hep-ph/0411018}
\end{flushright} 
\vspace{5mm} 
\begin{center} 
{\Large \bf \boldmath Next-to-Leading Order QCD Corrections to the Drell-Yan Cross 
Section in Models of TeV-Scale Gravity}\\
\vspace{5mm} 
{\bf 
Prakash Mathews$^{a,b}$,
V. Ravindran$^c$,
K.~Sridhar$^d$,
W.L. van Neerven$^e$
}\\ 
\end{center}
\vspace{10pt} 
\begin{flushleft}
{\it a) Institute for Particle Physics Phenomenology, 
University of Durham,\\ 
Durham DH1 3LE, UK\\

b) Theory Group, Saha Institute of Nuclear Physics, 1/AF Bidhan Nagar,\\ 
Kolkata 700 064, India\\
 
c) Harish-Chandra Research Institute, 
 Chhatnag Road, Jhunsi, Allahabad, India.\\

d) Department of Theoretical Physics, 
Tata Institute of Fundamental Research,\\   
Homi Bhabha Road, Mumbai 400 005, India. \\

e)Instituut-Lorentz, University of Leiden, PO Box 9506, 2300 RA Leiden, \\
The Netherlands.
} 
\end{flushleft}
 
\vspace{10pt} 
\begin{center}
{\bf ABSTRACT} 
\end{center} 
\vskip12pt 
The first results on next-to-leading order QCD corrections to graviton-induced
processes in hadron collisions in models of TeV-scale gravity are presented
focusing on the case of dilepton pair production in $\bar p p$ and $pp$
collisions.  Distributions in the invariant mass $Q$, the longitudinal 
fraction $x_F$, the rapidity $Y$ and the forward-backward asymmetry of the 
lepton pair are studied.  The quantitative impact of the QCD corrections for 
searches of extra dimensions at hadron colliders is investigated.  It turns 
out that at the LHC ($\sqrt{S}=14$ TeV) the $K$-factor is rather large 
($K=1.6$) for large invariant mass $Q$ of the lepton pair, indicating the 
importance of accounting for these QCD corrections in the experimental 
search for TeV-scale gravity. At the Tevatron, the K-factor 
does not substantially deviate from the Standard Model value. However, its 
inclusion is necessitated to make the cross-section stable with respect to 
scale variations.


\mysection{Introduction}
\noindent 
In the recent past, models of extra dimensions have been studied
as serious contenders for physics beyond the Standard Model. These
extra-dimensional models are attractive because they help address 
the hierarchy between the electroweak and the Planck scales.  In 
the model of Arkani-Hamed, Dimopoulos and Dvali (the ADD model)
\cite{add}
the Planck scale is reduced to a TeV and the hierarchy problem thereby
avoided by invoking a large magnitude of $d$ extra dimensions.

In the ADD model, the Standard Model (SM) fields are constrained to a
3-brane, while gravitons propagate in the $4+d$ dimensions.
Then the size of the extra dimensions is only constrained by the length
scales to which the gravitational inverse square law has been
experimentally tested, which are currently probing the sub-millimetre range.
For $d$ between 2 and 6, the size of the extra dimensions varies from
a millimetre to a Fermi. The relation between the
4-dimensional Planck scale $M_P$ and the scale $M_S$ in $(4+d)$-dimension is
\begin{equation}
M_P \approx M_S^{(d+2)} R^d\,,
\label{eq1}
\end{equation}
where $R$ is the compactification radius. In the ADD model, because $R$ is
large it is possible to lower $M_S$ down to a TeV and avoid the hierarchy
problem. An important consequence of the lowering of the Planck scale
is that quantum gravity effects could be tested at energies of
${\cal O}$(TeV).  So it opens up a new area of studies on
the effects of gravitons at present and future colliders.

The $4+d$ dimensional graviton manifests in 4-dimensions as a tower of massive
Kaluza-Klein (KK) modes. These KK modes interact with the SM particles confined
to the 3-brane via the energy-momentum tensor.  Each KK mode couples to the SM
particles with a coupling of the order of $1/M_P$, (we will use $\kappa$ in the
rest of the paper).  As can be seen the coupling of each KK mode to the SM 
particles are highly suppressed.  But the effective coupling is obtained after 
summing over all the KK modes and due to the high multiplicity of the KK modes
the effective interaction has a strength of $1/M_S$ \cite{grw,hlz}.
This enhanced coupling provides viable signatures of
the graviton KK modes at colliders. Both real graviton production
and the effects of virtual gravitons in various processes have
been studied in the literature \cite{rev} and have
yielded bounds on $M_S$ which are in the ball-park of a TeV.

These collider processes involving gravitons have, however, been studied
at the leading order (LO) in QCD \footnote{Recently NLO-QCD corrections
to $e^+ e^- \to$ hadrons have been studied in \cite{prs}}.  At hadron 
colliders such as present Tevatron and upcoming Large Hadron Collider (LHC), 
the theoretical uncertainties coming from the QCD effects due to initial 
state partons are quite sizable.  The sources of these uncertainties are 
two fold: renormalisation and factorisation scales and the parton 
distribution functions.  The scale uncertainties can come from the strong 
coupling constant as well as parton distribution functions.  Experience 
with next to leading order (NLO) contributions to SM processes 
strongly suggests that the LO corrections at hadronic colliders are quiet 
unreliable \cite{alel,hune}.  The standard Drell-Yan (DY) processes and 
Higgs production at hadron colliders, for example, not only get large 
corrections but the theoretical uncertainties get reduced 
significantly from NLO corrections.

It is with this motivation that the present paper presents the results 
of the computation of NLO-QCD corrections to the DY process, 
$P_1+P_2 \rightarrow \mu^+ \mu^- +X$, where $P_1$, $P_2$ are initial 
hadrons and $X$ is an arbitrary hadronic final state.  This process takes
place via the exchange of $\gamma$, $Z$ and graviton.  These NLO results 
are used to study the impact of the QCD correction for this process at 
the present Tevatron collider and the future LHC which is planned to 
operate at centre-of-mass energies around $14$ TeV.  This process had 
been considered earlier to LO \cite{jh,dy} in the ADD model.

The results we present here are for the ADD 
model, but since the QCD corrections, presented here, are model-independent 
they may equally be used for studying the Randall-Sundrum model of warped 
compactification \cite{rs}. 

This paper is organised as follows.  In Sec.~2 we describe additional 
graviton vertices needed to perform the NLO QCD corrections.  We study 
the distributions in the invariant mass $Q$ (Sec.~3),
the longitudinal fraction $x_F$ (Sec.~4), the rapidity $Y$ (Sec.~5) and
the forward-backward asymmetry (Sec.~6) of lepton pair.  Finally in Sec.~7
we present the discussion and summarise the results.  In Appendix A and B 
we present the detailed expressions needed for $x_F$ and $Y$ distributions
in Sec.~4 and Sec.~5 respectively.

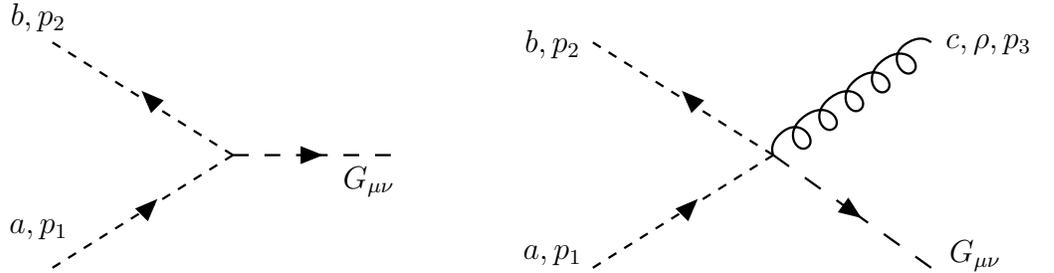
\begin{figure}[ht]
\SetScale{1.7}
\noindent
\begin{picture}(150,180)(0,0)
\ArrowLine(71,63)(72,64)
\ArrowLine(72,87)(71,88)
\DashLine(50,100)(90,75){2}
\DashLine(90,75)(50,50){2}
\DashArrowLine(90,75)(125,75){4}
\Text(80,100)[]{$a,p_1$}
\Text(80,180)[]{$b,p_2$}
\Text(205,118)[]{$G_{\mu \nu}$}
\end{picture}
\hspace*{1.5cm}
\SetScale{1.7}
\noindent
\begin{picture}(150,150)(0,0)
\ArrowLine(71,63)(72,64)
\ArrowLine(72,87)(71,88)
\DashLine(50,100)(90,75){2}
\DashLine(90,75)(50,50){2}
\Gluon(90,75)(125,100){3}{5}
\DashArrowLine(90,75)(125,50){4}
\Text(70,90)[]{$a,p_1$}
\Text(70,170)[]{$b,p_2$}
\Text(230,90)[]{$G_{\mu\nu}$}
\Text(235,170)[]{$c,\rho,p_3$}
\end{picture}
\vspace*{-2cm}
\caption{Ghost vertices}
\label{fig1}
\end{figure}

\mysection{The Model}
We work with the following action
\begin{eqnarray}
{\cal S}={\cal S}_{SM} -{\kappa \over 2} \int d^4x~
\Theta^{QCD}_{\mu \nu}(x) ~h^{\mu\nu}(x) ~,
\label{eq2}
\end{eqnarray}
where $S_{SM}$ is the Standard Model action, $h^{\mu\nu}$ is the
graviton field and $\kappa$ is the strength of the interaction.
The energy momentum tensor in QCD is given by
\begin{eqnarray}
\Theta_{\mu \nu}^{QCD}&=&-g_{\mu \nu} {\cal L}_{QCD}
-F_{\mu \rho}^a F_\nu^{a \rho}
-{1 \over \xi}g_{\mu \nu} \partial^\rho
(A_\rho^a \partial^\sigma A_\sigma^a)
\nonumber\\[2ex] &&
+{1 \over \xi}(A_\nu^a \partial_\mu(\partial^\sigma A^a_\sigma)
  +A_\mu^a \partial_\nu(\partial^\sigma A_\sigma^a))
+{i \over 4} \Big[
  \overline \psi \gamma_\mu (\overrightarrow \partial_\nu
-i g T^a A_\nu^a)\psi
\nonumber\\[2ex] &&
 -\overline \psi (\overleftarrow \partial_\nu
+i g T^a A_\nu^a)\gamma_\mu\psi
 +\overline \psi \gamma_\nu (\overrightarrow \partial_\mu
-i g T^a A_\mu^a)\psi
\nonumber\\[2ex] &&
 -\overline \psi (\overleftarrow \partial_\mu
+i g T^a A_\mu^a)\gamma_\nu\psi \Big]
+\partial_\mu \overline \omega^a (\partial_\nu \omega^a
-g f^{abc} A_\nu^c \omega^b)
\nonumber\\[2ex] &&
+\partial_\nu \overline \omega^a (\partial_\mu \omega^a
-g f^{abc} A_\mu^c \omega^b) ~.
\label{eq3}
\end{eqnarray}
In the above equation, $\xi$ is the gauge fixing parameter.
We work in the Feynman gauge in which the gauge parameter $\xi=1$.
We have displayed explicitly the ghost terms with the ghost fields
$\omega^a(x)$ since they contribute to our one-loop virtual corrections
to the process under study. The presence of the ghost fields introduces
two new vertices (see Fig. \ref{fig1}).\\
1) graviton-ghost-ghost vertex
\begin{eqnarray}
\Gamma_{\mu \nu}(p_1,p_2)=-i {\kappa \over 2} \delta^{ab}
C_{\mu \nu,\rho \sigma} ~p_1^\rho ~p_2^\sigma  ~,
\label{eq4}
\end{eqnarray}
2) graviton-ghost-ghost-gluon vertex
\begin{eqnarray}
\Gamma_{\mu \nu,\rho}(p_1,p_2)&=&{\kappa \over 2}g f^{abc}
C_{\mu \nu,\rho \sigma} ~p_2^\sigma ~,
\label{eq5}
\end{eqnarray}
where
\begin{eqnarray}
C_{\mu \nu,\rho \sigma} &=& g_{\mu\rho} g_{\nu\sigma} 
                          + g_{\mu\sigma} g_{\nu\rho}
                          - g_{\mu\nu}  g_{\rho\sigma} ~.
\label{eq6}
\end{eqnarray}
The $\mu,\nu$ indices refer to the graviton. The momenta $p_1,p_2$
and the colour indices $a,b$ correspond to the incoming and outgoing 
ghosts respectively. Finally $c, \rho, p_3$ indicate
the colour index, Lorentz index and momentum of the gluon. 
The other Feynman rules are given in Ref.~\cite{hlz}. \footnote
{The only exceptions are the Feynman rules for the 
fermion-anti-fermion-gauge 
boson-graviton vertex and the three gauge boson-graviton vertex which differ
from those of Ref.~\cite{hlz} by an overall sign.}

\mysection{Invariant lepton pair mass distribution $d\sigma/dQ^2$}
We start by considering $P_1,P_2$ scattering to leptonic final states,
say $\mu^+,\mu^-$  
\begin{eqnarray}
P_1(p_1)+P_2(p_2) \rightarrow \mu^+(l_1)+\mu^-(l_2)+X(P_X)  ~,
\label{eq7}
\end{eqnarray}
where $p_1,p_2$ are the momenta of incoming hadrons $P_1$ and $P_2$
respectively and $\mu^-,\mu^+$ are the outgoing leptons which have
the momenta $l_1,l_2$. 
The final inclusive hadronic state is denoted by $X$  
and carries the momentum $P_X$.
In the QCD improved parton model, the hadronic cross section
can be expressed in terms of partonic cross sections
convoluted with appropriate parton distribution functions as follows
\begin{eqnarray}
2 S~{d \sigma^{P_1 P_2} \over d Q^2}\left(\tau,Q^2\right)
&=&\sum_{ab={q,\overline q,g}} \int_0^1 dx_1
\int_0^1 dx_2~ f_a^{P_1}(x_1) ~
f_b^{P_2}(x_2)
\nonumber\\[2ex] &&
\times \int_0^1 dz \,\, 2 \hat s ~
{d \hat \sigma^{ab} \over d Q^2}\left(z,Q^2\right)
\delta(\tau-z x_1 x_2)\,.
\label{eq8}
\end{eqnarray}
The scaling variables are defined 
by $k_1 =x_1 p_1,k_2=x_2 p_2$ where $k_1,k_2$ are the momenta of
incoming partons.  
\begin{eqnarray}
(p_1+p_2)^2 &\equiv& S, \quad \quad \quad 
(k_1+k_2)^2 \equiv \hat s, \quad \quad \quad (l_1+l_2)^2=q.q \equiv Q^2,
\nonumber\\[2ex]
\tau&=&{Q^2 \over S}, 
\quad \quad \quad 
z={Q^2 \over \hat s }, 
\quad \quad \quad \tau=x_1 x_2 z.
\label{eq9}
\end{eqnarray}
The partonic cross section for the process
$a(k_1)+b(k_2) \rightarrow j(-q)+\displaystyle \sum_i^m X_i(-p_i)$ 
is given by
\begin{eqnarray}
2 \hat s ~ {d \hat \sigma^{ab} \over d Q^2} &=&
{1 \over 2 \pi} \sum_{jj'=\gamma,Z,G}
\int dPS_{m+1}~  |M^{ab \rightarrow jj'}|^2\cdot P_j(q)\cdot P^*_{j'}(q)\cdot 
{\cal L}^{jj' \rightarrow l^+l^-}(q)\,.
\label{eq10}
\end{eqnarray}
In the above equation, the sum over Lorentz indices between 
matrix element squared and the propagators is implicit through
a symbol ``dot product".
The $m+1$ body phase space is defined as
\begin{eqnarray}
\int dPS_{m+1}&=&\int \prod_i^m \Bigg({d^n p_i \over (2\pi)^n} 
2 \pi \delta^+(p_i^2)\Bigg) {d^nq\over (2\pi)^n} 2 \pi \delta^+(q^2-Q^2)
\nonumber\\[2ex]&&
\times (2 \pi)^n \delta^{(n)}(k_1+k_2+q+\sum_i^m p_i)\,, 
\label{eq11}
\end{eqnarray}
where $n$ is the space-time dimension.
The propagators are
\begin{eqnarray}
P_{\gamma}(q)&=&-{i \over Q^2} g_{\mu \nu}
\equiv  g_{\mu \nu} \tilde P_{\gamma}(Q^2)\,,
\nonumber\\
P_Z(q) &=& ~-{i \over 
(Q^2 -M_Z^2 - i M_Z \Gamma_Z)} g_{\mu\nu} \equiv g_{\mu\nu} \tilde P_Z(Q^2)
\label{eq12}
\end{eqnarray}
\begin{eqnarray}
P_{G}(q)&=&{\cal D}(Q^2) B_{\mu \nu \lambda \rho} (q)
\equiv  B_{\mu \nu \lambda \rho} (q)\tilde P_{G}(Q^2)\,,
\label{eq13}
\end{eqnarray}
where 
\begin{eqnarray}
B_{\mu \nu \rho \sigma}(q)&=&
\eta_{\mu \rho} \eta_{\nu \sigma}
+\eta_{\mu \sigma} \eta_{\nu \rho}
-{2 \over n-1} \eta_{\mu \nu} \eta_{\rho \sigma} ~,
\nonumber \\[2ex]
\eta_{\mu \nu}(q)&=&-g_{\mu \nu} + {q_\mu q_\nu \over Q^2} ~.
\label{eq14}
\end{eqnarray}
The summation of the virtual KK modes in the time-like propagators \cite{hlz}
leads to 
\begin{eqnarray}
{\cal D}(Q^2)= {Q^{d-2} R^d \over \Gamma(d/2) (4 \pi)^{d/2}} ~2~
I\left({M_S \over Q}\right) ~,
\end{eqnarray}
where the integral $I$ is regulated by an ultraviolet cutoff, presumably of 
the order of $M_S$ \cite{grw,hlz}, which sets the limit on the applicability 
of the effective theory.  For the DY case under consideration this consistency
would imply $Q<M_S$.  Further relating the gravitational coupling, the volume 
of extra dimension and the cutoff scale \cite{hlz}
\begin{eqnarray}
\kappa^2 R^d = 8 \pi (4 \pi)^{d/2} \Gamma(d/2) M_S^{-(d+2)}~,
\end{eqnarray}
we express the function ${\cal D}(Q^2)$ as
\begin{eqnarray}
{\cal D}(Q^2)=16 \pi~ \left({Q^{d-2} \over \kappa^2 M_S^{d+2}}\right)~
I\left({M_S \over Q}\right) ~.
\label{eq15}
\end{eqnarray}
The summation over the non-resonant KK modes yields
\begin{eqnarray}
I(\omega)&=&- \sum_{k=1}^{d/2-1} {1 \over 2 k} \omega^{2 k}
 -{1 \over 2} \log(\omega^2-1)\,, \qquad \qquad \qquad  d={\rm even}\,,
\label{eq16}
\end{eqnarray}
\begin{eqnarray}
I(\omega)&=&
-\sum_{k=1}^{(d-1)/2} {1\over 2 k-1} \omega^{2 k-1}
 +{1 \over 2} \log \left({\omega+1}\over{\omega-1}\right) \,,
\quad \quad  d={\rm odd}\,,
\label{eq17}
\end{eqnarray}
where $\omega=M_S/Q$.
The leptonic tensor involves the computation of square of
the matrix element for the process $\gamma/Z/G \rightarrow \mu^+ + \mu^- $ 
and the 2-body phase space integral
\begin{eqnarray}
{\cal L}^{jj'\rightarrow \mu^+\mu^-}(q)&=&
\int \prod_{i=1}^2 \Bigg({d^nl_i\over (2 \pi)^n}
2 \pi \delta^+(l_i^2) \Bigg)
(2 \pi)^n \delta^{(n)}(q -l_1-l_2) |M^{jj' \rightarrow \mu^+\mu^-}|^2\,.
\nonumber\\[2ex]
\label{eq18}
\end{eqnarray}
The leptonic part is easy to compute. It is equal to
\begin{eqnarray}
{\cal L}^{jj'\rightarrow \mu^+\mu^-}(q)&=&\eta_{\mu \nu}(q) L_{jj'}(Q^2) \,,
\quad \quad \quad jj'=\{\gamma\gamma,ZZ,\gamma Z\}\,,
\nonumber\\[2ex]
{\cal L}^{GG\rightarrow \mu^+\mu^-}(q)&=&B_{\mu \nu \rho \sigma}(q) L_{G}(Q^2) 
\,,
\label{eq19}
\end{eqnarray}
with 
\begin{eqnarray}
L_{\gamma\gamma}(Q^2)&=& Q^2 {2 \alpha \over 3}\,,
\qquad \qquad \qquad \qquad \qquad \qquad
L_{\gamma Z}(Q^2)= -Q^2 {2 \alpha g_e^V\over 3 c_w s_w}\,,
\nonumber\\[2ex]
L_{Z Z}(Q^2)&=& Q^2 {2 \alpha \over 3 c_w^2 s_w^2}
\Bigg(\left(g_e^V\right)^2 +\left(g_e^A\right)^2\Bigg)\,,
\quad \quad \quad
L_{GG}(Q^2)= Q^4 {\kappa^2\over 640 \pi}\,,
\label{eq20}
\end{eqnarray}
where $\alpha$ is the fine structure constant
\begin{eqnarray}
c_w &=& \cos \theta_W ~, \qquad \qquad s_w=\sin \theta_W ~,
\nonumber \\[2ex]
g_a^V &=& {1 \over 2 } T^3_a - s_w^2 Q_a ~,
\qquad
g_a^A=-{1 \over 2 } T^3_a ~,
\label{21}
\end{eqnarray}
and $Q_a$ is the electric charge of quarks and leptons.
Hence, we obtain
\begin{eqnarray}
2 S {d \sigma^{P_1P_2} \over d Q^2}(\tau,Q^2) &=&
{1 \over 2 \pi} \sum_{jj'=\gamma,Z,G}
\tilde P_j(Q^2)~ \tilde P_{j'}^*(Q^2)~ L_{jj'}(Q^2)~ W_{jj'}^{P_1 P_2}
(\tau,Q^2)\,.
\label{eq22}
\end{eqnarray}
The hadronic structure function is equal to
\begin{eqnarray}
W_{jj'}^{P_1P_2}(\tau,Q^2)&=&\sum_{ab,jj'}
\int dx_1 \int dx_2 ~
f_a^{P_1}(x_1) ~ 
f_b^{P_2}(x_2)
\nonumber\\[2ex]
&&\times \int dz ~
\delta(\tau-z x_1 x_2)~
\int dPS_{m+1}~ |M^{ab \rightarrow jj'}|^2~ T_{jj'}(q)~,
\label{eq23}
\end{eqnarray}
with 
\begin{eqnarray}
T_{jj'}&=&\eta_{\mu\nu}(q)\,, \quad \quad \quad jj'={\gamma\gamma,
\gamma Z,ZZ}\,,
\nonumber\\[2ex]
T_{GG}&=&B_{\mu\nu \rho \sigma}(q) \,.
\label{eq24}
\end{eqnarray}
The tensors $\eta_{\mu\nu}(q)$ and $B_{\mu\nu \rho \sigma}(q)$ are defined
in Eq. (\ref{eq14}).
To compute the $Q^2$ distribution of the di-lepton pairs, the matrix 
element
squared $|M^{ab \rightarrow jj'}|^2~ T_{jj'}(q)$ has to be substituted in 
Eq.(\ref{eq23}) provided the integrations over
$dPS_{m+1}$ and $dz$ are performed in a suitable frame.
We define the bare partonic coefficient function 
$\bar \Delta_{ab}^{jj'}(z,Q^2)$ as
\begin{eqnarray}
\bar \Delta_{ab}^{jj'}(z,Q^2)= C_{jj'}
\int dPS_{m+1}~ |M^{ab \rightarrow jj'}|^2~ T_{jj'}(q)~,
\label{eq25}
\end{eqnarray}
where 
\begin{eqnarray}
C_{jj'} &=& {1 \over e^2} \quad \quad \quad jj'=\gamma\gamma,ZZ,\gamma Z,
\nonumber\\
        &=& {1 \over Q^2 \kappa^2} \quad \quad jj'= GG.
\end{eqnarray}
There are two classes of processes that contribute
to the partonic cross section. The first one has only
a virtual photon or a Z boson whereas the second
one only contains a graviton in the intermediate state. Interestingly, the
interference between first and second class diagrams identically vanish
when the phase space integration is performed.   
In the case of the photon exchange process, it is easy to understand if 
we realise that there is no third rank tensor, say $S^{\mu \nu \rho}$
which can be constructed out of $g_{\mu \nu}$ and $q_\rho$ ($q.q\not=0$)
(which are the only tensor and vector available in our problem)
that satisfies $q_\mu S^{\mu \nu \rho}=0$ (as it should be for the theory
where we have gravity coupled to a conserved energy momentum tensor).
We found a similar argument for the Levi-Civita tensor which shows 
up in the case of the electro-weak vertices. Therefore also here
there are no interference terms.  
We could also verify this by explicit computation.
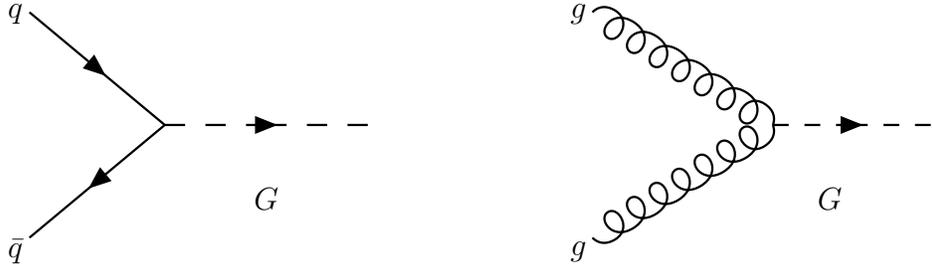
\begin{figure}[t]
\SetScale{1.7}
\noindent
\begin{picture}(150,150)(0,0)
\ArrowLine(50,100)(80,75)
\ArrowLine(80,75)(50,50) 
\DashArrowLine(80,75)(125,75){4}
\Text(80,80)[]{$\bar q$}
\Text(80,170)[]{$q$}
\Text(175,100)[]{$G$}
\end{picture}
\hspace*{1.8cm}
\SetScale{1.7}
\noindent
\begin{picture}(150,150)(0,0)
\Gluon(50,100)(90,75){3}{7}
\Gluon(90,75)(50,50){3}{7}
\DashArrowLine(90,75)(125,75){4}
\Text(80,80)[]{$g$}
\Text(80,170)[]{$g$}
\Text(175,100)[]{$G$}
\end{picture}
\vspace*{-2cm}
\caption{Born contributions:} 
\label{fig2}
\end{figure}
At LO the following process contribute when a photon or Z-boson 
appears in the intermediate state
\begin{eqnarray}
q + \bar q \rightarrow \gamma^*/Z^* \,.
\label{eq26}
\end{eqnarray}
In the case the Graviton appears in the intermediate state, we have two 
processes at LO (see Fig. \ref{fig2})
\begin{eqnarray}
q + \bar q \rightarrow G^*\,,  \quad \quad \quad
g + g \rightarrow G^* \,.
\label{eq27}
\end{eqnarray}
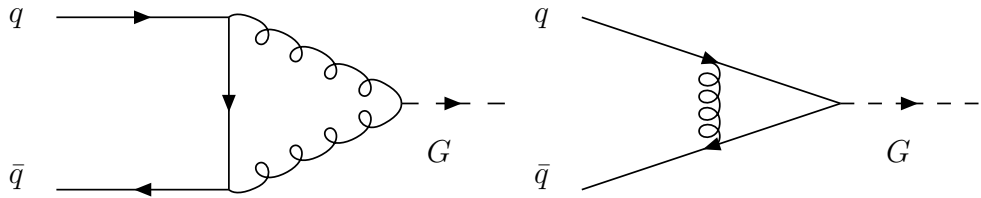
\begin{figure}[b]
\SetScale{1.3}
\noindent
\begin{picture}(150,150)(0,0)
\ArrowLine(50,100)(100,100)
\ArrowLine(100,100)(100,50)
\ArrowLine(100,50)(50,50)
\Gluon(100,100)(150,75){3}{4}
\Gluon(150,75)(100,50){3}{4}
\DashArrowLine(150,75)(180,75){4} 
\Text(50,70)[]{$\bar q$}
\Text(50,130)[]{$q$}
\Text(210,80)[]{$G$}
\end{picture}
\hspace*{1.3cm}
\SetScale{1.3}
\noindent
\begin{picture}(150,150)(0,0)
\ArrowLine(50,100)(125,75)
\ArrowLine(125,75)(50,50)
\Gluon(87,87)(87,62){3}{4}
\DashArrowLine(125,75)(165,75){4} 
\Text(50,70)[]{$\bar q$}
\Text(50,130)[]{$q$}
\Text(185,80)[]{$G$}
\end{picture}
\vspace*{-2cm}
\caption{Virtual corrections } 
\label{fig3}
\end{figure}
\begin{figure}[hbt]
\SetScale{1.2}
\noindent
\begin{picture}(150,150)(0,0)
\Gluon(50,100)(100,75){3}{4}
\Gluon(100,75)(50,50){3}{4}
\GlueArc(125,75)(25,0,180){3}{7}
\GlueArc(125,75)(25,180,360){3}{7}
\DashArrowLine(150,75)(180,75){4}
\Text(45,55)[]{$g$}
\Text(45,135)[]{$g$}
\Text(215,110)[]{$G$}
\end{picture}
\hspace*{1.3cm}
\SetScale{1.2}
\begin{picture}(150,150)(0,0)
\Gluon(50,100)(100,100){3}{4}
\Gluon(100,50)(100,100){3}{4}
\Gluon(100,50)(50,50){3}{4}
\Gluon(150,75)(100,50){3}{4}
\Gluon(100,100)(150,75){3}{4}
\DashArrowLine(150,75)(180,75){4}
\Text(45,55)[]{$g$}
\Text(45,135)[]{$g$}
\Text(215,110)[]{$G$}
\end{picture}
\vspace*{-2cm}
\caption{Virtual corrections, gluon loops.} 
\label{fig4}
\end{figure}
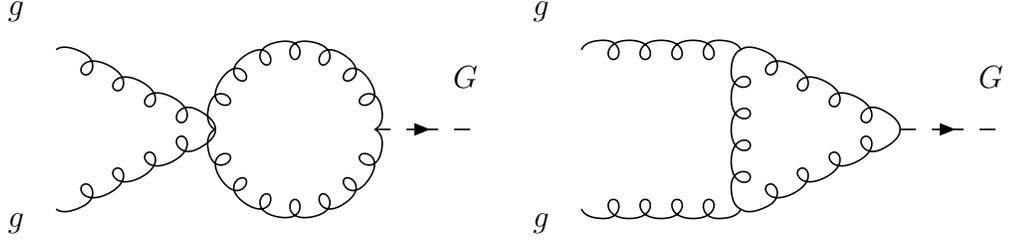
In NLO for the photon or Z-boson we have the 
following $2 \rightarrow 2$ reactions \cite{alel,hune,kupa}
\begin{eqnarray}
&&q + \bar q \rightarrow \gamma^*/Z^* + q\,, \quad \quad q+\bar q 
\rightarrow \gamma^*/Z^* + \mbox{one~~loop}\,,
\nonumber \\[2ex]
&&q + g \rightarrow \gamma^*/Z^* + q\,,
\qquad \bar q + g \rightarrow \gamma^*/Z^* + \bar q\,,
\label{eq28}
\end{eqnarray}
and for the graviton we have (see Figs. \ref{fig3}-\ref{fig9})
\begin{eqnarray}
&&q + \bar q \rightarrow G  + q\,, \quad \quad q+\bar q \rightarrow G + 
\mbox{one~~loop}\,,
\nonumber \\[2ex]
&&q + g \rightarrow G  + q\,,
\qquad \bar q + g \rightarrow G  + \bar q\,,
\nonumber \\[2ex]
&&g + g \rightarrow G + g \,, \quad \quad g+ g \rightarrow G+ 
\mbox{one~~loop}\,.
\label{eq29}
\end{eqnarray}
The cross sections beyond leading order involve the computation of
one loop virtual gluon corrections and real gluon bremsstrahlung 
contributions to leading order processes. We also include processes with
a gluon in the initial state. Since we are dealing with energy momentum
tensor coupled to gravity expressed in terms of renormalised fields and
masses, there is no overall ultraviolet renormalisation required.
In other words, the operator renormalisation constant for the
energy momentum operator is identical to unity to all orders in perturbation
theory.  But we encounter soft and collinear divergences
in our computation. We have used $n$-dimensional regularisation to
regulate both these divergences. To that order, we have defined 
$n=4+\varepsilon$ where $n$ is the space-time dimension.
With this procedure all divergences
appear as $1/\varepsilon^\alpha$ where $\alpha=1,2$.
The soft divergences coming from
virtual gluons and bremsstrahlung contributions cancel exactly according
to the Bloch-Nordsieck theorem.
The remaining collinear divergences are removed by mass factorisation 
which in our paper is performed in $\overline {MS}$ scheme.
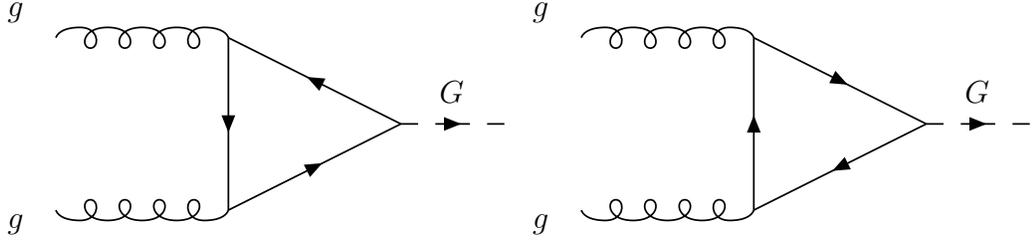
\begin{figure}[t]
\SetScale{1.3}
\noindent
\vspace{-0.5cm}
\begin{picture}(150,150)(0,0)
\Gluon(50,100)(100,100){3}{4}
\ArrowLine(100,100)(100,50)
\Gluon(100,50)(50,50){3}{4}
\ArrowLine(100,50)(150,75)
\ArrowLine(150,75)(100,100)
\DashArrowLine(150,75)(180,75){4}
\Text(50,60)[]{$g$}
\Text(50,140)[]{$g$}
\Text(215,110)[]{$G$}
\end{picture}
\hspace*{1.3cm}
\SetScale{1.3}
\begin{picture}(150,150)(0,0)
\Gluon(50,100)(100,100){3}{4}
\ArrowLine(100,50)(100,100)
\Gluon(100,50)(50,50){3}{4}
\ArrowLine(150,75)(100,50)
\ArrowLine(100,100)(150,75)
\DashArrowLine(150,75)(180,75){4}
\Text(50,60)[]{$g$}
\Text(50,140)[]{$g$}
\Text(215,110)[]{$G$}
\end{picture}
\vspace{-1.0cm}
\caption{Virtual corrections, quark loops.}
\label{fig5}
\end{figure}
\begin{figure}[b]
\SetScale{1.3}
\noindent
\begin{picture}(150,150)(0,0)
\Gluon(50,100)(100,100){3}{4}
\DashArrowLine(100,100)(100,50){2}
\Gluon(100,50)(50,50){3}{4}
\DashArrowLine(100,50)(150,75){2}
\DashArrowLine(150,75)(100,100){2}
\DashArrowLine(150,75)(180,75){4}
\Text(50,60)[]{$g$}
\Text(50,140)[]{$g$}
\Text(215,110)[]{$G$}
\end{picture}
\hspace*{1.3cm}
\SetScale{1.3}
\begin{picture}(150,150)(0,0)
\Gluon(50,100)(100,100){3}{4}
\DashArrowLine(100,50)(100,100){2}
\Gluon(100,50)(50,50){3}{4}
\DashArrowLine(150,75)(100,50){2}
\DashArrowLine(100,100)(150,75){2}
\DashArrowLine(150,75)(180,75){4}
\Text(50,60)[]{$g$}
\Text(50,140)[]{$g$}
\Text(215,110)[]{$G$}
\end{picture}
\vspace*{-2cm}
\caption{Virtual corrections, ghost loops.} 
\label{fig6}
\end{figure}
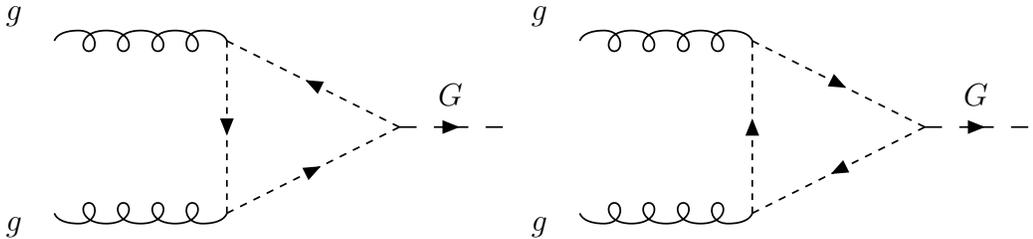
\begin{figure}[t]
\SetScale{1.7}
\noindent
\begin{picture}(150,150)(0,0)
\ArrowLine(50,100)(100,75)
\ArrowLine(100,75)(50,50)
\Gluon(75,87)(125,87){3}{7}
\DashArrowLine(100,75)(125,75){4}
\Text(80,80)[]{$\bar q$}
\Text(80,170)[]{$q$}
\Text(185,115)[]{$G$}
\Text(185,160)[]{$g$}
\end{picture}
\hspace*{1.3cm}
\SetScale{1.7}
\begin{picture}(150,150)(0,0)
\ArrowLine(50,100)(100,75)
\ArrowLine(100,75)(50,50)
\Gluon(75,62)(125,62){3}{7}
\DashArrowLine(100,75)(125,75){4}
\Text(80,80)[]{$\bar q$}
\Text(80,170)[]{$q$}
\Text(195,140)[]{$G$}
\Text(195,90)[]{$g$}
\end{picture}
\\
\SetScale{1.7}
\noindent
\begin{picture}(150,150)(0,0)
\ArrowLine(50,100)(80,75)
\ArrowLine(80,75)(50,50)
\Gluon(80,75)(110,100){3}{4}
\DashArrowLine(80,75)(110,50){4}
\Text(80,80)[]{$\bar q$}
\Text(80,170)[]{$q$}
\Text(165,90)[]{$G$}
\Text(165,175)[]{$g$}
\end{picture}
\hspace*{1.3cm}
\SetScale{1.7}
\begin{picture}(150,150)(0,0)
\ArrowLine(50,100)(75,75)
\ArrowLine(75,75)(50,50)
\Gluon(75,75)(100,75){3}{3}
\DashArrowLine(100,75)(125,50){3}
\Gluon(100,75)(125,100){3}{4}
\Text(80,80)[]{$\bar q$}
\Text(80,170)[]{$q$}
\Text(190,90)[]{$G$}
\Text(195,180)[]{$g$}
\end{picture}
\vspace{-2cm}
\caption{Real emission, $q ~\bar q \rightarrow g ~G$.} 
\label{fig7}
\end{figure}
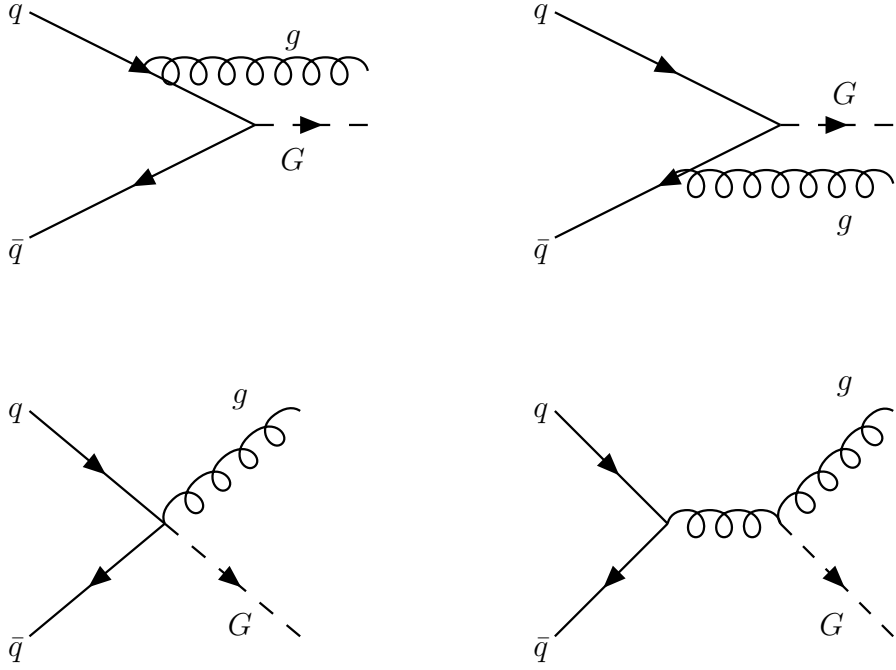
The Drell-Yan coefficient function after mass factorisation,
denoted by $\Delta_{ab}^i(z,Q^2,\mu_F^2)$, is computed by
\begin{eqnarray}
\bar \Delta_{ab}^i(z,Q^2,1/\varepsilon)=\sum_{c,d}
\Gamma_{ca}(z,\mu_F^2,1/\varepsilon) 
\otimes \Gamma_{db}(z,\mu_F^2,1/\varepsilon) \otimes 
\Delta_{cd}^i(z,Q^2,\mu_F^2)\,,
\label{eq30}
\end{eqnarray}
where $\bar \Delta_{ab}^i(z,Q^2,1/\varepsilon)$ is the bare partonic
coefficient function before mass factorisation is carried out. Further
we have dropped the double index $jj'$ because of the vanishing 
interference terms and replace it by the single index $i$ instead.
The factorisation scale is given by $\mu_F$ and 
$\otimes$ is the convolution symbol defined as
\begin{eqnarray}
f(z)\otimes g(z)= \int_z^1~{dy \over y} f(y)\,g\left({z \over y}\right)\,,
\label{eq31}
\end{eqnarray}
and the kernel
$\Gamma_{cd}(z)$ in the ${\overline {MS}}$ scheme is given by
\begin{eqnarray}
\Gamma_{cd}(z,\mu_F)&=&\delta_{c d} \delta(1-z)+a_s {1 \over \varepsilon} 
\Gamma_{c d}^{(1)}(z,\mu_F)
\nonumber\\
&=&\delta_{c d} \delta(1-z)+ a_s {1 \over \varepsilon}
\left({\mu_F^2 \over \mu^2}\right)^
{\varepsilon/2} P_{cd}^{(0)}(z)\,,
\label{eq32}
\end{eqnarray}
where $P_{cd}^{(0)}(z)$ are the leading order Altarelli-Parisi 
splitting functions \cite{alpa},
\begin{eqnarray}
\Delta_{ab}^i&=&\Delta^{(0),i}_{ab} +
a_s\, \Delta_{ab}^{(1),i}\,.
\label{eq33}
\end{eqnarray}
For convenience we define $a_s$ as
\begin{eqnarray}
a_s={\alpha_s(\mu_R^2) \over 4 \pi}\,.
\label{eq34}
\end{eqnarray}
Expanding Eq. (\ref{eq30}) up to order $a_s$, we find
\begin{eqnarray}
\bar \Delta_{q \bar q}^{\gamma/Z}&=&\Delta^{(0)\gamma/Z}_
{q \bar q}+a_s {2 \over \varepsilon} \Gamma_{qq}^{(1)} \otimes 
\Delta^{(0)\gamma/Z}_{q \bar q} +a_s \Delta^{(1)\gamma/Z}_{q \bar q}\,,
\nonumber\\[2ex]
\bar \Delta_{q g}^{\gamma/Z}&=&
a_s {1 \over \varepsilon} \Gamma_{qg}^{(1)} \otimes \Delta^{(0)\gamma/Z}_
{q \bar q} +a_s \Delta^{(1)\gamma/Z}_{q g}\,,
\nonumber\\[2ex]
\bar \Delta_{q \bar q}^G&=&\Delta^{(0)G}_{q \bar q}
+a_s {2 \over \varepsilon} \Gamma_{qq}^{(1)} \otimes \Delta^{(0)G}_{q \bar q}
+a_s \Delta^{(1)G}_{q \bar q}\,,
\nonumber\\[2ex]
\bar \Delta_{q g}^G&=&
a_s {1 \over \varepsilon}\left( \Gamma_{qg}^{(1)} \otimes 
\Delta^{(0)G}_{q \bar q}
+\Gamma_{gq}^{(1)} \otimes \Delta^{(0)G}_{g g} \right)
+a_s \Delta^{(1)G}_{q g}\,,
\nonumber\\[2ex]
\bar \Delta_{g g}^G&=&\Delta^{(0)G}_{g g}
+a_s {2 \over \varepsilon} \Gamma_{gg}^{(1)} \otimes \Delta^{(0)G}_{g g}
+a_s \Delta^{(1)G}_{g g}\,.
\label{eq35}
\end{eqnarray}
%
%
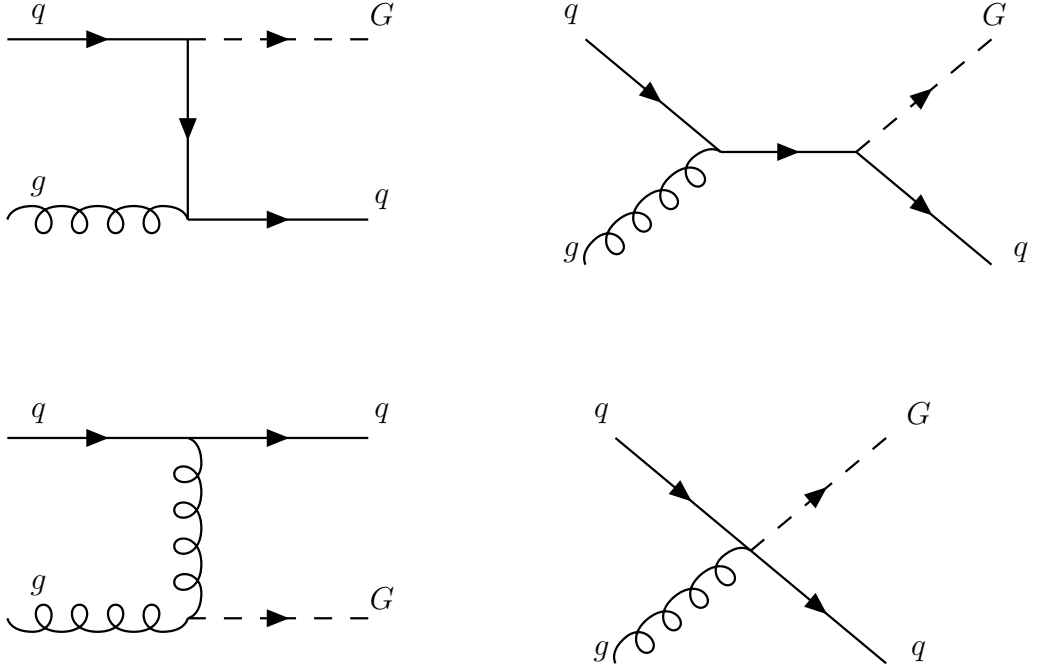
\begin{figure}[hbt]
\SetScale{1.7}
\noindent
\begin{picture}(150,150)(0,0)
\ArrowLine(40,100)(80,100)
\DashArrowLine(80,100)(120,100){4}
\ArrowLine(80,100)(80,60)
\ArrowLine(80,60)(120,60)
\Gluon(40,60)(80,60){3}{4}
\Text(210,180)[]{$G$}
\Text(80,180)[]{$q$}
\Text(210,110)[]{$q$}
\Text(80,115)[]{$g$}
\end{picture}
\hspace*{1.4cm}
\SetScale{1.7}
\begin{picture}(150,150)(0,0)
\ArrowLine(50,100)(80,75)
\ArrowLine(80,75)(110,75)
\ArrowLine(110,75)(140,50)
\Gluon(50,50)(80,75){3}{4}
\DashArrowLine(110,75)(140,100){4}
\Text(80,180)[]{$q$}
\Text(240,180)[]{$G$}
\Text(250,90)[]{$q$}
\Text(80,90)[]{$g$}
\end{picture}
\\
\SetScale{1.7}
\begin{picture}(150,150)(0,0)
\ArrowLine(40,100)(80,100)
\ArrowLine(80,100)(120,100)
\Gluon(80,100)(80,60){3}{4}
\Gluon(80,60)(40,60){3}{4}
\DashArrowLine(80,60)(120,60){4}
\Text(210,180)[]{$q$}
\Text(80,180)[]{$q$}
\Text(210,110)[]{$G$}
\Text(80,115)[]{$g$}
\end{picture}
\hspace*{1.8cm}
\SetScale{1.7}
\begin{picture}(150,150)(0,0)
\ArrowLine(50,100)(80,75)
\ArrowLine(80,75)(110,50)
\Gluon(50,50)(80,75){3}{4}
\DashArrowLine(80,75)(110,100){4}
\Text(200,180)[]{$G$}
\Text(80,180)[]{$q$}
\Text(200,90)[]{$q$}
\Text(80,90)[]{$g$}
\end{picture}
\vspace*{-2cm}
\caption{Real emission, $q ~g \rightarrow q ~G$.}
\label{fig8}
\end{figure}
From the above expressions one can compute the coefficient function
$\Delta_{ab}^i(z,Q^2,\mu_F)$ from the bare
$\overline \Delta_{ab}^i(z,Q^2,\mu_F,1/\varepsilon)$ and the known 
Altarelli-Parisi
kernels $P_{ab}^{(0)}$. Finally we have to fold these finite 
$\Delta_{ab}^i(z,Q^2,\mu_F)$
with the appropriate partonic distribution functions to arrive at the $Q^2$ 
distribution 
for the DY pair. For completeness we present the results below
\vspace{0.5cm}
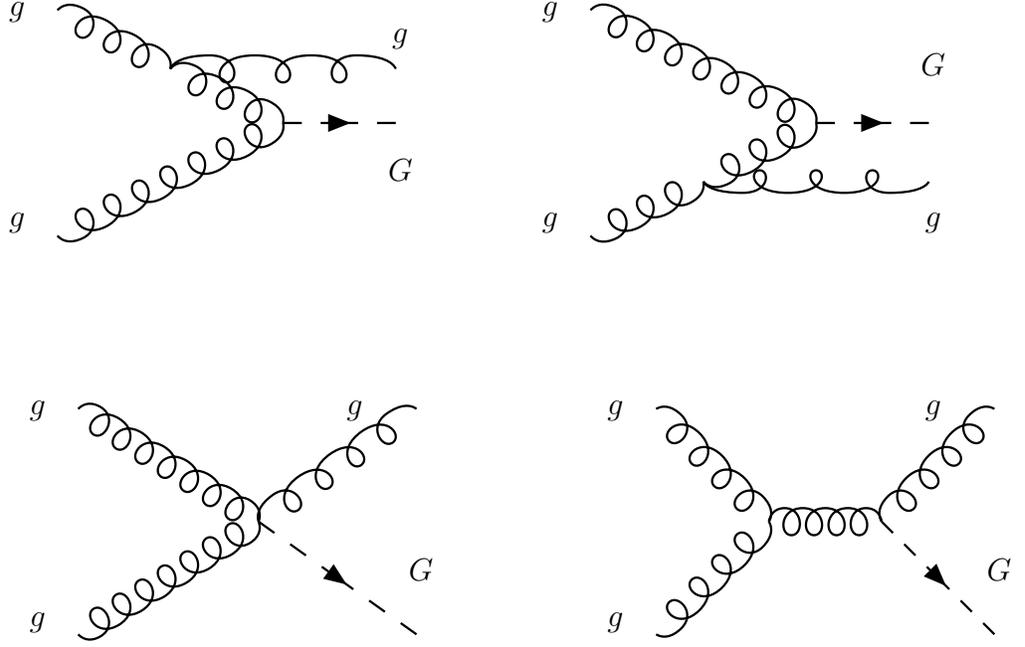
\begin{figure}[ht]
\SetScale{1.7}
\noindent
\begin{picture}(150,150)(0,0)
\Gluon(50,100)(75,87){3}{3}
\Gluon(75,87)(100,75){3}{3}
\Gluon(100,75)(50,50){3}{7}
\Gluon(75,87)(125,87){3}{3}
\DashArrowLine(100,75)(125,75){4}
\Text(70,90)[]{$g$}
\Text(70,170)[]{$g$}
\Text(215,110)[]{$G$}
\Text(215,160)[]{$g$}
\end{picture}
\hspace*{1.4cm}
\SetScale{1.7}
\begin{picture}(150,150)(0,0)
\Gluon(50,100)(100,75){3}{7}
\Gluon(100,75)(75,62){3}{3}
\Gluon(75,62)(50,50){3}{3}
\Gluon(125,62)(75,62){2.5}{3}
\DashArrowLine(100,75)(125,75){4}
\Text(70,90)[]{$g$}
\Text(70,170)[]{$g$}
\Text(215,150)[]{$G$}
\Text(215,90)[]{$g$}
\end{picture}
%
\\
\vspace*{1cm}
\SetScale{1.7}
\noindent
\begin{picture}(150,150)(0,0)
\Gluon(50,100)(90,75){3}{7}
\Gluon(90,75)(50,50){3}{7}
\Gluon(90,75)(125,100){3}{4}
\DashArrowLine(90,75)(125,50){4}
\Text(70,90)[]{$g$}
\Text(70,170)[]{$g$}
\Text(215,110)[]{$G$}
\Text(190,170)[]{$g$}
\end{picture}
\hspace*{2cm}
\SetScale{1.7}
\begin{picture}(150,150)(0,0)
\Gluon(50,100)(75,75){3}{4}
\Gluon(75,75)(50,50){3}{4}
\Gluon(75,75)(100,75){3}{4}
\DashArrowLine(100,75)(125,50){4}
\Gluon(100,75)(125,100){3}{4}
\Text(70,90)[]{$g$}
\Text(70,170)[]{$g$}
\Text(215,110)[]{$G$}
\Text(190,170)[]{$g$}
\end{picture}
\vspace{-3.0cm}
\caption{Real emission, $g ~g \to g ~G$.} 
\label{fig9}
\end{figure}
\begin{eqnarray}
2 S~{d \sigma^{P_1P_2} \over dQ^2}(\tau,Q^2)&=&
\sum_q{\cal F}_{SM,q} \int_0^1~ {d x_1}~ \int_0^1 
~{dx_2}~ \int_0^1~ dz~ \delta(\tau-z x_1 x_2)
\nonumber\\[2ex]&&
\times \Bigg[ H_{q \bar q}(x_1,x_2,\mu_F^2) \Big(
\Delta_{q \bar q}^{(0),\gamma/Z}(z,Q^2,\mu_F^2)
 +a_s \Delta_{q \bar q}^{(1),\gamma/Z}(z,Q^2,\mu_F^2)\Big)
\nonumber\\[2ex] &&
+\Big( H_{q g}(x_1,x_2,\mu_F^2)+H_{g q}(x_1,x_2,\mu_F^2)\Big)
 a_s \Delta_{q g}^{(1),\gamma/Z}(z,\mu_F^2) \Bigg]
\nonumber\\[2ex]
&&+\sum_q{\cal F}_{G} \int_0^1~ {d x_1 }~ \int_0^1 
~{dx_2}~ \int_0^1~ dz~ \delta(\tau-z x_1 x_2)
\nonumber\\[2ex]&&
\times \Bigg[ H_{q \bar q}(x_1,x_2,\mu_F^2) \Big(
\Delta_{q \bar q}^{(0),G}(z,Q^2,\mu_F^2)
 +a_s \Delta_{q \bar q}^{(1),G}(z,Q^2,\mu_F^2)\Big)
\nonumber\\[2ex] &&
+\Big( H_{q g}(x_1,x_2,\mu_F^2)+H_{g q}(x_1,x_2,\mu_F^2)\Big)
 a_s \Delta_{q g}^{(1),G}(z,Q^2,\mu_F^2) 
\nonumber\\[2ex]&&
+ H_{g g}(x_1,x_2,\mu_F^2) \Big(
\Delta_{g g}^{(0),G}(z,Q^2,\mu_F^2)
 +a_s \Delta_{g g}^{(1),G}(z,Q^2,\mu_F^2)\Big) \Bigg]\,.
\nonumber\\[2ex]
\label{eq36}
\end{eqnarray}
The constants ${\cal F}_{SM,q},{\cal F}_{G}$ are given by
\begin{eqnarray}
{\cal F}_{SM,q}&=&{4 \alpha^2 \over 3 Q^2} \Bigg[Q_q^2
-{2 Q^2 (Q^2-M_Z^2) \over  
\left((Q^2-M_Z^2)^2+M_Z^2 \Gamma_Z^2\right) c_w^2 s_w^2}
Q_q g_e^V g_q^V 
\nonumber\\[2ex]
&&+{Q^4 \over  \left((Q^2-M_Z^2)^2+M_Z^2 \Gamma_Z^2\right)
c_w^4 s_w^4}\Big((g_e^V)^2+(g_e^A)^2\Big)\Big((g_q^V)^2+(g_q^A)^2\Big)
\Bigg]\,,
\nonumber\\[2ex]
\label{eq37}
\\[2ex]
{\cal F}_{G}&=&{\kappa^4 Q^6 \over 320 \pi^2 }|{\cal D}(Q^2)|^2\,.
\label{eq38}
\end{eqnarray}
The renormalised partonic distributions are 
\begin{eqnarray}
H_{q \bar q}(x_1,x_2,\mu_F^2)&=&
f_q^{P_1}(x_1,\mu_F^2)~ 
f_{\bar q}^{P_2}(x_2,\mu_F^2)
+f_{\bar q}^{P_1}(x_1,\mu_F^2)~ 
f_q^{P_2}(x_2,\mu_F^2)\,,
\nonumber
\\[2ex]
H_{g q}(x_1,x_2,\mu_F^2)&=&
f_g^{P_1}(x_1,\mu_F^2) ~
\Big(f_q^{P_2}(x_2,\mu_F^2)
+f_{\bar q}^{P_2}(x_2,\mu_F^2)\Big)\,,
\nonumber
\\[2ex]
H_{q g}(x_1,x_2,\mu_F^2)&=&
H_{g q}(x_2,x_1,\mu_F^2)\,,
\nonumber
\\[2ex]
H_{g g}(x_1,x_2,\mu_F^2)&=&
f_g^{P_1}(x_1,\mu_F^2)~ 
f_g^{P_2}(x_2,\mu_F^2)\,.
\label{eq39}
\end{eqnarray}
The LO coefficient functions are
\begin{eqnarray}
\Delta^{(0),\gamma/Z}_{q\bar q}&=&{2 \pi \over N} \delta(1-z)\,,
\nonumber \\[2ex]
\Delta^{(0),G}_{q \bar q}&=&{\pi \over 8 N} \delta(1-z)\,,
\nonumber \\[2ex]
\Delta^{(0),G}_{gg}&=&{\pi \over 2 (N^2-1)}  \delta(1-z)\,,
\label{eq40}
\end{eqnarray}
and the NLO results are given by
\begin{eqnarray}
\Delta^{(1)\gamma/Z}_{q \bar q}&=&\left({2 \pi \over N}\right)
4~ C_F
\Bigg[ \Big(-4+2 \zeta(2)\Big)\delta(1-z)+
2 {1 \over (1-z)_+} \ln{\left(Q^2 \over \mu_F^2\right)}
\nonumber \\[2ex] &&
+4 \left({\ln(1-z) \over 1-z}\right)_+
+{3 \over 2} \delta(1-z)
\ln\left({Q^2 \over \mu_F^2}\right)
-(1+z) \ln\left({Q^2 (1-z)^2 \over \mu_F^2 z}\right)
\nonumber \\[2ex]&&
-2 {\ln(z) \over 1-z} \Bigg]\,,
\nonumber \\[2ex]
\Delta^{(1)\gamma/Z}_{q (\bar q)g}&=&\left({2 \pi \over N}\right) T_F
\Bigg[ 2 (1-2 z+2 z^2) \ln\left({Q^2 (1-z)^2 \over \mu_F^2 z}\right)
+1+6 z-7 z^2 \Bigg]\,,
\nonumber \\[2ex]
\Delta^{(1)G}_{q \bar q}&=&\left({\pi \over 8 N}\right) 4 C_F
\Bigg[ \Big(-5+2 \zeta(2)\Big)\delta(1-z)+
2 {1 \over (1-z)_+} \ln\left({Q^2 \over \mu_F^2}\right)
\nonumber \\[2ex]
&&
+4 \left({\ln(1-z) \over 1-z}\right)_+
+{3 \over 2} \delta(1-z)
\ln\left({Q^2 \over \mu_F^2}\right)
-(1+z) \ln\left({Q^2 (1-z)^2 \over \mu_F^2 z}\right)
\nonumber \\[2ex] &&
-2 {\ln(z) \over 1-z}+ {4 \over 3 z} - {4 z^2 \over 3}\Bigg]\,,
\nonumber \\[2ex]
\Delta^{(1)G}_{q (\bar q) g}&=&\left({\pi \over 8 N}\right)  T_F
\Bigg[ 2 (-7+{8 \over z} +2 z +2 z^2) \ln\left({Q^2 (1-z)^2 \over 
\mu_F^2 z}\right)
\nonumber \\[2ex]
&&+9-{12 \over z} +18 z-7 z^2 \Bigg]\,,
\nonumber \\[2ex]
\Delta^{(1)G}_{g g}&=&\left({\pi \over 2 (N^2-1)}\right) C_A
\Bigg[ \Big(-{203 \over 9}+8 \zeta(2)\Big)\delta(1-z)+
8 {1 \over (1-z)_+} \ln\left({Q^2 \over \mu_F^2}\right)
\nonumber \\[2ex]
&&+16 \left({\ln(1-z) \over 1-z}\right)_+
+{22 \over 3} \delta(1-z)
\ln\left({Q^2 \over \mu_F^2}\right)
+8 (-2+{1\over z}+z-z^2) 
\nonumber \\[2ex]
&&\times \ln\left({Q^2 (1-z)^2 \over \mu_F^2 z}\right)
-8 {\ln(z) \over (1-z)}
-2-{22 \over 3 z}+2 z+{22 z^2 \over 3}\Bigg]
\nonumber\\[2ex]
&&+\left({\pi \over 2 (N^2-1)}\right) n_f T_F \Bigg[\left({70 \over 9}-
{8 \over 3} \ln\left({Q^2 \over \mu_F^2}\right) \right) \delta(1-z)\Bigg]\,.
\label{eq41}
\end{eqnarray}
For $SU(N)$ the colour factors
in the above equations are
\begin{eqnarray}
C_F={N^2 -1 \over 2 N}\,, \quad \quad C_A=N, \quad \quad T_F=1/2 \,,
\label{eq42}
\end{eqnarray}
and $n_f$ is the number of flavours.
The "plus" functions appearing in the above results are the distributions
which satisfy the following equation
\begin{eqnarray}
f_+(z)&=&\left (\frac{\ln^i(1-z)}{1-z}\right )_+ \,,\qquad 
f(z)=\left (\frac{\ln^i(1-z)}{1-z}\right ) \,,
\nonumber\\[2ex]
\int_0^1 dz~f_+(z)~ g(z) &=& \int_0^1 dz~f(z) \Big(g(z)-g(1)\Big)\,,
\label{eq43}
\end{eqnarray}
where $g(z)$ is any well behaved function in the region $0\le z \le 1$.

\mysection{Differential cross section with respect to $x_F$}
In this section we compute the $x_F$-distribution of the di-muon pair 
up to NLO
in QCD with the $\gamma,Z$ and the graviton in the intermediate state. 
The NLO corrections to processes containing an intermediate photon were 
already calculated in \cite{alel}, \cite{kube}-\cite{rine}.
The variable $x_F$ is defined as 
\begin{eqnarray}
x_F={2 (p_1-p_2)\cdot q \over S}\,,
\label{eq44}
\end{eqnarray}
where $p_1,p_2$ are the momenta of incoming hadrons and $q$ is
the sum of final state muon momenta, $q=l_1+l_2$. In the CM frame of the
hadrons (CMH)
\begin{eqnarray}
x_F={2 q^3_{CMH} \over \sqrt{S}} \,,
\label{eq45}
\end{eqnarray}
where $q_{CMH}^3$ is the third component of $q$ in the CM frame of Hadrons.
The hadronic $x_F$ distribution
can be obtained by introducing the identity 
\begin{eqnarray}
\int dx_F \delta\left(x_F-{2 (p_1-p_2)\cdot q \over S}\right) =1\,,
\label{eq46}
\end{eqnarray}
in Eq.(\ref{eq8}) and bringing the measure $dx_F$ to the left. 
Hence we find
\begin{eqnarray}
2 S {d \sigma^{P_1P_2} \over d Q^2 dx_F}(\tau,x_F,Q^2)&=&
{1 \over 2 \pi}\sum_{jj'=\gamma,Z,G}\!\!\!
\tilde P_j(Q^2)~ \tilde P_{j'}^*(Q^2)~ L_{jj'}(Q^2)~
{ dW_{jj'}^{P_1 P_2}\over dx_F}(\tau,x_F,Q^2)\,.
\nonumber\\[2ex]
\label{eq47}
\end{eqnarray}
The hadronic structure function is defined as
\begin{eqnarray}
{dW_{jj'}^{P_1P_2}\over dx_F}(\tau,x_F,Q^2)&=&\sum_{ab,jj'}
\int dx_1 \int dx_2~
f_a^{P_1}(x_1)~
f_b^{P_2}(x_2)
\nonumber\\[2ex]
&&\times \int dz ~\int dPS_{m+1} ~|M^{ab \rightarrow jj'}|^2 ~T_{jj'}(q)
~\delta(\tau-z x_1 x_2)
\nonumber\\[2ex]
&&\times \delta\left(x_F-{2 (p_1-p_2)\cdot q \over S}\right)\,.
\label{eq48}
\end{eqnarray}
We start by computing the Born contribution.
This involves the computation of the matrix element 
squared for process $a(k_1)+b(k_2)\rightarrow
jj'(-q)$ and the $(0+1)$-body phase space integration 
as well as the $dz$ integration.  
The $(0+1)$-body phase space is computed using Eq. (\ref{eq11}) and
one finds
\begin{eqnarray}
\int dPS_{0+1} ={2 \pi \over Q^2} \delta(z-1)\,.
\label{eq49}
\end{eqnarray}
To compute the Born process $a(k_1)+b(k_2) \rightarrow jj'(-q)$,
we choose the following parameterisation for the momenta
\begin{eqnarray}
k_1={\sqrt{\hat s} \over 2}(1,0,\cdot\cdot\cdot,0,1)\,, \quad 
k_2={\sqrt{\hat s} \over 2}(1,0,\cdot\cdot\cdot,0,-1)\,, \quad 
-q=\sqrt{\hat s} (1,0,\cdot\cdot\cdot,0,0)\,.
\label{eq50}
\end{eqnarray}
Subsequently we express the partonic momenta in terms of the hadronic 
momenta using the scaling
variables $x_1,x_2$ as $k_1=x_1 p_1,k_2=x_2 p_2$
\begin{eqnarray}
\int dPS_{0+1} \int dz \delta\left(x_F-{2 (p_1-p_2)\cdot q \over S}\right)
\delta(\tau-z x_1 x_2)&=&{2 \pi \over Q^2} \int dz 
\delta(x_F-x_1+x_2)
\nonumber\\[2ex]&&
\times \delta(z-1) \delta (\tau-z x_1 x_2)\,.
\nonumber\\[2ex]
\label{eq51}
\end{eqnarray}
We choose to work with the variables $x_1^0,x_2^0$ which are defined 
through 
\begin{eqnarray}
x_F\equiv x_1^0-x_2^0\,, \quad \quad \quad \tau \equiv x_1^0 x_2^0\,.
\label{eq52}
\end{eqnarray}
Solving the above equations, we find
\begin{eqnarray}
x_1^0={1 \over 2} (x_F+\sqrt{x_F^2+4 \tau})\,,
\quad \quad
x_2^0={1 \over 2} (-x_F+\sqrt{x_F^2+4 \tau})\,.
\label{eq53}
\end{eqnarray}
After rewriting the arguments of the delta functions in terms of the 
variables $x_1^0,x_2^0$, and performing the $z$ integration using 
$\delta(\tau-z x_1 x_2)$, we arrive at a simple looking expression for Eq.
(\ref{eq51}) as
\begin{eqnarray}
&&\quad \quad \quad {2 \pi \over Q^2} \int dz \delta(z-1) \delta(x_F-x_1+x_2)
\delta (\tau-z x_1 x_2)
|M^{ab \rightarrow jj'}|^2\cdot T_{jj'} =
\nonumber\\[2ex]
&&\quad \quad \quad \quad \quad \quad \quad  
{2 \pi \over Q^2 }{ \delta(x_1-x_1^0)\delta(x_2-x_2^0)
\over x_1^0+x_2^0} |M^{ab \rightarrow jj'}|^2\cdot T_{jj'}|_{z=1}\,.
\label{eq54}
\end{eqnarray}
Finally, the Born matrix element squared 
\begin{eqnarray}
|M^{ab \rightarrow jj'}|^2\cdot T_{jj'} \,,
\label{eq55}
\end{eqnarray}
is computed using the parameterisations given 
in Eq. (\ref{eq50}). It is then substituted  
in Eq. (\ref{eq48}) to arrive at the 
leading order $x_F$ distribution.
Our next task is to compute the next to leading order contributions
to the Born $x_F$ distribution.
This involves the computation of the matrix element squared for
the processes $a(k_1)+b(k_2) \rightarrow jj'(-q)+c(-k)$  
and $1+1$ phase space integration.
Since the integral
on the right hand side of Eq. (\ref{eq47}) is Lorentz invariant, we can work
in the CM frame of the incoming partons.  In this frame the momenta 
of the particles are parametrised as
\begin{eqnarray}
k_1&=&{\sqrt{ \hat s }\over 2} (1,0,\cdot \cdot \cdot,0,1)\,,\quad \quad \quad
k_2={\sqrt{\hat s} \over 2} (1,0,\cdot \cdot \cdot,0,-1)\,,
\nonumber\\[2ex] 
-q&=&{\sqrt {\hat s} \over 2} (1+z,0,\cdot \cdot \cdot,
-(1-z)\sin\theta,-(1-z)\cos\theta)\,, 
\nonumber\\[2ex] 
-k&=&{\sqrt {\hat s} \over 2} (1-z,0,\cdot \cdot \cdot,
(1-z)\sin\theta,(1-z)\cos\theta)\,.
\label{eq56}
\end{eqnarray}
In this frame, the $1+1$ phase space becomes
\begin{eqnarray}
\int dPS_{1+1}={1 \over 8 \pi} \left({Q^2 \over 4 \pi}
 \right)^{{\varepsilon/2}}
{1 \over \Gamma\left(1+{\varepsilon/2}\right)}
z^{-\varepsilon/2} (1-z)^{1+\varepsilon}
\int_0^1 dy \left(y (1-y)\right)^{\varepsilon/2}\,,
\label{eq57}
\end{eqnarray}
where $y$ is related to $\cos\theta$ by $\cos\theta=2 y -1$.
The $x_F$ delta function becomes
\begin{eqnarray}
\delta\left(x_F-{2 (p_1-p_2)\cdot q \over S}\right)
=\delta\left(x_F-x_1+z x_2+y (1-z) (x_1+x_2) \right)\,.
\label{eq58}
\end{eqnarray}
We can perform $y$ as well as $z$ integrations by writing the delta 
functions in terms of the variables $\xo,\xt$ as
\begin{eqnarray}
\delta\left(x_F-x_1+z x_2+y (1-z) (x_1+x_2) \right)
&=&{x_1 x_2 \over (x_1+x_2)(x_1 x_2-x_1^0 x_2^0)}\,
\delta\left(y-y^*
\right)\,,
\nonumber \\[2ex]
\delta(\tau-z x_1 x_2)&=&{1 \over x_1 x_2} \delta
\left(z-z^*\right)\,,
\label{eq59}
\end{eqnarray}
where 
\begin{eqnarray}
y^*={x_2 (x_1-x_1^0)(x_1+x_2^0) 
 \over (x_1 x_2-x_1^0 x_2^0) (x_1+x_2)}\,, \quad \quad \quad 
z^*={x_1^0 x_2^0 \over x_1 x_2}\,.
\label{eq60}
\end{eqnarray}
In other words
\begin{eqnarray}
\int\!\!\! &&\!\!\!dPS_{1+1} \int dz \delta(\tau-z x_1 x_2) \delta\left(x_F
-{2 (p_1-p_2)\cdot q \over S}\right)
|M^{ab \rightarrow jj'}|^2\cdot T_{jj'} =
\nonumber\\[2ex]&&
\quad \quad \quad {1 \over 8 \pi}
\left({Q^2 \over 4 \pi}\right)^{\varepsilon/2}
{1 \over \Gamma\left(1+{\varepsilon/2}\right)}
{ (x_1^0 x_2^0)^{-{\varepsilon \over 2}}
 (x_1+x_2)^{-\varepsilon } \over x_1 x_2 (x_1+x_2) }
\nonumber\\[2ex]&&
\quad \quad \quad \times \left( (x_1-x_1^0) (x_1+x_2^0) (x_2-x_2^0) 
(x_2+x_1^0)\right)^{\varepsilon/2}
|M^{ab \rightarrow jj'}|^2\cdot T_{jj'}|_{y=y^*,z=z^*}\,.
\nonumber\\
\label{eq61}
\end{eqnarray}
Notice that the matrix elements $|M^{ab \rightarrow jj'}|^2\cdot T_{jj'}$
are evaluated for $y$ and $z$ given by the delta function constraints given in 
Eq. (\ref{eq59}) and Eq. (\ref{eq60}).
To this order, one has to include the virtual corrections to the Born 
processes as well.  
We use the same formula given in Eq. (\ref{eq49}) with the one loop corrected 
matrix elements. 
The sum of real emission contributions and the one loop corrections to
the Born processes is free of the soft singularities as expected. 
Therefore the result
contains only the collinear divergences which can be removed by the standard 
mass factorisation.
This is done by redefining the parton densities using the 
Altarelli-Parisi kernels as follows
\begin{eqnarray}
f_a^{P}(z)=\sum_b \Gamma_{ab}^{-1} \otimes f_b^{P}(z,\mu_F^2)\,,
\label{eq62}
\end{eqnarray}
which implies
\begin{eqnarray}
f_q^P(z)&=&f_q^P(z,\mu_F^2)
-a_s {1 \over \varepsilon}\Gamma_{q q}^{(1)} \otimes f_q^P (z,\mu_F^2)
-a_s{1 \over \varepsilon} \Gamma_{q g}^{(1)} \otimes 
f_g^P (z,\mu_F^2)\,,
\nonumber\\[2ex]
f_{\bar q}^P(z)&=&f_{\bar q}^P(z,\mu_F^2)
-a_s{1 \over \varepsilon} \Gamma_{\bar q \bar q}^{(1)} \otimes 
f_{\bar q}^P (z,\mu_F^2)
-a_s{1 \over \varepsilon} \Gamma_{\bar q g}^{(1)} \otimes f_{g}^P
(z,\mu_F^2)\,,
\nonumber\\[2ex]
f_g^P(z)&=&f_g^P(z,\mu_F^2)
-a_s{1 \over \varepsilon} n_f \left(\Gamma_{g q}^{(1)} \otimes 
f_q^P (z,\mu_F^2)
+\Gamma_{g \bar q}^{(1)} \otimes f_{\bar q}^P 
(z,\mu_F^2)\right)
\nonumber\\[2ex]&&
-a_s{1 \over \varepsilon} \Gamma_{g g}^{(1)} \otimes 
f_g^ P (z,\mu_F^2)\,.
\label{eq63}
\end{eqnarray} 
Finally we arrive at
\begin{eqnarray}
&&2 S {d \sigma^{P_1P_2} \over d Q^2 dx_F}(\tau,x_F,Q^2)=
\nonumber\\[2ex] && \quad \quad
\sum_{i=q}{\cal F}_{SM,q}
\Bigg( C_{q\overline q}^{SM}(\xo,\xt,\mu_F^2)
+C_{q g}^{SM}(\xo,\xt,\mu_F^2)
+C_{g q}^{SM}(\xo,\xt,\mu_F^2)\Bigg)
\nonumber\\[2ex] && \quad \quad 
+\sum_{i=q}{\cal F}_{G}
\Bigg( C_{q\overline q}^G(\xo,\xt,\mu_F^2)
+C_{q g}^G(\xo,\xt,\mu_F^2)
+C_{g q}^G(\xo,\xt,\mu_F^2)\Bigg)
\nonumber\\[2ex] && \quad \quad
+{\cal F}_{G} ~C_{gg}^G(\xo,\xt,\mu_F^2)\,,
\label{eq64}
\end{eqnarray}
where
\begin{eqnarray}
C_{ab}^{SM}(\xo,\xt,\mu_F^2)&=&C_{ab}^{SM,(0)}(\xo,\xt,\mu_F^2)
+a_s~C_{ab}^{SM,(1)}(\xo,\xt,\mu_F^2)\,,
\nonumber\\[2ex] 
C_{ab}^{G}(\xo,\xt,\mu_F^2)&=&C_{ab}^{G,(0)}(\xo,\xt,\mu_F^2)
+a_s~C_{ab}^{G,(1)}(\xo,\xt,\mu_F^2)\,.
\label{eq65}
\end{eqnarray}
We have presented $C_{ab}^{SM,(0)}$, $C_{ab}^{SM,(1)}$,
$C_{ab}^{G,(0)}$ and $C_{ab}^{G,(1)}$ in appendix A.

\mysection{Differential cross section with respect to $Y$}
In this section we compute the rapidity distribution of di-lepton pair up to
NLO in QCD with $\gamma,Z$ and graviton in the intermediate state.  
The NLO corrections to processes containing an intermediate photon were
already calculated in \cite{alel}, \cite{kube}-\cite{rine}.
We define $Y$ as
\begin{eqnarray}
Y={1 \over 2}\log\left({p_2\cdot q \over p_1 \cdot q}\right)\,.
\label{eq66}
\end{eqnarray}
In the CM frame of the hadron ($CMH$), this is nothing but 
\begin{eqnarray}
Y={1 \over 2}\log\left({q^0_{CMH} -q^3_{CMH}\over 
     q^0_{CMH} +q^3_{CMH}}\right)\,.
\label{eq67}
\end{eqnarray}
With this definition the distribution in $Y$
can be computed by introducing the identity
\begin{eqnarray}
\int dY \delta\left(Y-
{1 \over 2}\log\left({p_2\cdot q \over p_1 \cdot q}\right)\right) =1\,,
\label{eq68}
\end{eqnarray}
in Eq. (\ref{eq8}) and bringing the measure $dY$ to the left.
Hence we find
\begin{eqnarray}
\!\!\!\!2 S~ {d \sigma^{P_1P_2} \over d Q^2 dY}(\tau,Y,Q^2)
 \!\!\!&=&\!\!\!
{1 \over 2 \pi} \sum_{jj'=\gamma,Z,G}
\tilde P_j(Q^2)~ \tilde P_{j'}^*(Q^2)~ L_{jj'}(Q^2)~
{ dW_{jj'}^{P_1 P_2}\over dY} (\tau,Y,Q^2)\,.
\label{eq69}
\end{eqnarray}
The hadronic structure function equals
\begin{eqnarray}
{dW_{jj'}^{P_1P_2}\over dY}(\tau,Y,Q^2)&=&\sum_{ab,jj'}
\int dx_1 \int dx_2~
f_a^{P_1}(x_1)~f_b^{P_2}(x_2)
\nonumber\\[2ex]
&&\times \int dz~ \int dPS_{m+1}~ |M^{ab \rightarrow jj'}|^2~ T_{jj'}(q)
~\delta(\tau-z x_1 x_2)
\nonumber\\[2ex]
&&\times \delta\left(Y-
{1 \over 2}\log\left({p_2\cdot q \over p_1 \cdot q}\right)\right)\,.
\label{eq70}
\end{eqnarray}
We first compute the matrix element squared for $a+b \rightarrow
jj'$ and then substitute this in Eq. (\ref{eq69}). Then we perform the
$(0+1)$-body phase space integration as well as $z$ integration.  
We follow the same steps as for the $x_F$ distribution.
Using Eq. (\ref{eq49}) for the $(0+1)$-body phase space and
performing $z$ integration, we obtain
\begin{eqnarray}
\int dPS_{0+1} \int dz \delta\left(Y-
{1 \over 2}\log\left({p_2\cdot q \over p_1 \cdot q}\right)\right)
\delta(\tau-z x_1 x_2)
\!\!\!&=&\!\!\!{2 \pi \over Q^2} \int dz 
\delta\left(Y-{1 \over 2 }
\log\left({x_1 \over x_2}\right)\right)
\nonumber\\[2ex]&&
\times \delta(z-1) \delta (\tau-z x_1 x_2)\,.
\label{eq71}
\end{eqnarray}
Instead of working with the variables $Y,\tau$, we choose to work
with $x_1^0,x_2^0$ which are defined as
\begin{eqnarray}
Y={1 \over 2} \log \left({x_1^0 \over x_2^0}\right)
\,, \quad \quad \quad \tau=x_1^0 x_2^0\,.
\label{eq72}
\end{eqnarray}
Solving the above equations, we obtain
\begin{eqnarray}
x_1^0=\sqrt{\tau} e^{Y}\,,
\quad \quad
x_2^0=\sqrt{\tau} e^{-Y}\,.
\label{eq73}
\end{eqnarray}
We can perform the $z$ integration in Eq. (\ref{eq71}) using 
$\delta(\tau-z x_1 x_2)$ after rewriting
the remaining delta functions in terms of the variables 
$x_1^0,x_2^0$. We then get 
\begin{eqnarray}
&&{2 \pi \over Q^2} \int dz \delta(z-1) \delta\left(Y-{1 \over 2}
\log\left({x_1 \over x_2}\right)\right)
\delta (\tau-z x_1 x_2)
|M^{ab \rightarrow jj'}|^2~ T_{jj'}=
\nonumber\\[2ex]&&
\quad \quad \quad \quad {2 \pi \over Q^2 } \delta(x_1-x_1^0)\delta(x_2-x_2^0)
|M^{ab \rightarrow jj'}|^2~ T_{jj'}|_{z=1}\,.
\label{eq74}
\end{eqnarray}
Next we evaluate the NLO correction to the Born processes.
This involves the computation of $1+1$ phase space for the process
$a(k_1)+b(k_2) \rightarrow jj'(-q)+c(-k)$.  Since the integral
on the right hand side of Eq. (\ref{eq70}) is Lorentz invariant, we can work
in the CM frame of the incoming partons. 
We follow again the derivation of $x_F$ distribution.
The $Y$ delta function in this frame becomes
\begin{eqnarray}
\delta\left(Y-
{1 \over 2}\log\left({p_2\cdot q \over p_1 \cdot q}\right)\right)
=\delta\left(Y-{1 \over 2} \log\left({x_1 (1-y (1-z))
\over x_2 (z+y(1-z))}\right)\right)\,.
\label{eq75}
\end{eqnarray}
The two delta functions reduce to
\begin{eqnarray}
\delta\left(Y-{1 \over 2} \log\left({x_1 (1-y (1-z))
\over x_2 (z+y(1-z))}\right)\right)
&=&{2 x_1 x_2 x_1^0 x_2^0 (x_1 x_2 +x_1^0 x_2^0)\over 
(x_1 x_2 -x_1^0 x_2^0)(x_1 x_2^0+x_1^0 x_2)^2}
\delta\left(y-y^* 
\right)\,,
\nonumber \\[2ex]
\delta(\tau-z x_1 x_2)&=&{1 \over x_1 x_2} \delta
\left(z-z^*
\right)\,,
\label{eq76}
\end{eqnarray}
where 
\begin{eqnarray}
y^*={x_2 x_2^0 (x_1+x_1^0)(x_1-x_1^0)
 \over (x_1 x_2^0+x_2 x_1^0) (x_1 x_2-x_1^0 x_2^0)}\,,
\quad \quad
z^*={\xo\xt \over x_1 x_2}\,.
\label{eq77}
\end{eqnarray}
Using Eqs. (\ref{eq76}) and (\ref{eq77}) we arrive at
\begin{eqnarray}
&&\int dPS_{1+1} 
\int dz~ \delta(\tau-z x_1 x_2)~ \delta\left(Y
-{1 \over 2}~\log\left({p_2\cdot q \over p_1 \cdot q}\right)\right)
|M^{ab \rightarrow jj'}|^2~ T_{jj'}
~~=~~
\nonumber\\[2ex]
&&\quad \quad \quad {1 \over 8 \pi}
\left({Q^2 \over 4 \pi}\right)^{\varepsilon/2}
{1 \over \Gamma\left(1+{\varepsilon/2}\right)}
{ 2 x_1^0 x_2^0 (x_1 x_2 +x_1^0 x_2^0 ) \over
x_1 x_2 (x_1 x_2^0 + x_2 x_1^0)^2 }
(x_1 x_2^0+x_1^0 x_2)^{-\varepsilon}
\nonumber\\[2ex]&&
\quad \quad \quad \left( (x_1-x_1^0) (x_1+x_1^0) (x_2-x_2^0) (x_2+x_2^0)
\right)^{\varepsilon/2}
|M^{ab \rightarrow jj'}|^2~ T_{jj'}|_{y=y^*,z=z^*}\,.
\label{eq78}
\end{eqnarray}
We substitute Eq. (\ref{eq78}) in Eq. (\ref{eq70}) to 
obtain the real emission contributions to the $Y$ distribution.  The
virtual corrections to the Born processes can be obtained using
Eq. (\ref{eq49}) with the one loop corrected matrix elements. After adding
the real emission contributions and virtual corrections to the Born processes,
all soft singularities cancel. The remaining  
collinear divergences are removed by mass factorisation
using Eq. (\ref{eq63}). Finally we arrive at the finite results for 
the $Y$ distribution 
\begin{eqnarray}
2 S {d \sigma^{P_1P_2} \over d Q^2 dY}(\tau,Y,Q^2)&=&
\sum_{i=q}{\cal F}_{SM,q}
\Bigg( D_{q\overline q}^{SM}(\xo,\xt,\mu_F^2)
+D_{q g}^{SM}(\xo,\xt,\mu_F^2)
\nonumber\\[2ex] &&
+D_{g q}^{SM}(\xo,\xt,\mu_F^2)\Bigg)
+\sum_{i=q}{\cal F}_{G}
\Bigg( D_{q\overline q}^G(\xo,\xt,\mu_F^2)
\nonumber\\[2ex] &&
+D_{q g}^G(\xo,\xt,\mu_F^2)
+D_{g q}^G(\xo,\xt,\mu_F^2)\Bigg)
\nonumber\\[2ex] &&
+{\cal F}_{G} ~D_{gg}^G(\xo,\xt,\mu_F^2)\,,
\label{eq79}
\end{eqnarray}
where
\begin{eqnarray}
D_{ab}^{SM}(\xo,\xt,\mu_F^2)&=&D_{ab}^{SM,(0)}(\xo,\xt,\mu_F^2)
+a_s~D_{ab}^{SM,(1)}(\xo,\xt,\mu_F^2)\,,
\nonumber\\[2ex] 
D_{ab}^{G}(\xo,\xt,\mu_F^2)&=&D_{ab}^{G,(0)}(\xo,\xt,\mu_F^2)
+a_s~D_{ab}^{G,(1)}(\xo,\xt,\mu_F^2)\,.
\label{eq80}
\end{eqnarray}
We have presented $D_{ab}^{SM,(0)}$, $D_{ab}^{SM,(1)}$
$D_{ab}^{G,(0)}$ and $D_{ab}^{G,(1)}$ in appendix B.

\mysection{Forward Backward asymmetry $A_{FB}$}

In this section, we compute the forward backward asymmetry of the di-lepton 
pair.  The NLO correction to $A_{FB}$ in the SM were calculated in \cite{afb}.
The forward backward asymmetry is computed using the distribution define by
\begin{eqnarray}
\label{eqn6.1}
2 S~{d \delta \sigma^{P_1 P_2} \over d Q^2}\left(\tau,Q^2\right)
&=& \left( \int_0^1 - \int_{-1}^0\right) d\cos\theta^* \,\,
2 S~{d \sigma^{P_1 P_2} \over d Q^2 d\cos\theta^*}\left(\tau,Q^2,\cos\theta^*\right)\,,
\label{eq81}
\end{eqnarray}
where $\cos \theta^*$ is the angle between the final state lepton (say $l_1$) 
momentum and the initial state hadron (say $p_1$) momentum in the CM frame 
of the lepton pair.  In general it can be written as
\begin{eqnarray}
\label{eqn6.2}
\cos\theta^* = {p_1 \cdot (l_1-l_2) \over p_1 \cdot (l_1+l_2)}\,,
\end{eqnarray}
The computation of the above distribution is quite similar to that of
$d\sigma/dQ^2$.  In the QCD improved parton model, we find
\begin{eqnarray}
\label{eqn6.3}
2 S~{d \sigma^{P_1 P_2} \over d Q^2 d\cos\theta^*}\left(\tau,Q^2\right)
&=&\sum_{ab={q,\overline q,g}} \int_0^1 dx_1
\int_0^1 dx_2~ f_a^{P_1}(x_1) ~
f_b^{P_2}(x_2)
\nonumber\\[2ex] &&
\times \int_0^1 dz \,\, 2 \hat s ~
{d \hat \sigma^{ab} \over d Q^2d\cos \theta^*}\left(z,Q^2\right)
\delta(\tau-z x_1 x_2)\,.
\end{eqnarray}
We compute the partonic cross section using
\begin{eqnarray}
\label{eqn6.4}
2 \hat s ~ {d \hat \sigma^{ab} \over d Q^2 d\cos\theta^*} &=&
{1 \over 2 \pi} \sum_{jj'=\gamma,Z,G}
\int dPS_{m+1}~  |M^{ab \rightarrow jj'}|^2\cdot P_j(q)\cdot P^*_{j'}(q)\cdot
\nonumber\\[2ex]&&
{\cal L}^{jj' \rightarrow l^+l^-}(q,\cos\theta^*)\,,
\end{eqnarray}
where ${\cal L}^{jj' \rightarrow l^+l^-}(q,\cos\theta^*)$ can be computed
using  
\begin{eqnarray}
\label{eqn6.5}
{\cal L}^{jj'\rightarrow \mu^+\mu^-}(q,\cos\theta^*)&=&
\int \prod_{i=1}^2 \Bigg({d^nl_i\over (2 \pi)^n}
2 \pi \delta^+(l_i^2) \Bigg)
(2 \pi)^n \delta^{(n)}(q -l_1-l_2) 
\nonumber\\[2ex] &&
\times \delta \left(\cos\theta^*-{p_1 \cdot (l_1-l_2) \over p_1 \cdot (l_1+l_2)}\right)
|M^{jj' \rightarrow \mu^+\mu^-}|^2\,.
\end{eqnarray}
The computation of the leptonic part using the above formula is straight forward.
The hadronic part again involves the computation of various processes
that contribute to $Q$ distribution.  Since we are looking at the 
angular distributions which are "odd" in $\cos \theta^*$, the contributions
come mainly from interferences.  
The non-vanishing contribution in the standard model sector 
comes from the interference of photon mediated processes 
with $Z$-boson mediated processes.  We also find that
non-vanishing contributions come from the interference of standard model diagrams
with the graviton exchange diagrams.  These inference diagrams are absent in the
computation of $Q,X_F$ and rapidity distributions because they are odd in 
$\cos \theta^*$.
We have regularised all the divergences using dimensional regularisation.  
The remaining mass singularities are removed
by the mass factorisation.  To the end, we find
\begin{eqnarray}
\label{eqn6.6}
2 S~{d \delta \sigma^{P_1P_2} \over dQ^2}(\tau,Q^2)&=&
\sum_q{\delta \cal F}_{SM,q} \int_0^1~ {d x_1}~ \int_0^1
~{dx_2}~ \int_0^1~ dz~ \delta(\tau-z x_1 x_2)
\nonumber\\[2ex]&&
\times \Bigg[ \delta H_{q \bar q}(x_1,x_2,\mu_F^2) \Big(
\delta \Delta_{q \bar q}^{(0),\gamma Z}(z,Q^2,\mu_F^2)
 +a_s \delta \Delta_{q \bar q}^{(1),\gamma Z}(z,Q^2,\mu_F^2)\Big)
\nonumber\\[2ex] &&
+\delta H_{q g}(x_1,x_2,\mu_F^2)
 \Big(a_s \delta \Delta_{q g}^{(1),\gamma Z}(z,\mu_F^2) \Big)
\nonumber\\[2ex] &&
+\delta H_{g q}(x_1,x_2,\mu_F^2)
 \Big(a_s \delta \Delta_{g q}^{(1),\gamma Z}(z,\mu_F^2) \Big)
\Bigg]
\nonumber\\[2ex]
&&+\sum_q{\delta \cal F}_{G} \int_0^1~ {d x_1 }~ \int_0^1
~{dx_2}~ \int_0^1~ dz~ \delta(\tau-z x_1 x_2)
\nonumber\\[2ex]&&
\times \Bigg[ \delta H_{q \bar q}(x_1,x_2,\mu_F^2) \Big(
\delta \Delta_{q \bar q}^{(0),G}(z,Q^2,\mu_F^2)
 +a_s \delta \Delta_{q \bar q}^{(1),G}(z,Q^2,\mu_F^2)\Big)
\nonumber\\[2ex] &&
+\delta H_{q g}(x_1,x_2,\mu_F^2)
 \Big(a_s\delta \Delta_{q g}^{(1),G}(z,\mu_F^2) \Big)
\nonumber\\[2ex] &&
+\delta H_{g q}(x_1,x_2,\mu_F^2)\Big)
 \Big(a_s\delta \Delta_{g q}^{(1),G}(z,\mu_F^2)\Big)
 \Bigg]\,.
\end{eqnarray}
The constants $\delta {\cal F}_{SM,q},\delta {\cal F}_{G}$ are given by
\begin{eqnarray}
\label{eqn6.7}
\delta {\cal F}_{SM,q}&=&{2 \alpha^2 } \Bigg[{(Q^2-M_Z^2) \over
\left((Q^2-M_Z^2)^2+M_Z^2 \Gamma_Z^2\right)c_w^2 s_w^2} Q_q Q_e g_q^A g_e^A
\nonumber\\[2ex]
&&+{2 Q^2 \over
\left((Q^2-M_Z^2)^2+M_Z^2 \Gamma_Z^2\right) c_w^4 s_w^4}
g_q^V g_e^V g_q^A g_e^A
\Bigg]\,,
\\[2ex]
\label{eqn6.8}
\delta {\cal F}_{G}&=&{\alpha \kappa^2 Q^2 \over 4 \pi }|{\cal D}(Q^2)|
\Bigg[Q_q Q_e +
{Q^2 (Q^2-M_Z^2) \over
\left((Q^2-M_Z^2)^2+M_Z^2 \Gamma_Z^2\right)c_w^2 s_w^2}
g_q^V g_e^V \Bigg] \,.
\end{eqnarray}
The renormalised incoming partonic fluxes are defined by
\begin{eqnarray}
\label{eqn6.9}
\delta H_{q \bar q}(x_1,x_2,\mu_F^2)&=&
f_q^{P_1}(x_1,\mu_F^2)~
f_{\bar q}^{P_2}(x_2,\mu_F^2)
-f_{\bar q}^{P_1}(x_1,\mu_F^2)~
f_q^{P_2}(x_2,\mu_F^2)\,,
\nonumber
\\[2ex]
\delta H_{g q}(x_1,x_2,\mu_F^2)&=&
f_g^{P_1}(x_1,\mu_F^2) ~
\Big(f_q^{P_2}(x_2,\mu_F^2)
-f_{\bar q}^{P_2}(x_2,\mu_F^2)\Big)\,,
\nonumber
\\[2ex]
\delta H_{q g}(x_1,x_2,\mu_F^2)&=&
\delta H_{g q}(x_2,x_1,\mu_F^2)\,.
\label{eq}
\end{eqnarray}
The LO coefficient functions corresponding to Eq. (\ref{eqn6.6}) are 
\begin{eqnarray}
\label{eqn6.10}
\delta \Delta^{(0),\gamma Z}_{q\bar q}&=&{2 \pi \over N} \delta(1-z)\,,
\nonumber \\[2ex]
\delta \Delta^{(0),G}_{q \bar q}&=&{\pi \over 8 N} \delta(1-z)\,.
\end{eqnarray}
The NLO contributions are given by
\begin{eqnarray}
\label{eqn6.11}
\delta \Delta^{(1)\gamma Z}_{q \bar q}&=&
\Delta^{(1)\gamma/Z}_{q \bar q}+{2 \pi \over N} C_F 
\Bigg[4 (1+z) \ln(z) +4 (1-z) \Bigg]\,,
\nonumber\\[2ex]
\delta \Delta^{(1)\gamma Z}_{q g}&=&
\Delta^{(1)\gamma/Z}_{q g}+{2 \pi \over N} T_f
\Bigg[4 (1-z) \ln(z) \Bigg]\,,
\nonumber\\[2ex]
\delta \Delta^{(1)\gamma Z}_{g q}&=&
-\Delta^{(1)\gamma/Z}_{g q}+{2 \pi \over N} T_f
\Bigg[8 z^2 \ln(z) +2 (1+2 z -3 z^2)\Bigg]\,,
\nonumber\\[2ex]
\delta \Delta^{(1) G}_{q \bar q}&=&
{1 \over 16}\Delta^{(1) G}_{q \bar q}+{\pi \over 8 N} C_F
\Bigg[-4 (1+z) \ln(z) -2 \delta (1-z)\Bigg]\,,
\nonumber\\[2ex]
\delta \Delta^{(1) G}_{q g}&=&
{1 \over 16}\Delta^{(1) G}_{q g}+{\pi \over 8 N} T_f
\Bigg[-4 (1-2 z) \ln(z) +4 (1-z)\Bigg]\,,
\nonumber\\[2ex]
\delta \Delta^{(1) G}_{g q}&=&
-{1 \over 16}\Delta^{(1) G}_{g q}+{\pi \over 8 N} T_f
\Bigg[8 (2-z+z^2) \ln(z) +2 (1+2 z-3 z^2)\Bigg]\,.
\end{eqnarray}
The coefficient functions $\Delta_{ab}^{(1),i}$ with $i=\gamma,Z,G$ correspond
to the ones found in the cross section for the invariant lepton pair distribution
Eq. (\ref{eq36}). They can be found in Eq. (\ref{eq41}).
The forward backward asymmetry is defined by
\begin{eqnarray}
\label{eqn6.12}
A_{FB}=\Bigg[2 S~{d \sigma^{P_1 P_2} \over d Q^2}\left(\tau,Q^2\right)\Bigg]^{-1}
2 S~{d \delta \sigma^{P_1 P_2} \over d Q^2}\left(\tau,Q^2\right)\,.
\end{eqnarray}
The impact of the NLO correction is discussed in the next section.

\mysection{Discussions}

In this section, the effect of the NLO QCD
corrections on various distributions such as the invariant 
mass $Q$, the longitudinal moment fraction $x_F$ and
the rapidity $Y$ of the lepton pair are presented. We have chosen
to present these distributions for the
LHC ($\sqrt{S}=14~{\rm TeV}$) and for the Run II of Tevatron
($\sqrt{S}=1.96~{\rm TeV}$). We want to emphasise that we are not
analysing the existing Tevatron data to extract bounds on $M_S$,
which will require doing a full hadron-level simulation,
but are only interested in gauging the impact of the QCD corrections on
the bound.

The standard model parameters which enter our analysis are 
$\alpha=1/137.03604$, $M_Z=91.1876$ GeV, $\Gamma_Z=2.4952~{\rm GeV}$ and 
$\sin^2\theta_W=0.227$.  For the ADD model parameters, we choose 
$M_S=1.5 (2)$ TeV and $d=2 (3)$ for the Tevatron (LHC) respectively.  
For the parton density sets we adopt in 
leading order MRST 2001 LO ($\Lambda=0.1670~{\rm GeV}$) and in 
next-to-leading order the MRST 2001 NLO ($\Lambda=0.2390$ GeV) \cite{mrst}. 
For LHC and the Tevatron we choose the range $150~{\rm GeV}<Q<1100~{\rm GeV}$. 
Further we have for the LHC the ranges $-0.7<x_F< 0.7$ and $-2.7<Y<2.7$ at 
$Q=700$ GeV.  A similar analysis is performed for the Tevatron. Here we 
choose for the $x_F$ and $Y$ distributions the regions $-0.7< x_F< 0.7$ and 
$-0.9 <Y< 0.9$ at $Q=400~ {\rm GeV}$ respectively.  The renormalisation
scale $\mu_R$ and factorisation scale $\mu_F$ are taken to be equal.  
Finally the factorisation 
scale $\mu_F$ is chosen to be $\mu_F=Q$ unless mentioned otherwise.\\[2mm]
Let us first study the invariant lepton pair mass distributions
\begin{eqnarray}
\label{eqn7.1}
{d \sigma^I(Q) \over dQ}\,,
\end{eqnarray}
for $I=SM$ (standard model) and $I=GR$ (gravity) and the sum of both 
mechanisms $SM+GR$. 

Starting our discussion with the LHC we have plotted the distribution
in Eq.~(\ref{eqn7.1}) for $150~{\rm GeV}<Q<1100~{\rm GeV}$ in Fig.~10a.  
The standard model dominates the cross section for $Q<600~{\rm GeV}$ 
but for $Q>600~ {\rm GeV}$ the graviton mediated process takes over 
and overwhelms the standard model result.  The dominance of the graviton 
becomes very clear in the longitudinal momentum fraction cross section 
Fig.~10b
\begin{eqnarray}
\label{eqn7.2}
{d \sigma^I(Q,x_F) \over {dQ~dx_F}}\Bigg |_{Q=700~{\rm GeV}}\,,
\end{eqnarray}
and the rapidity distribution in Fig.~10c.
\begin{eqnarray}
\label{eqn7.3}
{d \sigma^I(Q,Y) \over {dQ~dY}}\Bigg |_{Q=700~{\rm GeV}}\,.
\end{eqnarray}
In a large invariant mass bin like $Q=700~{\rm GeV}$, the effect of the
graviton is clearly discernible in both the $x_F$ and $Y$  distributions 
especially for the central $x_F$ and $Y$ regions.  At large absolute 
values for both variables the cross section for the graviton mediated 
processes is again smaller than the standard model result.

In Fig.~11a we have studied the K-factor for the invariant lepton pair
mass distribution defined by 
\begin{eqnarray}
\label{eqn7.4}
K^I=\Bigg[ {d \sigma_{LO}^I(Q) \over dQ} \Bigg]^{-1}
       \Bigg[ {d \sigma_{NLO}^I(Q) \over dQ} \Bigg] \,,
\end{eqnarray}
where NLO stands for the next-to-leading order corrected cross section.
A similar definitions exist for the longitudinal momentum fraction in 
Fig.~11b
\begin{eqnarray}
\label{eqn7.5}
K^I=\Bigg[ {d \sigma_{LO}^I(Q,x_F) \over {dQ~dx_F}} \Bigg]^{-1}
       \Bigg[ {d \sigma_{NLO}^I(Q,x_F) \over {dQ~dx_F}} \Bigg]\Bigg 
|_{Q=700~{\rm GeV}} \,,
\end{eqnarray}
and the rapidity in Fig.~11c
\begin{eqnarray}
\label{eqn7.6}
K^I=\Bigg[ {d \sigma_{LO}^I(Q,Y) \over {dQ~dY}} \Bigg]^{-1}
       \Bigg[ {d \sigma_{NLO}^I(Q,Y) \over {dQ~dY}} \Bigg]\Bigg
|_{Q=700~{\rm GeV}}\,.
\end{eqnarray}
Such a definition is useful for $Q$, $x_F$ and $Y$ distributions because
there is no interference between the standard model and the graviton mediated 
processes.  We find that $K^{SM}$ is moderate for all values of $Q$ in 
Fig.~11a.  On the other hand $K^{GR}$ is much larger than the standard model 
value for $K^{SM}$.  In particular for $Q>700~{\rm GeV}$, $K^{GR}$ overwhelms 
the standard model result $K^{SM}$ completely so that the whole $K^{SM+GR}$ 
can be attributed to $K^{GR}$.  This is because the graviton mediated processes
show up already at the Born level (see Eq.~(3.35)) and at the LHC the gluon 
flux is quite large which lead in NLO to big effects mainly due to small $x$ 
terms in the coefficient function $\Delta_{gg}^{(1)G}$ (Eq.~(\ref{eq41})).
These effects are also visible in Fig.~11b for the $x_F$ and in Fig.~11c for 
the $Y$-distributions.  If the choice of parameters $M_S=2~{\rm TeV}$ and 
$d=3$ is reasonable then for $Q>700~{\rm GeV}$ the effect of the graviton will 
be observed in view of the cross section which amounts to about $10^{-3}$ pico 
barn.

In Fig.~12a, Fig.~12b, and Fig.~12c, we have plotted the scale variations 
of the various distributions for both LO and NLO cross sections. For this 
we define $R^I$ for the invariant lepton mass distribution in Fig.~12a as 
\begin{eqnarray}
\label{eqn7.7}
R_{LO}^I&=&\Bigg[ {d \sigma_{LO}^I(Q,\mu=\mu_0) \over dQ} \Bigg]^{-1}
       \Bigg[ {d \sigma_{LO}^I(Q,\mu) \over dQ } \Bigg]\Bigg |_{Q=700~
       {\rm GeV}}\,,
\\[2ex]
\label{eqn7.8}
R_{NLO}^I&=&\Bigg[ {d \sigma_{NLO}^I(Q,\mu=\mu_0) \over dQ} \Bigg]^{-1}
       \Bigg[ {d \sigma_{NLO}^I(Q,\mu) \over dQ} \Bigg]\Bigg |_{Q=700~
       {\rm GeV}}\,,
\end{eqnarray}
where $\mu_0$ is a fixed scale which is chosen to be $\mu_0=Q$.
For the longitudinal fraction of the lepton pair we have in Fig.~12b
\begin{eqnarray}
\label{eqn7.9}
R_{LO}^I&=&\Bigg[ {d \sigma_{LO}^I(Q,x_F,\mu=\mu_0) \over {dQ~dx_F}} \Bigg]^{-1}
       \Bigg[ {d \sigma_{LO}^I(Q,x_F,\mu) \over {dQ~dx_F} } \Bigg]\Bigg
|_{Q=700~{\rm GeV}, x_F=0}\,,
\\[2ex]
\label{eqn7.10}
R_{NLO}^I&=&\Bigg[ {d \sigma_{NLO}^I(Q,x_F,\mu=\mu_0) \over {dQ~dx_F}} 
            \Bigg]^{-1}
       \Bigg[ {d \sigma_{NLO}^I(Q,x_F,\mu) \over {dQ~dx_F}} \Bigg]\Bigg
|_{Q=700~{\rm GeV}, x_F=0}\,.
\end{eqnarray}
Finally we have plotted the scale variation for the rapidity in Fig.~12c
\begin{eqnarray}
\label{eqn7.11}
R_{LO}^I&=&\Bigg[ {d \sigma_{LO}^I(Q,Y,\mu=\mu_0) \over dQ dY} \Bigg]^{-1}
       \Bigg[ {d \sigma_{LO}^I(Q,\mu) \over dQ  dY} \Bigg]\Bigg
|_{Q=700~{\rm GeV}, Y=0}\,,
\\[2ex]
\label{eqn7.12}
R_{NLO}^I&=&\Bigg[ {d \sigma_{NLO}^I(Q,Y,\mu=\mu_0) \over dQ dY} \Bigg]^{-1}
       \Bigg[ {d \sigma_{NLO}^I(Q,\mu) \over dQ dY} \Bigg]\Bigg
|_{Q=700~{\rm GeV}, Y=0}\,.
\end{eqnarray}
In Fig.~12a we observe that for the invariant lepton mass distribution the 
scale variation is appreciably reduced for $\mu<\mu_0$ while going from LO 
to NLO.  For $\mu>\mu_0$ the same is happening but here the effect is not 
very big.  However the inclusion of the standard model into gravity makes 
the situation better in particular for $\mu<\mu_0$. This holds for LO as 
well for NLO.  Apparently the gluon initiated processes in the case of the 
gravity show a larger scale variation than occurs in the case of the 
standard model. If we look at Eq.~(\ref{eqn7.9},\ref{eqn7.10}) plotted in 
Fig.~12b the features are the same. There is an improvement in scale 
variation while going form LO to NLO.  The same holds for Eq.~(\ref{eqn7.11},
\ref{eqn7.12}) plotted in Fig.~12c. 

Now we turn our attention to the Tevatron.  In Fig.~13a we show the 
result for the invariant mass distribution of the lepton pair. Here 
the graviton mediated process becomes larger than the standard model 
result at $Q>700~{\rm GeV}$.  In the $x_F$-distributions (Fig.~13b) 
and the $Y$-distributions (Fig.~13c) the gravity effects are small. 
The picture for 
the $K$-factor differs from the one observed at the LHC. In Fig.~14a 
the $K$-factor for the graviton is larger than the standard model result 
at $Q<600~{\rm GeV}$. However the value for the cross section is smaller 
than that of the standard model. Therefore $K^{SM}$ is close to $K^{SM+GR}$. 
At larger values for $Q$ the cross section for the graviton is larger than 
the standard model result but $K^{GR}<K^{SM}$ so that also here $K^{SM}$ is 
close to $K^{SM+GR}$. The last feature is different from that observed for 
the LHC in Fig.~11a where $K^{SM+GR}>>K^{SM}$. This is because the gluon 
flux does not dominate the processes at the Tevatron quite contrary to what 
is observed for the LHC.  For the $x_F$ in Fig.~14b and $Y$-distributions 
in Fig.~14c which are taken at $Q=400~{\rm GeV}$, the $K^{SM}$ is close to 
$K^{SM+GR}$.  Finally we look at the scale variation in Fig.~15a. For 
$\mu<\mu_0$ there is a better improvement when we 
go from LO to NLO than that observed for the LHC. On the other hand the 
difference between the graviton mediated process and the graviton plus the 
standard model becomes less. Also this is an indication that the gluon flux 
is less important for the Tevatron than for the LHC. The same feature is 
observed at $Q=400~{\rm GeV}$ for the $x_F$ distribution in Fig.~15b and 
the $Y$-distribution in Fig.~15c.
Because the full $K$-factor does not
differ very much from $K^{SM}$, it may appear that it is sufficient to
use the SM K-factor in extracting bounds on $M_S$ at the Tevatron. Indeed,
existing bounds on $M_S$ from the dilepton production process
at the Tevatron \cite{tev} are obtained using a
constant $K$-factor of 1.3. These bounds range from 1.4 to 1 TeV for d=2
to d=6. While the inclusion of the NLO corrections
computed here may not change the Tevatron bounds significantly, it
will certainly stabilise the cross-section with respect to scale
variations. It is for this reason that the inclusion of the NLO corrections
becomes important even for the Tevatron.

Finally we consider the forward backward asymmetry defined by
\begin{eqnarray}
A_{FB}=\Bigg[2 S~{d \sigma^{P_1 P_2} \over d Q^2}
\left(\tau,Q^2\right)\Bigg]^{-1} 2 S~{d \delta \sigma^{P_1 P_2} \over d Q^2}
\left(\tau,Q^2\right).
\label{eqn7.13}
\end{eqnarray}
This asymmetry is a result of the interferences of the parity violating 
structure of $Z$ exchange with the photon \cite{afb} and 
with the graviton initiated process. The dominant contribution to $A_{FB}$ 
comes from the $q \bar q$ subprocess and hence is important at the Tevatron.  
At LHC for the definition of scattering angle (Eq.~\ref{eqn6.2}) the 
$A_{FB}$ will be negligible \cite{afb} since the LO contribution vanishes.
For the Tevatron the plots are made in Fig.~16 for the parameters $d=2$ 
and $M_S=1.5$ TeV in the region $20~{\rm GeV} < Q <800$ GeV.
Here we have plotted $A_{FB}$ for the SM and the SM+GR at NLO. The effects of 
extra dimensions show up at $Q>300$ GeV. In this region the graviton
contribution is slightly larger than the SM where the latter shows   
a constant behaviour.

In summary, we have computed the cross sections $d\sigma/dQ$,
$d^2\sigma/dQdx_F$, $d^2\sigma/dQdY$ and the forward backward asymmetry 
$A_{FB}$ up to next to leading order. The standard model result was already 
given 25 years ago but now we also included all subprocesses due to the 
graviton in the context of TeV-scale gravity models.  Our main conclusion 
is that the NLO QCD corrections are very significant at the LHC because of 
the large incident gluon flux at the LHC. At the Tevatron where the gluon 
flux is small, the NLO effects are also quite small. But, significantly, 
at both colliders inclusion of the NLO QCD corrections help stabilise the 
cross-section with respect to scale variations. The extraction of bounds 
from both colliders will, therefore, require the inclusion of these NLO 
corrections.\\[3mm]

\noindent
{\it Acknowledgements}: 
VR would like to thank R.~K.~Ellis for useful discussion.  The work of PM 
and KS is partially supported by the Indo-French Centre for the Promotion of 
Advanced Research, New Delhi, India (IFCPAR Project No.~2904-2).  PM's work 
is also supported in part by the Commonwealth Scholarship Commission, UK 
(INCF-2004-92) and Board of Research in Nuclear Sciences, India (BRNS Project 
No.~2002/37/23).


\appendix
\mysection*{Appendix A}
\setcounter{section}{1}
Here we present the $x_F$-distributions which were too long to be published
in section 4. The Born contributions (LO) to the coefficient functions are 
\begin{eqnarray}
C_{q\overline q}^{SM,(0)}(\xo,\xt)\!\!&=&\!\!{2 \pi \over N} 
{H_{q\overline q} (\xo,\xt,\mu_F^2) \over \xo+\xt}\,,
\nonumber\\[2ex]
C_{q\overline q}^{G,(0)}(\xo,\xt)\!\!&=&\!\! {\pi \over 8 N} 
{H_{q\overline q}(\xo,\xt,\mu_F^2)\over \xo+\xt}\,,
\nonumber\\[2ex]
C_{g g}^{G,(0)}(\xo,\xt)\!\!&=&\!\! {\pi \over 2 (N^2-1)} 
{H_{gg}(\xo,\xt,\mu_F^2)\over \xo+\xt}\,.
\label{eqa1}
\end{eqnarray}
For the NLO corrections we start with processes with a photon or Z-boson
in the intermediate state. The corrections are given by \cite{alel}, 
\cite{kube}-\cite{rine}
\begin{eqnarray}
C_{q\overline q}^{SM,(1)}(\xo,\xt)\!\!&=&\!\!
{2 \pi \over N} C_F \Bigg\{\int dx_1~ {H_{q\overline q}(x_1,\xt,\mu_F^2) \over (\xo+\xt)}
\Bigg[-2~ {\xo \over x_1^2}+{2 \over x_1} +
\Bigg(-2~ {\xo \over x_1^2}-{2 \over x_1} \Bigg)~{\cal L}_{a_1}
\nonumber\\[2ex]&&
+{4 \over (x_1-\xo)}~ {\cal L}_{c_1} \Bigg]
+\int dx_1~ {H_{q\overline q,1}(x_1,\xt,\mu_F^2) 
  \over (x_1-\xo)(\xo+\xt)}
\Big[4~ {\cal L}_{b_1}\Big]
\nonumber\\[2ex]&&
+\int dx_1~ \int dx_2~ {H_{q\overline q,1}(x_1,x_2,\mu_F^2) 
  \over (x_1-\xo)(\xo+\xt)}
\Bigg[-2~ {\xt \over x_2^2}-{2 \over x_2} \Bigg]
\nonumber\\[2ex]&&
+\int dx_1 \int dx_2{2~H_{q\overline q,12}(x_1,x_2,\mu_F^2) 
  \over (x_1-\xo)(x_2-\xt)(\xo+\xt)}
\nonumber\\[2ex]&&
+{H_{q\overline q}(\xo,\xt,\mu_F^2) 
  \over (\xo+\xt)} \Bigg[\Big(3+2~{\cal L}(\xo,\xt)\Big)
\log\left({Q^2 \over \mu_F^2}\right)
+\left ({\cal L}(\xo,\xt)\right )^2
\nonumber\\[2ex]&&
+6~ \zeta(2) -8 \Bigg]
\nonumber\\[2ex]&&
+\int dx_1 \int dx_2 {H_{q\overline q}(x_1,x_2,\mu_F^2) \over
(x_1+x_2)(x_1+\xt)(x_2+\xo)(\xo+\xt)}
\nonumber\\[2ex]&&
\times \Bigg[{2 \xo \xt \over x_1^2} (x_2+\xo+\xt)
+{2 \over x_1}~ \Big(\xo^2+\xo \xt-\xt^2
+x_2~ (\xo+\xt)\Big)\Bigg]
\nonumber\\[2ex]&&
+1 \leftrightarrow 2 \Bigg\}\,,
\label{eqa2}
\\[2ex]
C_{qg}^{SM,(1)}(\xo,\xt)\!\!&=&\!\! {2 \pi \over N} T_f \Bigg\{
\int dx_2~{H_{qg}(\xo,x_2,\mu_F^2) \over (\xo+\xt)}
\Bigg[4~{\xt \over x_2^2}-4~{\xt^2 \over x_2^3}+
\Bigg({2 \over x_2}-4~{\xt \over x_2^2}
+4~{\xt^2 \over x_2^3}\Bigg)
{\cal L}_{a_2}
\Bigg]
\nonumber\\[2ex]&&
+\int dx_1~\int dx_2~{H_{qg,1}(x_1,x_2,\mu_F^2) \over (x_1-\xo)(\xo+\xt)}
\Bigg[{2 \over x_2}-4~{\xt \over x_2^2}+4~{\xt^2 \over x_2^3}\Bigg]
\nonumber\\[2ex]&&
+\int dx_1 \int dx_2 {H_{qg}(x_1,x_2,\mu_F^2)
\over (x_1+x_2)^2(x_1+\xt)(\xo+\xt)}
\Bigg[ -{2 \xo \xt^2  \over x_1^2 x_2} (\xo+\xt)
\nonumber\\[2ex]&&
+{2 \xt \over x_2^2 x_1}~ \Big(-2 \xo^2 \xt 
+ \xt x_2~ (2 x_2-\xt)
-\xo~ (-2 x_2^2 +x_2\xt +2 \xt^2)\Big)
\nonumber\\[2ex]&&
+{2 \xo \over x_2^3}~ \Big(\xt~ (5 x_2^2 -2 x_2 \xt -2 \xt^2)
-\xo~ (x_2^2 -2 x_2 \xt +2 \xt^2)\Big)
\nonumber\\[2ex]&&
-8{\xt^3\over x_2^2} + 8 {\xt^2 \over x_2}
+4 \xt-2 x_2
+{2 x_1 \over x_2^3}~ \Big(\xo~ (x_2^2 +2 x_2 \xt-2 \xt^2)
\nonumber\\[2ex]&&
+2 x_1^2 \Big( -2 {\xt^2 \over x_2^3} +2 {\xt \over x_2^2}-{1 \over x_2}\Big)
-(2 x_2^3 -5 x_2^2 \xt +2 x_2 \xt^2+2 \xt^3)\Big)\Bigg] \Bigg\}\,,
\nonumber\\[2ex]
\label{eqa3}
\\[2ex]
C_{gq}^{SM,(1)}(\xo,\xt)\!\!&=&\!\! C_{qg}^{SM,(1)}(\xo,\xt)
|_{1 \leftrightarrow 2}\,.
\label{eqa4}
\end{eqnarray}
We have introduced the following abbreviations
\begin{eqnarray}
H_{ab,12}(x_1,x_2,\mu_F^2)&=&H_{ab}(x_1,x_2,\mu_F^2)
-H_{ab}(\xo,x_2,\mu_F^2)-H_{ab}(x_1,\xt,\mu_F^2)
\nonumber\\[2ex]&&
+H_{ab}(\xo,\xt,\mu_F^2)\,,
\nonumber\\[2ex]
H_{ab,1}(x_1,z,\mu_F^2)&=&H_{ab}(x_1,z,\mu_F^2)
-H_{ab}(\xo,z,\mu_F^2)\,,
\nonumber\\[2ex]
H_{ab,2}(z,x_2,\mu_F^2)&=&H_{ab}(z,x_2,\mu_F^2)
-H_{ab}(z,\xt,\mu_F^2)\,,
\label{eqa5}
\end{eqnarray}
\begin{eqnarray}
{\cal L}_{a_1}&=&\ln\left(
{Q^2 (\xo+\xt) (1-\xt)(x_1-\xo) \over
\mu_F^2 (x_1+\xt) \xo \xt}
\right)\,,
\quad \quad
{\cal L}_{b_1}=\ln\left(
{Q^2 (1-\xt)(x_1-\xo) \over
\mu_F^2  \xo \xt}
\right)\,,
\nonumber\\[2ex]
{\cal L}_{c_1}&=&\ln\left(
{ \xo+\xt \over
x_1+\xt}
\right)\,,
\quad \quad
{\cal L}(\xo,\xt)=\ln\left(
{ (1-\xo) (1-\xt) \over
\xo \xt}
\right)\,.
\label{eqa6}
\end{eqnarray}
The ${\cal L}_{a_2}$, ${\cal L}_{b_2}$ and ${\cal L}_{c_2}$ can be obtained
from ${\cal L}_{a_1}$, ${\cal L}_{b_1}$ and ${\cal L}_{c_1}$
by using $1 \leftrightarrow 2$ symmetry.
The NLO corrections with a graviton in the intermediate state are given by
\begin{eqnarray}
C_{q\overline q}^{G,(1)}(\xo,\xt)\!\!&=&\!\! {\pi \over 8 N} C_F \Bigg\{
\int dx_1~ {H_{q\overline q}(x_1,\xt,\mu_F^2) \over (\xo+\xt)}
\Bigg[-2~ {\xo \over x_1^2}+{2 \over x_1} +
\Bigg(-2~ {\xo \over x_1^2}-{2 \over x_1} \Bigg)~{\cal L}_{a_1}
\nonumber\\[2ex]&&
+{4 \over (x_1-\xo)}~ {\cal L}_{c_1} \Bigg]
+\int dx_1~ {H_{q\overline q,1}(x_1,\xt,\mu_F^2) 
  \over (x_1-\xo)(\xo+\xt)}
\Big[4~ {\cal L}_{b_1}\Big]
\nonumber\\[2ex]&&
+\int dx_1~ \int dx_2~ {H_{q\overline q,1}(x_1,x_2,\mu_F^2) 
  \over (x_1-\xo)(\xo+\xt)}
\Bigg[-2~ {\xt \over x_2^2}-{2 \over x_2} \Bigg]
\nonumber\\[2ex]&&
+\int dx_1 \int dx_2{2~H_{q\overline q,12}(x_1,x_2,\mu_F^2) 
  \over (x_1-\xo)(x_2-\xt)(\xo+\xt)}
\nonumber\\[2ex]&&
+{H_{q\overline q}(\xo,\xt,\mu_F^2) 
  \over (\xo+\xt)} \Bigg[\Big(3+2~{\cal L}(\xo,\xt)\Big)
\log\left({Q^2 \over \mu_F^2}\right)
+\left ({\cal L}(\xo,\xt)\right )^2
\nonumber\\[2ex]&&
+6~ \zeta(2) -10 \Bigg]
\nonumber\\[2ex]&&
+\int dx_1 \int dx_2 {H_{q\overline q}(x_1,x_2,\mu_F^2) 
\over (x_1+x_2)^3(x_1+\xt)(x_2+\xo)(\xo+\xt)}
\nonumber\\[2ex]&&
\times \Bigg[
{1 \over x_1^2}~ \Big(2 x_2^3 \xo \xt 
+(\xo+\xt)~ \Big(2 x_2^2 \xo \xt 
+8 x_2 \xo \xt^2 
+8 \xo^2 \xt^2\Big)\Big)
\nonumber\\[2ex]&&
+{1 \over x_1}~ \Big(2 x_2^3 (\xo+\xt)
+2 x_2^2~ \Big(\xo^2+11 \xo \xt+7 \xt^2\Big)
\nonumber\\[2ex]&&
+4 x_2 \xt ~\Big(11 \xo^2 + 7 \xo \xt -4 \xt^2\Big)
+8 \xo \xt~ \Big(3 \xo^2 +\xo \xt-2 \xt^2\Big)\Big)
\nonumber\\[2ex]&&
+{8 x_1^3 \over \xt}~ (\xo +\xt)
+x_1^2~\Bigg(28 \xo -16 {\xo^2 \over \xt} +44 \xt +x_2~
\Bigg(8-16 {\xo \over \xt}
\nonumber\\[2ex]&&
  +24~ {\xt \over \xo} \Bigg)\Bigg)
+x_1~ \Bigg(-50 \xo^2 +8 x_2^3~ \left({1 \over \xo}+{1 \over \xt}\right)
  +16~ {\xo^3 \over \xt} +8 \xo \xt 
\nonumber\\[2ex]&&
+66 \xt^2
  +2 x_2~ \Bigg(-7 \xo +8 {\xo^2 \over \xt}-7\xt+8 {\xt^2 \over \xo}\Bigg)
\Bigg) 
+2~ \Big(8 \xo^3 -7 \xo^2 \xt 
\nonumber\\[2ex]&&
-7 \xo \xt^2 +8 \xt^3 \Big)\Bigg]
+ 1\leftrightarrow 2 \Bigg\}\,,
\label{eqa7}
\\[2ex]
C_{g g}^{G,(1)}(\xo,\xt)\!\!&=&\!\!{\pi \over 2 (N^2-1)} C_A \Bigg\{
\int dx_1~ {H_{gg}(x_1,\xt,\mu_F^2) \over (\xo+\xt)}
\Bigg[\Bigg(-4~ {\xo^2 \over x_1^3}+4{\xo \over x_1^2} 
-{8 \over x_1}+{4 \over \xo} \Bigg)~{\cal L}_{a_1}
\nonumber\\[2ex]&&
+{4 \over (x_1-\xo)}~ {\cal L}_{c_1} \Bigg]
+\int dx_1~ {H_{gg,1}(x_1,\xt,\mu_F^2) 
  \over (x_1-\xo)(\xo+\xt)}
\Big[4~ {\cal L}_{b_1}\Big]
\nonumber\\[2ex]&&
+\int dx_1 \int dx_2 {H_{gg,1}(x_1,x_2,\mu_F^2) 
  \over (x_1-\xo)(\xo+\xt)}
\Bigg[\Bigg(-4~ {\xt^2 \over x_2^3}+4{\xt \over x_2^2} 
-{8 \over x_2}+{4 \over \xt} \Bigg) \Bigg]
\nonumber\\[2ex]&&
+\int dx_1 \int dx_2{2~H_{gg,12}(x_1,x_2,\mu_F^2) 
  \over (x_1-\xo)(x_2-\xt)(\xo+\xt)}
\nonumber\\[2ex]&&
+{H_{gg}(\xo,\xt,\mu_F^2) 
  \over (\xo+\xt)} \Bigg[\Bigg({11\over 3}+2~{\cal L}(\xo,\xt)\Bigg)
\log\left({Q^2 \over \mu_F^2}\right)
+\left ({\cal L}(\xo,\xt)\right )^2
\nonumber\\[2ex]&&
+6~ \zeta(2) -{203 \over 18} \Bigg] \Bigg\}
\nonumber\\[2ex]&&
+{n_f T_f H_{gg}(\xo,\xt,\mu_F^2) \over (\xo+\xt)}
\Bigg[-{4 \over 3} \log\left( {Q^2 \over \mu_F^2}\right)
+{35\over 9} \Bigg]
\nonumber\\[2ex]&&
+C_A \Bigg\{\int dx_1 \int dx_2 {H_{gg}(x_1,x_2,\mu_F^2) 
\over (x_1+x_2)^3(x_1+\xt)(x_2+\xo)(\xo+\xt)}
\nonumber\\[2ex]&&
\times\Bigg[
{1 \over x_1^3}~ \Bigg(4 x_2^3 \xo^2 \xt
   +4 x_2^2~ \Big(\xo^2 \xt^2+\xo^3\xt\Big)
   +4 x_2\Big(\xo^2 \xt^3+\xo^3 \xt^2\Big)\Bigg)
\nonumber\\[2ex]&&
+{1 \over x_1^2}~\Bigg(4 x_2^3(\xo^2-\xo\xt)
    +x_2^2~ \Big(4 \xo^3
    +12 \xo^2\xt-4\xo\xt^2\Big)
\nonumber\\[2ex]&&
    +x_2~\Big(-4 \xo \xt^3 +16 \xo^3\xt+12 \xo^2\xt^2\Big)
    +8\xo^2\xt^3 
    +8 \xo^3 \xt^2\Bigg)
\nonumber\\[2ex]&&
+{1 \over x_1}~ \Bigg(x_2^3~ (-4 \xo+8 \xt)
    +x_2~\Big (-32 \xo \xt^2
-16 \xo^2\xt+12 \xo^3 + 8 \xt^3\Big)
\nonumber\\[2ex]&&
    +x_2^2~\Big(-32 \xo\xt + 8 \xo^2-16\xt^2\Big)
    +{1\over 2 x_2}~\Big(12 \xo^2 \xt^3 +12 \xo^3 \xt^2\Big)
\nonumber\\[2ex]&&
    +8 \xo \xt^3+20 \xo^2 \xt^2 +12 \xo^3 \xt\Bigg)
+{4 x_1^4 \over \xt}
+x_1^3~ \Bigg(12-8 {\xo\over \xt}\Bigg)
\nonumber\\[2ex]&&
+x_1^2~ \Bigg(4 {\xo^2 \over \xt}-28 \xt\Bigg)
+x_1~ \Bigg(-64 \xo \xt
      -12 \xo^2 -{4 \xo^3 \over \xt}
      -24 \xt^2 
\nonumber\\[2ex]&&
      +x_2~ \Bigg(-4 {\xt^2 \over \xo}-4\xo-4 {\xo^2 \over \xt}
             -4\xt\Bigg)
      +x_2^2~\Bigg(-8 {\xt \over \xo} -12 {\xo \over \xt}+40\Bigg)
\nonumber\\[2ex]&&
      +4 x_2^3 \Bigg({1 \over \xo}-{1\over \xt}\Bigg)\Bigg)
-14 \xo \xt^2 
-14 \xo^2 \xt +2 \xo^3+2\xt^3\Bigg]
\!+\!1 \leftrightarrow 2 \Bigg\}\,,
\nonumber\\[2ex]
\label{eqa8}
\\[2ex]
C_{q g}^{G,(1)}(\xo,\xt)\!\!&=&\!\! {\pi \over 8 N} T_f\Bigg\{
\int dx_2~{H_{qg}(\xo,x_2,\mu_F^2) \over (\xo+\xt)}
\Bigg[4~{\xt \over x_2^2}-4~{\xt^2 \over x_2^3}+
\Bigg({2 \over x_2}-4~{\xt \over x_2^2}+4~{\xt^2 \over x_2^3}\Bigg)
{\cal L}_{a_2}\Bigg]
\nonumber\\[2ex]&&
+\int dx_1~\int dx_2~{H_{qg,1}(x_1,x_2,\mu_F^2) \over (x_1-\xo)(\xo+\xt)}
\Bigg[{2 \over x_2}-4~{\xt \over x_2^2}+4~{\xt^2 \over x_2^3}\Bigg]
\nonumber\\[2ex]&&
+\int dx_1~{H_{qg}(x_1,\xt,\mu_F^2) \over (\xo+\xt)}
~\Bigg[8~{\xo \over x_1^2} +\Bigg(8 {\xo \over x_1^2}
-{16 \over x_1}+{16 \over \xo} \Bigg) {\cal L}_{a_1}\Bigg]
\nonumber\\[2ex]&&
+\int dx_1~\int dx_2~{H_{qg,2}(x_1,x_2,\mu_F^2) \over (x_2-\xt)(\xo+\xt)}
~\Bigg[8 {\xo\over x_1^2}-{16 \over x_1}+{16 \over \xo}\Bigg]
\nonumber\\[2ex]&&
+\int dx_1 \int dx_2 {H_{qg}(x_1,x_2,\mu_F^2)
\over (x_1+x_2)^2(x_1+\xt)(x_2+\xo)(\xo+\xt)}
\nonumber\\[2ex]&&
\times \Bigg[
{1 \over x_1^2} \Bigg(
 -2 \xo \xt^3 -2 \xo^2 \xt^2 +x_2^2~\Big (-8 \xo \xt\Big)
\nonumber\\[2ex]&&
 +x_2~\Big (-8 \xo^2 \xt -8 \xo \xt^2\Big)
 +{1 \over x_2}~\Big (-2 \xo^3 \xt^2 -2 \xo^2 \xt^3\Big)\Bigg)
\nonumber\\[2ex]&&
+{1\over x_1}~ \Bigg(-2 \xt^3 -12 \xo^2 \xt -14 \xo \xt^2
   +x_2^2~ \Big(16 \xt-8 \xo\Big)
\nonumber\\[2ex]&&
   +x_2~\Big (4 \xt^2 -8 \xo^2-20 \xo \xt\Big)
   +{1 \over x_2}~\Big(-6 \xo^2\xt^2 -6 \xo \xt^3\Big)
\nonumber\\[2ex]&&
   +{1\over x_2^2}~\Big(-4 \xo^3 \xt^2-4 \xo^2 \xt^3\Big)\Bigg)
+{16 x_1^3 \over \xt}
+x_1^2 ~\Bigg(14-16 {\xo \over \xt} 
\nonumber\\[2ex]&&
   + x_2~\Bigg({16 \over \xt}- {16 \over \xo}\Bigg)
   +{1 \over x_2}~\Bigg (4  \xt-2 \xo\Bigg)
   +{1 \over x_2^2}~\Big(-4 \xt^2+4 \xo \xt\Big)
\nonumber\\[2ex]&&
   +{1 \over x_2^3}~\Big(-4 \xo \xt^2\Big)\Bigg)
+x_1~\Bigg (-6 \xt +8 {\xo^2 \over \xt} -34 \xo
   +x_2^2~\Bigg({8 \over \xt}-{8 \over \xo}\Bigg)
\nonumber\\[2ex]&&
   +x_2~\Bigg(36-24 {\xt \over \xo}\Bigg)
   +{1 \over x_2}~\Big (-4 \xt^2-14 \xo^2-2 \xo \xt\Big)
\nonumber\\[2ex]&&
   +{1 \over x_2^2}~\Big(-4 \xt^3 +4 \xo^2 \xt-8 \xo\xt^2\Big)
   +{1 \over x_2^3}~\Big(-4 \xo^2 \xt^2-4\xo \xt^3\Big)\Bigg)
\nonumber\\[2ex]&&
+x_2^2~\Bigg(22-8 {\xt \over \xo}\Bigg)
+x_2~ \Bigg(28 \xt-18 \xo -8 {\xt^2 \over \xo}\Bigg)
+{1 \over x_2}~\Big(-8 \xt^3 
\nonumber\\[2ex]&&
-2 \xo^3 
-2 \xo^2 \xt-12 \xo \xt^2\Big)
+{1 \over x_2^2}~\Big(4 \xo^3 \xt-8 \xo^2 \xt^2 -12 \xo\xt^3\Big)
\nonumber\\[2ex]&&
+{1 \over x_2^3}~\Big(-4 \xo^3 \xt^2-4 \xo^2 \xt^3\Big)
-8 \xt^2 -10 \xo^2-18 \xo \xt
\Bigg]\Bigg\}\,,
\label{eqa9}
\\[2ex]
C_{g q}^{G,(1)}(\xo,\xt)\!\!&=&\!\!
C_{q g}^G(\xo,\xt) |_{ 1 \leftrightarrow 2}\,.
\label{eqa10}
\end{eqnarray}

\mysection*{Appendix B}
\setcounter{section}{2}
Here we present the $Y$-distributions which were too long to be published
in section 5. The Born contributions (LO) to the coefficient functions are
\begin{eqnarray}
D_{q\overline q}^{SM,(0)}(\xo,\xt)\!\!&=&\!\!{2 \pi \over N} 
{H_{q\overline q} (\xo,\xt,\mu_F^2)}\,,
\nonumber\\[2ex]
D_{q\overline q}^{G,(0)}(\xo,\xt)\!\!&=&\!\!{\pi \over 8 N} 
{H_{q\overline q} (\xo,\xt,\mu_F^2)}\,,
\nonumber\\[2ex]
D_{g g}^{G,(0)}(\xo,\xt)\!\!&=&\!\!{\pi \over 2 (N^2-1)} 
{H_{gg}(\xo,\xt,\mu_F^2)}\,.
\label{eqb1}
\end{eqnarray}
For the NLO corrections we start with processes with a photon or Z-boson
in the intermediate state. The corrections are given by \cite{alel}, 
\cite{kube}-\cite{rine}
\begin{eqnarray}
D_{q \overline q}^{SM,(1)}(\xo,\xt)\!\!&=&\!\! {2 \pi \over N} C_F \Bigg\{
\int dx_1~ {H_{q\overline q}(x_1,\xt,\mu_F^2)}
\Bigg[-2~ {\xo \over x_1^2}+{2 \over x_1} +
\Bigg(-2~ {\xo \over x_1^2}-{2 \over x_1} \Bigg)~{\cal K}_{a_1}
\nonumber\\[2ex]&&
+{4 \over (x_1-\xo)}~ {\cal K}_{c_1} \Bigg]
+\int dx_1~ {H_{q\overline q,1}(x_1,\xt,\mu_F^2) 
  \over (x_1-\xo)}~
\Big[4~ {\cal K}_{b_1}\Big]
\nonumber\\[2ex]&&
+\int dx_1~ \int dx_2~ {H_{q\overline q,1}(x_1,x_2,\mu_F^2) 
  \over (x_1-\xo)}
\Bigg[-2~ {\xt \over x_2^2}-{2 \over x_2} \Bigg]
\nonumber\\[2ex]&&
+\int dx_1 \int dx_2{2~H_{q\overline q,12}(x_1,x_2,\mu_F^2) 
  \over (x_1-\xo)(x_2-\xt)}
\nonumber\\[2ex]&&
+{H_{q\overline q}(\xo,\xt,\mu_F^2) 
  } \Bigg[\Bigg(3+2~{\cal K}(\xo,\xt)\Bigg)
\log\left({Q^2 \over \mu_F^2}\right)
+\left ({\cal K}(\xo,\xt)\right )^2
\nonumber\\[2ex]&&
+6~ \zeta(2) -8 \Bigg]
\nonumber\\[2ex]&&
+\int dx_1 \int dx_2 {H_{q\overline q}(x_1,x_2,\mu_F^2)
\over (x_1+\xo)(x_2+\xt)(x_1 \xt+\xo x_2)^2}
\Bigg[{1 \over x_1^2} \Bigg( 4 \xo^4 \xt^2 
\nonumber\\[2ex]&&
+ x_2^2 \Big(2 \xo^4\Big) 
+x_2 \Big(4 \xo^4 \xt\Big)\Bigg)
+{1 \over x_1} \Bigg( 4 \xo^3 \xt^2 +x_2^2 \Big(4 \xo^3\Big)
+x_2 \Big(12 \xo^3 \xt\Big)\Bigg)
\nonumber\\[2ex]&&
+x_1 \Big(4 \xo \xt^2\Big) 
+x_1^2 \Big(4 \xt^2\Big)
+2 \xo^2\xt^2 \Bigg]
+1 \leftrightarrow 2 \Bigg\}\,,
\label{eqb2}
\\[2ex]
D_{qg}^{SM,(1)}(\xo,\xt)\!\!&=&\!\! {2 \pi \over N} T_f\! \Bigg \{
\int dx_2~{H_{qg}(\xo,x_2,\mu_F^2)}
\Bigg[4~{\xt \over x_2^2}-4~{\xt^2 \over x_2^3}+
\Bigg({2 \over x_2}-4~{\xt \over x_2^2}+4~{\xt^2 \over x_2^3}\Bigg)
{\cal K}_{a_2}
\nonumber\\[2ex]&&
+\int dx_1~\int dx_2~{H_{qg,1}(x_1,x_2,\mu_F^2) \over (x_1-\xo)}
\Bigg[{2 \over x_2}-4~{\xt \over x_2^2}+4~{\xt^2 \over x_2^3}\Bigg]
\nonumber\\[2ex]&&
+\int dx_1 \int dx_2 {H_{q g}(x_1,x_2,\mu_F^2)
\over (x_1+\xo)(x_1 \xt+\xo x_2)^3}
\Bigg[ {1 \over x_1^2}~ \Bigg(-4{ \xo^5 \xt^3 \over x_2}\Bigg)
\nonumber\\[2ex]&&
+{1 \over x_1}~\Bigg(4 \xo^4 \xt^2 
-4 {\xo^4 \xt^3\over x_2}
    -8~{ \xo^4 \xt^4\over x_2^2}\Bigg)
+x_1^3~ \Bigg( -2~{ \xt^3\over x_2}  +4~ {\xt^4 \over x_2^2}
\nonumber\\[2ex]&&
-4 {\xt^5 \over x_2^3} \Bigg)
+x_1^2~ \Bigg(2 \xo \xt^2 
+8{\xo \xt^3 \over x_2} 
    -4{\xo \xt^4 \over x_2^2} -8{\xo \xt^5 \over x_2^3}\Bigg)
\nonumber\\[2ex]&&
+x_1~ \Bigg(8 \xo^2 \xt^2 
+2 x_2 \xo^2 \xt
    +16 {\xo^2 \xt^3 \over x_2}
     -16 {\xo^2 \xt^4 \over x_2^2} 
     -8{\xo^2 \xt^5 \over x_2^3}\Bigg)
\nonumber\\[2ex]&&
+8 \xo^3 \xt^2 
+2 x_2^2 \xo^3
+4{\xo^3 \xt^3 \over x_2}
-8 {\xo^3 \xt^4 \over x_2^2}-8 {\xo^3 \xt^5 \over x_2^3}
\Bigg] \Bigg\}\,,
\label{eqb3}
\\[2ex]
D_{gq}^{SM,(1)}(\xo,\xt)\!\!&=&\!\! D_{qg}^{SM}(\xo,\xt)|_
{1 \leftrightarrow 2}\,.
\label{eqb4}
\end{eqnarray}
We have introduced the following abbreviations
\begin{eqnarray}
{\cal K}_{a_1}&=&\ln\left(
{2 Q^2 (1-\xt)(x_1-\xo) \over
\mu_F^2 (x_1+\xo)  \xt}
\right)\,,
\quad \quad
{\cal K}_{b_1}=\ln\left(
{Q^2 (1-\xt)(x_1-\xo) \over
\mu_F^2  \xo \xt}
\right)\,,
\nonumber\\[2ex]
{\cal K}_{c_1}&=&\ln\left(
{ 2 \xo \over
x_1+\xo}
\right)\,,
\quad \quad
{\cal K}(\xo,\xt)=\ln\left(
{ (1-\xo) (1-\xt) \over
\xo \xt}
\right)\,.
\label{eqb5}
\end{eqnarray}
The ${\cal K}_{a_2}$, ${\cal K}_{b_2}$ and ${\cal K}_{c_2}$ can be obtained
from ${\cal K}_{a_1}$, ${\cal K}_{b_1}$ and ${\cal K}_{c_1}$
by using $1 \leftrightarrow 2$ symmetry.
The NLO corrections with a graviton in the intermediate state are given by
\begin{eqnarray}
D_{q \overline q}^{G,(1)}(\xo,\xt)\!\!&=&\!\! {\pi \over 8 N} C_F \Bigg\{
\int dx_1~ {H_{q\overline q}(x_1,\xt)}
\Bigg[-2~ {\xo \over x_1^2}+{2 \over x_1} +
\Bigg(-2~ {\xo \over x_1^2}-{2 \over x_1} \Bigg)~{\cal K}_{a_1}
\nonumber\\[2ex]&&
+{4 \over (x_1-\xo)}~ {\cal K}_{c_1} \Bigg]
+\int dx_1~ {H_{q\overline q,1}(x_1,\xt,\mu_F^2) 
  \over (x_1-\xo)}
\Big[4~ {\cal K}_{b_1}\Big]
\nonumber\\[2ex]&&
+\int dx_1~ \int dx_2~ {H_{q\overline q,1}(x_1,x_2,\mu_F^2) 
  \over (x_1-\xo)}
\Bigg[-2~ {\xt \over x_2^2}-{2 \over x_2} \Bigg]
\nonumber\\[2ex]&&
+\int dx_1 \int dx_2{2~H_{q\overline q,12}(x_1,x_2,\mu_F^2) 
  \over (x_1-\xo)(x_2-\xt)}
\nonumber\\[2ex]&&
+{H_{q\overline q}(\xo,\xt,\mu_F^2) 
  } \Bigg[\Bigg(3+2~{\cal K}(\xo,\xt)\Bigg)
\log\left({Q^2 \over \mu_F^2}\right)
+\left ({\cal K}(\xo,\xt)\right )^2
\nonumber\\[2ex]&&
+6~ \zeta(2) -10 \Bigg]
\nonumber\\[2ex]&&
+\int dx_1 \int dx_2 { H_{q \overline q}(x_1,x_2,\mu_F^2)
\over (x_1+\xo)(x_2+\xt) (x_1 \xt+\xo x_2)^4}
\Bigg[ {1 \over x_1^2}~ \Big(16 \xo^6 \xt^4
\nonumber\\[2ex]&&
   +2 x_2^4 \xo^6
   +4 x_2^3 \xo^6 \xt
   +4 x_2^2 \xo^6 \xt^2
   +16 x_2 \xo^6 \xt^3\Big)
\nonumber\\[2ex]&&
+{1\over x_1}~\Big(16 \xo^5 \xt^4 
   +4 x_2^4 \xo^5
   +16 x_2^3 \xo^5 \xt
   +60 x_2^2 \xo^5 \xt^2
   +72 x_2 \xo^5 \xt^3 \Big)
\nonumber\\[2ex]&&
+4 x_1^4 \xt^4
+60 x_1^3 \xo \xt^4
+x_1^2 ~\Big(126 \xo^2 \xt^4 
   +16 x_2^4 \xo^2
   +16 x_2^3 \xo^2 \xt
\nonumber\\[2ex]&&
   +4 x_2^2 \xo^2 \xt^2\Big)
+x_1 \Big(64 \xo^3 \xt^4 
   +16 x_2^4 \xo^3
   +72 x_2^3 \xo^3 \xt
   +64 x_2^2 \xo^3 \xt^2
\nonumber\\[2ex]&&
   +16 x_2 \xo^3 \xt^3 \Big)
+4\xo^4 \xt^4 \Bigg]
+1 \leftrightarrow 2 \Bigg\}\,,
\label{eqb6}
\\[2ex]
D_{g g}^{G,(1)}(\xo,\xt)\!\!&=&\!\! {\pi \over 2 (N^2-1)} C_A\Bigg\{
\int dx_1~ {H_{gg}(x_1,\xt,\mu_F^2) }
\Bigg[\Bigg(-4~ {\xo^2 \over x_1^3}+4{\xo \over x_1^2} 
-{8 \over x_1}+{4 \over \xo} \Bigg)~{\cal K}_{a_1}
\nonumber\\[2ex]&&
+{4 \over (x_1-\xo)}~ {\cal K}_{c_1} \Bigg]
+\int dx_1~ {H_{gg,1}(x_1,\xt,\mu_F^2) 
  \over (x_1-\xo)}
\Big[4~ {\cal K}_{b_1}\Big]
\nonumber\\[2ex]&&
+\int dx_1 \int dx_2 {H_{gg,1}(x_1,x_2,\mu_F^2) 
  \over (x_1-\xo)}
\Bigg[\Bigg(-4~ {\xt^2 \over x_2^3}+4{\xt \over x_2^2} 
-{8 \over x_2}+{4 \over \xt} \Bigg) \Bigg]
\nonumber\\[2ex]&&
+\int dx_1 \int dx_2{2~H_{gg,12}(x_1,x_2,\mu_F^2) 
  \over (x_1-\xo)(x_2-\xt)}
\nonumber\\[2ex]&&
+{H_{gg}(\xo,\xt,\mu_F^2) 
  } \Bigg[\Bigg({11\over 3}+2~{\cal K}(\xo,\xt)\Bigg)
\log\left({Q^2 \over \mu_F^2}\right)
+\left ({\cal K}(\xo,\xt)\right )^2
\nonumber\\[2ex]&&
+6~ \zeta(2) -{203 \over 18} \Bigg] \Bigg\}
\nonumber\\[2ex]&&
+{n_f T_f H_{gg}(\xo,\xt,\mu_F^2) }
\Bigg[-{4 \over 3} \log\left( {Q^2 \over \mu_F^2}\right)
+{35\over 9} \Bigg]
\nonumber\\[2ex]&&
+C_A \Bigg\{\int dx_1 \int dx_2 {H_{g g}(x_1,x_2,\mu_F^2)
\over (x_1+\xo)(x_2+\xt) (x_1 \xt+\xo x_2)^4}
\Bigg[
{1 \over x_1^3} \Big( 4 x_2^4 \xo^7
\nonumber\\[2ex]&&
   +8 x_2^3 \xo^7 \xt
   +8 x_2^2 \xo^7 \xt^2
   +8 x_2 \xo^7 \xt^3 \Big)
+{1 \over x_1^2} \Big(
   16 \xo^6 \xt^4
   +16 x_2^3 \xo^6 \xt
\nonumber\\[2ex]&&
   +32 x_2^2 \xo^6 \xt^2
   +24 x_2 \xo^6 \xt^3 \Big)
+{ 1\over x_1} \Bigg(
   40 \xo^5 \xt^4
   +4 x_2^4 \xo^5
   +8 x_2^3 \xo^5 \xt
\nonumber\\[2ex]&&
   -16 x_2^2 \xo^5 \xt^2
   +12 {\xo^5 \xt^5 \over x_2} \Bigg)
+4 {x_1^5 \xt^4 \over \xo}
+4 x_1^4 \xt^4
+x_1^3 (-20 \xo \xt^4 
\nonumber\\[2ex]&&
+12 x_2^3 \xo\xt)
+x_1^2 \Big(-48 \xo^2 \xt^4
    +16 x_2^4 \xo^2
    -28 x_2^2 \xo^2 \xt^2 \Big)  
\nonumber\\[2ex]&&
+x_1\Bigg(-76 \xo^3 \xt^4
    +8 {x_2^5 \xo^3 \over \xt}
    +4 x_2^4 \xo^3
    -40 x_2^3 \xo^3 \xt
    -56 x_2^2 \xo^3 \xt^2
\nonumber\\[2ex]&&
    -32 x_2 \xo^3 \xt^3 \Bigg)
-8 \xo^4 \xt^4 \Bigg] 
+1 \leftrightarrow 2 \Bigg\}\,,
\label{eqb7}
\\[2ex]
D_{q g}^{G,(1)}(\xo,\xt)\!\!&=&\!\! {\pi \over 8 N} T_f \Bigg\{
\int dx_2~{H_{qg}(\xo,x_2,\mu_F^2) )}
\Bigg[4~{\xt \over x_2^2}-4~{\xt^2 \over x_2^3}+
\Bigg({2 \over x_2}-4~{\xt \over x_2^2}+4~{\xt^2 \over x_2^3}\Bigg)
{\cal K}_{a_2}
\nonumber\\[2ex]&&
+\int dx_1~\int dx_2~{H_{qg,1}(x_1,x_2,\mu_F^2) \over (x_1-\xo)}
\Bigg[{2 \over x_2}-4~{\xt \over x_2^2}+4~{\xt^2 \over x_2^3}\Bigg]
\nonumber\\[2ex]&&
+\int dx_1~{H_{qg}(x_1,\xt,\mu_F^2) }
\Bigg[8~{\xo \over x_1^2} +\Bigg(8 {\xo \over x_1^2}
-{16 \over x_1}+{16 \over \xo} \Bigg) {\cal K}_{a_1}\Bigg]
\nonumber\\[2ex]&&
+\int dx_1~\int dx_2~{H_{qg,2}(x_1,x_2,\mu_F^2) \over (x_2-\xt)}
\Bigg[8 {\xo\over x_1^2}-{16 \over x_1}+{16 \over \xo}\Bigg]
\nonumber\\[2ex]&&
+\int dx_1 \int dx_2 {H_{q g}(x_1,x_2,\mu_F^2)
\over (x_1+\xo)(x_2+\xt) (x_1 \xt+\xo x_2)^3}
\Bigg[
{1 \over x_1^2} \Bigg(
-4 \xo^5 \xt^3
\nonumber\\[2ex]&&
-8 x_2^3 \xo^5
-16 x_2^2 \xo^5 \xt
-16 x_2 \xo^5 \xt^2
-4{\xo^5\xt^4 \over x_2} \Bigg)
+{1 \over x_1} \Bigg(
-32 \xo^4 \xt^3
\nonumber\\[2ex]&&
+8 x_2^3 \xo^4
-8 x_2^2 \xo^4 \xt
-44 x_2 \xo^4 \xt^2
-12 {\xo^4 \xt^4 \over x_2}
-8 {\xo^4 \xt^5 \over x_2^2}\Bigg)
\nonumber\\[2ex]&&
+x_1^4 \Bigg(16 {\xt^3 \over \xo}
        +32{x_2 \xt^2 \over \xo} \Bigg)
+x_1^3\Bigg(-2 \xt^3 +32 x_2^2 \xt +16 x_2 \xt^2
     +2{\xt^4 \over x_2}
\nonumber\\[2ex]&&
     -4{\xt^6 \over x_2^3} \Bigg)
+x_1^2 \Bigg(-30 \xo \xt^3
     +16 x_2^3 \xo
     -78 x_2 \xo \xt^2
     +4 {\xo \xt^4 \over x_2}
     -12 {\xo \xt^5\over x_2^2}
\nonumber\\[2ex]&&
     -8 {\xo \xt^6 \over x_2^3}\Bigg)
+x_1 \Bigg(-32 \xo^2 \xt^3
    -30 x_2^2 \xo^2 \xt
    -30 x_2 \xo^2 \xt^2
    -32{\xo^2 \xt^4 \over x_2}
\nonumber\\[2ex]&&
    -24 {\xo^2 \xt^5 \over x_2^2}
    -8 {\xo^2 \xt^6 \over x_2^3} \Bigg)
-52 \xo^3 \xt^3
+2 x_2^3 \xo^3
+10 x_2^2 \xo^3 \xt
-36{\xo^3 \xt^4 \over x_2}
\nonumber\\[2ex]&&
-16 {\xo^3 \xt^5 \over x_2^2}
-8 {\xo^3 \xt^6 \over x_2^3}\Bigg]\,,
\label{eqb8}
\\[2ex]
D_{gq}^{G,(1)}(\xo,\xt)\!\!&=&\!\! D_{qg}^{G,(1)}(\xo,\xt)|_
{1 \leftrightarrow 2}\,.
\label{eqb9}
\end{eqnarray}



\centerline{\bf \large{Figure Captions}}
\begin{description}

\item[Fig. 10a]
The invariant di-lepton pair mass distribution $d\sigma/dQ$ for $M_s=2~
{\rm TeV}$ and $d=3$ at $\sqrt{S}=14~{\rm TeV}$ (LHC). Standard model 
(dotted line), gravity (long dashed line), standard model plus gravity 
(solid line).
\item[Fig. 10b]
The double differential cross section $d^2\sigma/dQ~dx_F$ where $x_F$ is the 
longitudinal momentum fraction of the di-lepton pair. The plot is given for 
$M_s=2~{\rm TeV}$ and $d=3$ at $\sqrt{S}=14~{\rm TeV}$ and $Q=700~{\rm GeV}$ 
(LHC).  Standard model (dotted line), gravity (long dashed line), standard 
model plus gravity (solid line).
\item[Fig. 10c]
The double differential cross section $d^2\sigma/dQ~dY$ where $Y$ is the 
rapidity of the di-lepton pair. The plot is given for $M_s=2~{\rm TeV}$
and $d=3$ at $\sqrt{S}=14~{\rm TeV}$ and $Q=700~{\rm GeV}$ (LHC). 
Standard model (dotted line), gravity (long dashed line), standard model 
plus gravity (solid line).
\item[Fig. 11a]
The $K^I$-factor for the cross section $d\sigma/dQ$ at $M_s=2~{\rm TeV}$
and $d=3$. The plot is made for the LHC  ($\sqrt{S}=14~{\rm TeV}$). 
Standard model (dotted line), gravity (long dashed line), standard model 
plus gravity (solid line).
\item[Fig. 11b]
The $K^I$-factor for the double differential cross section $d^2\sigma/dQ~dx_F$ 
at $M_s=2~{\rm TeV}$ and $d=3$. The plot is made for the LHC  with $\sqrt{S}
=14~{\rm TeV}$ and $Q=700~{\rm GeV}$. Standard model (dotted line), gravity 
(long dashed line), standard model plus gravity (solid line).
\item[Fig. 11c]
The $K^I$-factor for the double differential cross section $d^2\sigma/dQ~dY$ 
at $M_s=2~{\rm TeV}$ and $d=3$. The plot is made for the LHC  with $\sqrt
{S}=14~{\rm TeV}$ and $Q=700~{\rm GeV}$. Standard model (dotted line), gravity 
(long dashed line), standard model plus gravity (solid line).
\item[Fig. 12a]
The ratio $R^I$ Eq. (\ref{eqn7.7}) for the cross section $d\sigma/dQ$ at 
$M_s=2~{\rm TeV}$ and $d=3$. The plot is made for the LHC  ($\sqrt{S}=14~
{\rm TeV}$).  Gravity alone in LO (dot-dashed line), standard model plus 
gravity in LO (short dashed line), Gravity alone in NLO (long dashed line), 
standard model plus gravity in NLO (solid line),
\item[Fig. 12b]
The ratio $R^I$ Eq. (\ref{eqn7.8}) for the cross section $d^2\sigma/dQ~dx_F$ 
at $M_s=2~{\rm TeV}$ and $d=3$. The plot is made for the LHC  with 
$\sqrt{S}=14~{\rm TeV}$), $Q=700~{\rm GeV}$ and $x_F=0$.  Gravity alone in LO 
(dot-dashed line), standard model plus gravity in LO (short dashed line),
Gravity alone in NLO (long dashed line), standard model plus gravity in NLO 
(solid line),
\item[Fig. 12c]
The ratio $R^I$ Eq. (\ref{eqn7.9}) for the cross section $d^2\sigma/dQ~dY$ at
$M_s=2~{\rm TeV}$ and $d=3$. The plot is made for the LHC  with 
$\sqrt{S}=14~{\rm TeV}$), $Q=700~{\rm GeV}$ and $Y=0$.  Gravity alone in LO 
(dot-dashed line), standard model plus gravity in LO (short dashed line),
Gravity alone in NLO (long dashed line), standard model plus gravity in NLO
(solid line),
\item[Fig. 13a]
The invariant di-lepton pair mass distribution $d\sigma/dQ$ for $M_s=2~
{\rm TeV}$ and $d=2$ at $\sqrt{S}=1.96~{\rm TeV}$ (TEVATRON). Standard 
model (dotted line), gravity (long dashed line), standard model plus 
gravity (solid line).
\item[Fig. 13b]
The double differential cross section $d^2\sigma/dQ~dx_F$ where $x_F$ is the
longitudinal momentum fraction of the di-lepton pair. The plot is given for
$M_s=2~{\rm TeV}$ and $d=2$ at $\sqrt{S}=1.96~{\rm TeV}$ and $Q=400~{\rm GeV}$ 
(TEVATRON).  Standard model (dotted line), gravity (long dashed line), 
standard model plus gravity (solid line).
\item[Fig. 13c]
The double differential cross section $d^2\sigma/dQ~dY$ where $Y$ is the 
rapidity of the di-lepton pair. The plot is given for $M_s=2~{\rm TeV}$
and $d=2$ at $\sqrt{S}=1.96~{\rm TeV}$ and $Q=400~{\rm GeV}$ (TEVATRON). 
Standard model (dotted line), gravity (long dashed line), standard model 
plus gravity (solid line).
\item[Fig. 14a]
The $K^I$-factor for the cross section $d\sigma/dQ$ at $M_s=2~{\rm TeV}$
and $d=2$. The plot is made for the TEVATRON  ($\sqrt{S}=1.96~{\rm TeV}$).
Standard model (dotted line), gravity (long dashed line), standard model 
plus gravity (solid line).
\item[Fig. 14b]
The $K^I$-factor for the double differential cross section $d^2\sigma/dQ~dx_F$
at $M_s=2~{\rm TeV}$ and $d=2$. The plot is made for the TEVATRON with 
$\sqrt{S}=1.96~{\rm TeV}$ and $Q=400~{\rm GeV}$. Standard model (dotted line),
gravity (long dashed line), standard model plus gravity (solid line).
\item[Fig. 14c]
The $K^I$-factor for the double differential cross section $d^2\sigma/dQ~dY$
at $M_s=2~{\rm TeV}$ and $d=2$. The plot is made for the TEVATRON with 
$\sqrt{S}=1.96~{\rm TeV}$ and $Q=400~{\rm GeV}$. Standard model (dotted line),
gravity (long dashed line), standard model plus gravity (solid line).
\item[Fig. 15a]
The ratio $R^I$ Eq. (\ref{eqn7.7}) for the cross section $d\sigma/dQ$ at
$M_s=2~{\rm TeV}$ and $d=2$. The plot is made for the TEVATRON ($\sqrt{S}
=1.96~{\rm TeV}$).  Gravity alone in LO (dot-dashed line), standard model plus 
gravity in LO (short dashed line), Gravity alone in NLO (long dashed line), 
standard model plus gravity in NLO (solid line),
\item[Fig. 15b]
The ratio $R^I$ Eq. (\ref{eqn7.8}) for the cross section $d^2\sigma/dQ~dx_F$ at
$M_s=2~{\rm TeV}$ and $d=2$. The plot is made for the TEVATRON  with
$\sqrt{S}=1.96~{\rm TeV}$, $Q=400~{\rm GeV}$ and $x_F=0$.  Gravity alone in LO 
(dot-dashed line), standard model plus gravity in LO (short dashed line),
Gravity alone in NLO (long dashed line), standard model plus gravity in NLO
(solid line),
\item[Fig. 15c]
The ratio $R^I$ Eq. (\ref{eqn7.9}) for the cross section $d^2\sigma/dQ~dY$ at
$M_s=2~{\rm TeV}$ and $d=2$. The plot is made for the TEVATRON  with
$\sqrt{S}=1.96~{\rm TeV}$, $Q=400~{\rm GeV}$ and $Y=0$.  Gravity alone in LO 
(dot-dashed line), standard model plus gravity in LO (short dashed line),
Gravity alone in NLO (long dashed line), standard model plus gravity in NLO
(solid line),
\item[Fig. 16]
The forward backward asymmetry $A_{FB}$ in NLO for the TEVATRON
$\sqrt{S}=1.96~{\rm TeV}$. The parameters are $M_s=1.5~{\rm TeV}$ and $d=2$.
Standard model (dotted line), standard model plus gravity (solid line).
                                                                       
\end{description}



\begin{figure}[htb]
\vspace{1mm}
\centerline{\epsfig{file=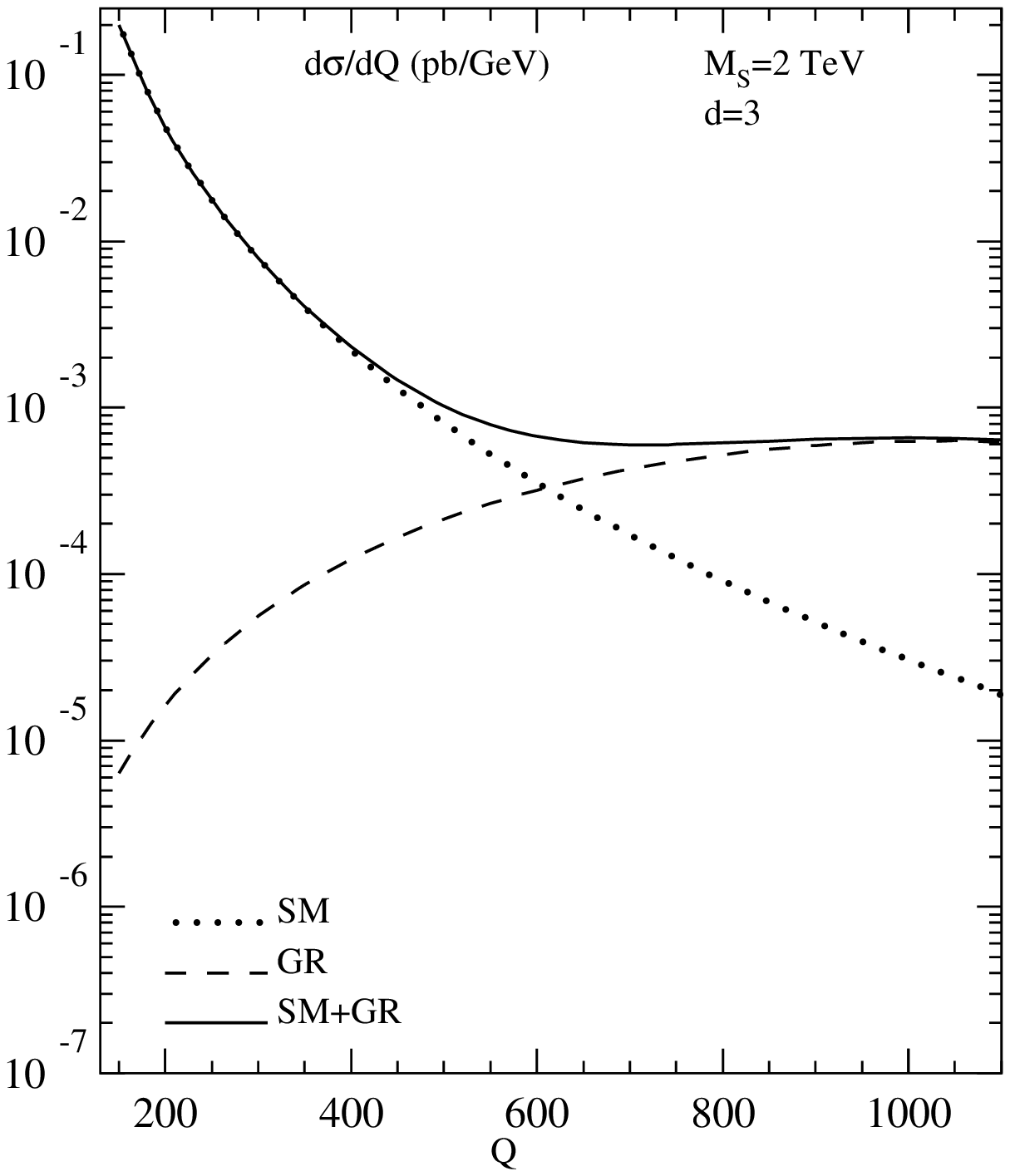,width=15cm,height=16cm,angle=0}}
\vspace{5mm}
\centerline{\bf Fig.~10a}
\label{fig10a}
\end{figure}
                                                                                
\eject
                                                                                
\begin{figure}[htb]
\vspace{1mm}
\centerline{\epsfig{file=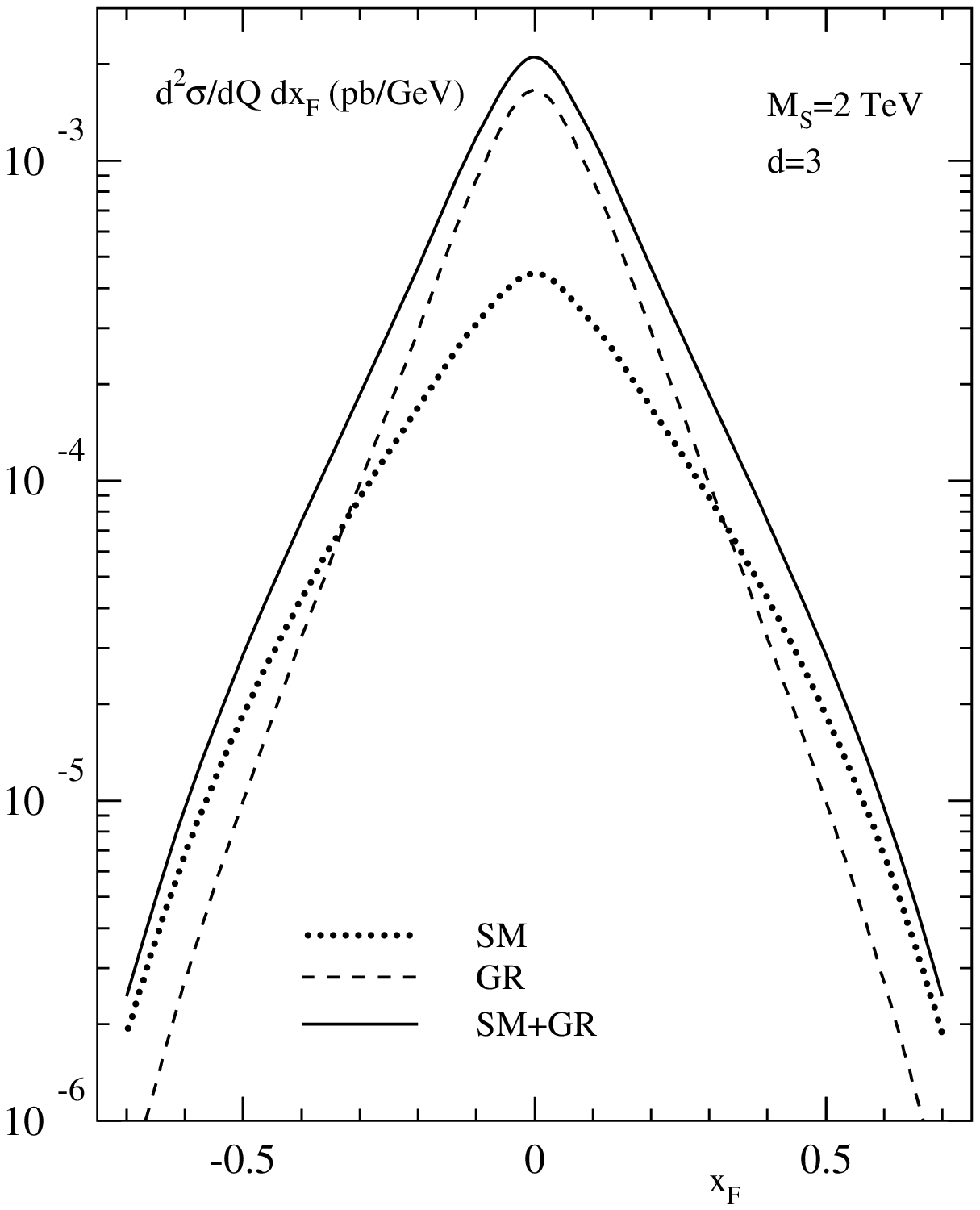,width=15cm,height=16cm,angle=0}}
\vspace{5mm}
\centerline{\bf Fig.~10b}
\label{fig10b}
\end{figure}
                                                                                
\eject
                                                                                
\begin{figure}[htb]
\vspace{1mm}
\centerline{\epsfig{file=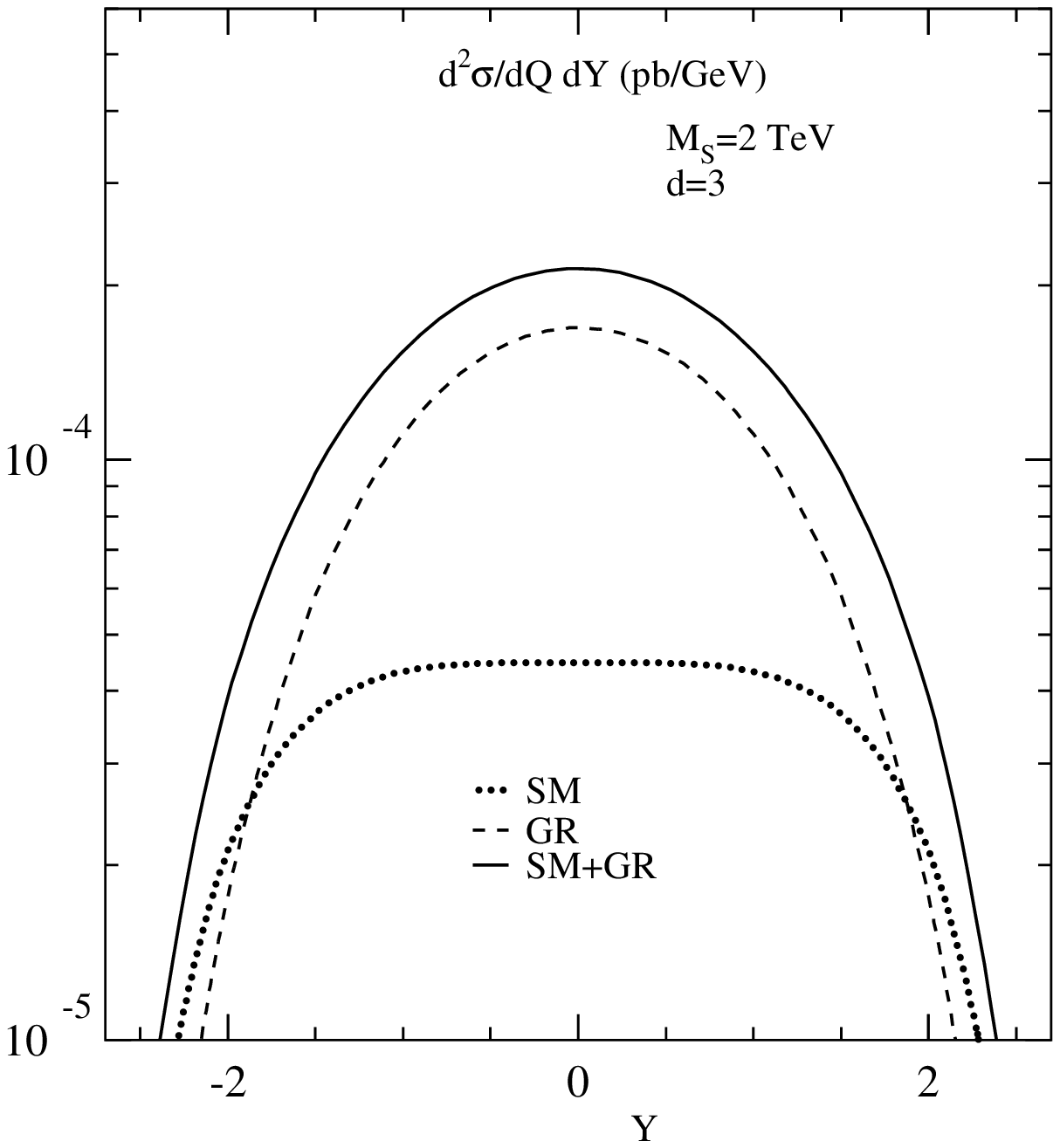,width=15cm,height=16cm,angle=0}}
\vspace{5mm}
\centerline{\bf Fig.~10c}
\label{fig10c}
\end{figure}
                                                                                
\eject
                                                                                
\begin{figure}[htb]
\vspace{1mm}
\centerline{\epsfig{file=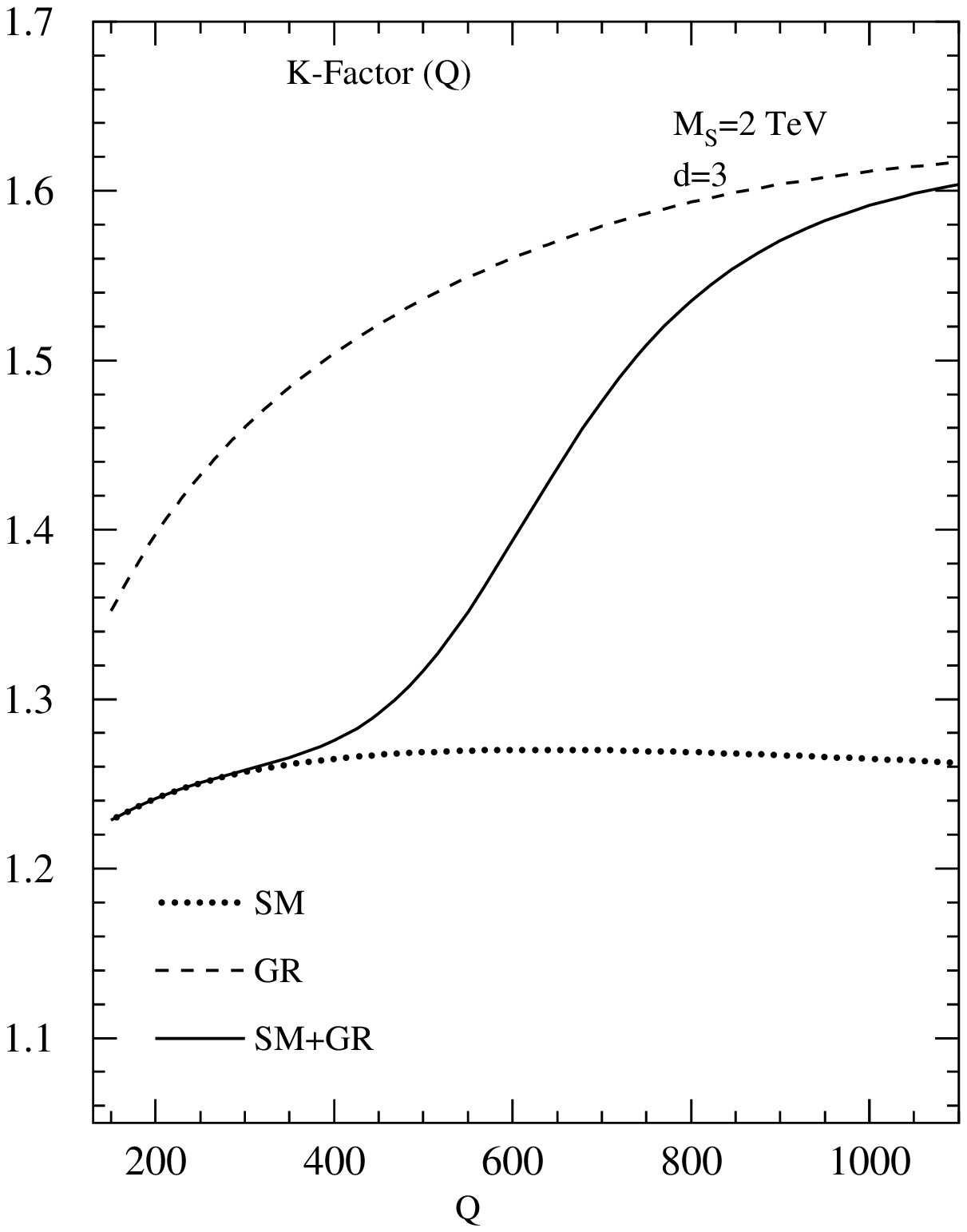,width=15cm,height=16cm,angle=0}}
\vspace{5mm}
\centerline{\bf Fig.~11a}
\label{fig11a}
\end{figure}
                                                                                
\eject
                                                                                
\begin{figure}[htb]
\vspace{1mm}
\centerline{\epsfig{file=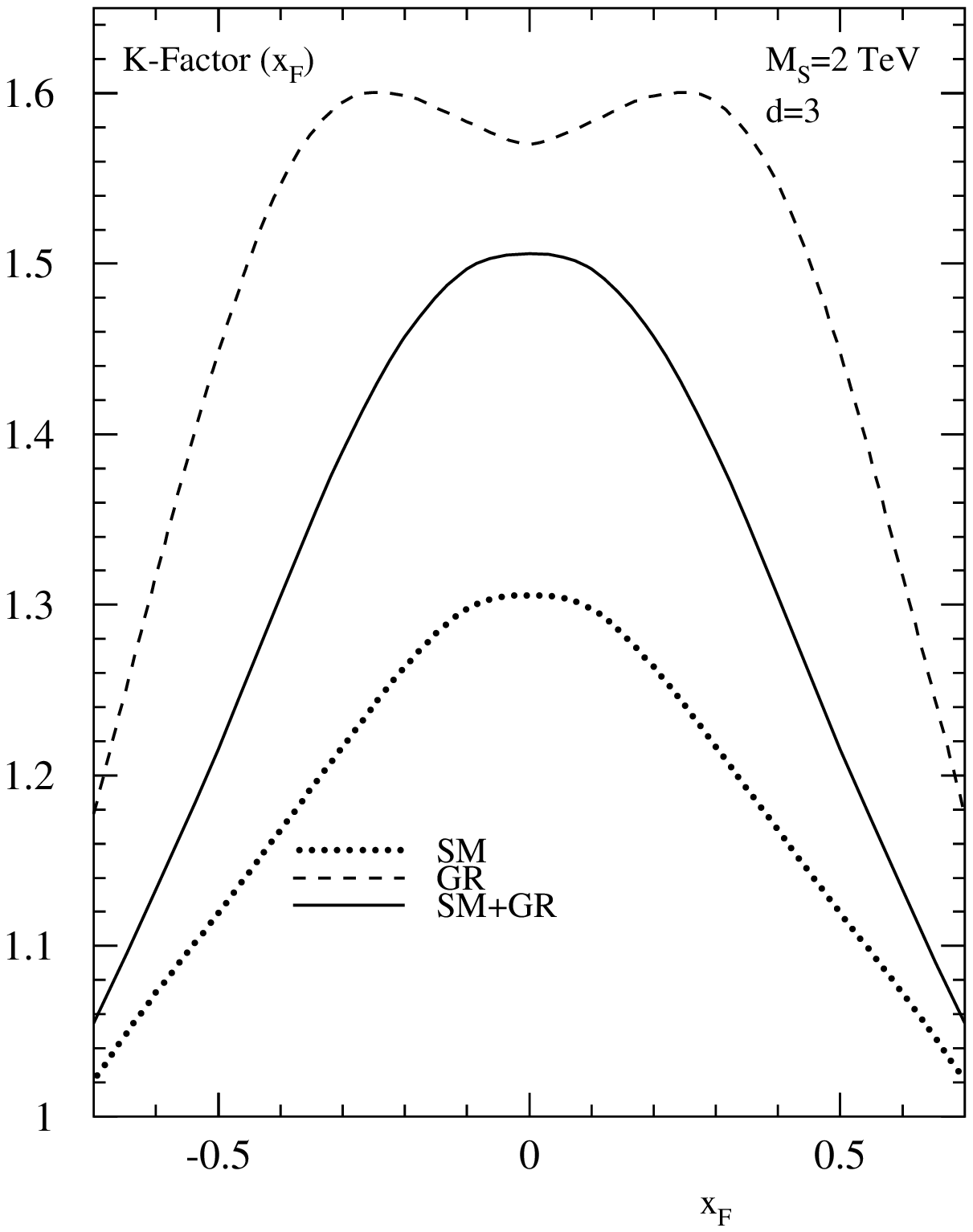,width=15cm,height=16cm,angle=0}}
\vspace{5mm}
\centerline{\bf Fig.~11b}
\label{fig11b}
\end{figure}
                                                                                
\eject
                                                                                
\begin{figure}[htb]
\vspace{1mm}
\centerline{\epsfig{file=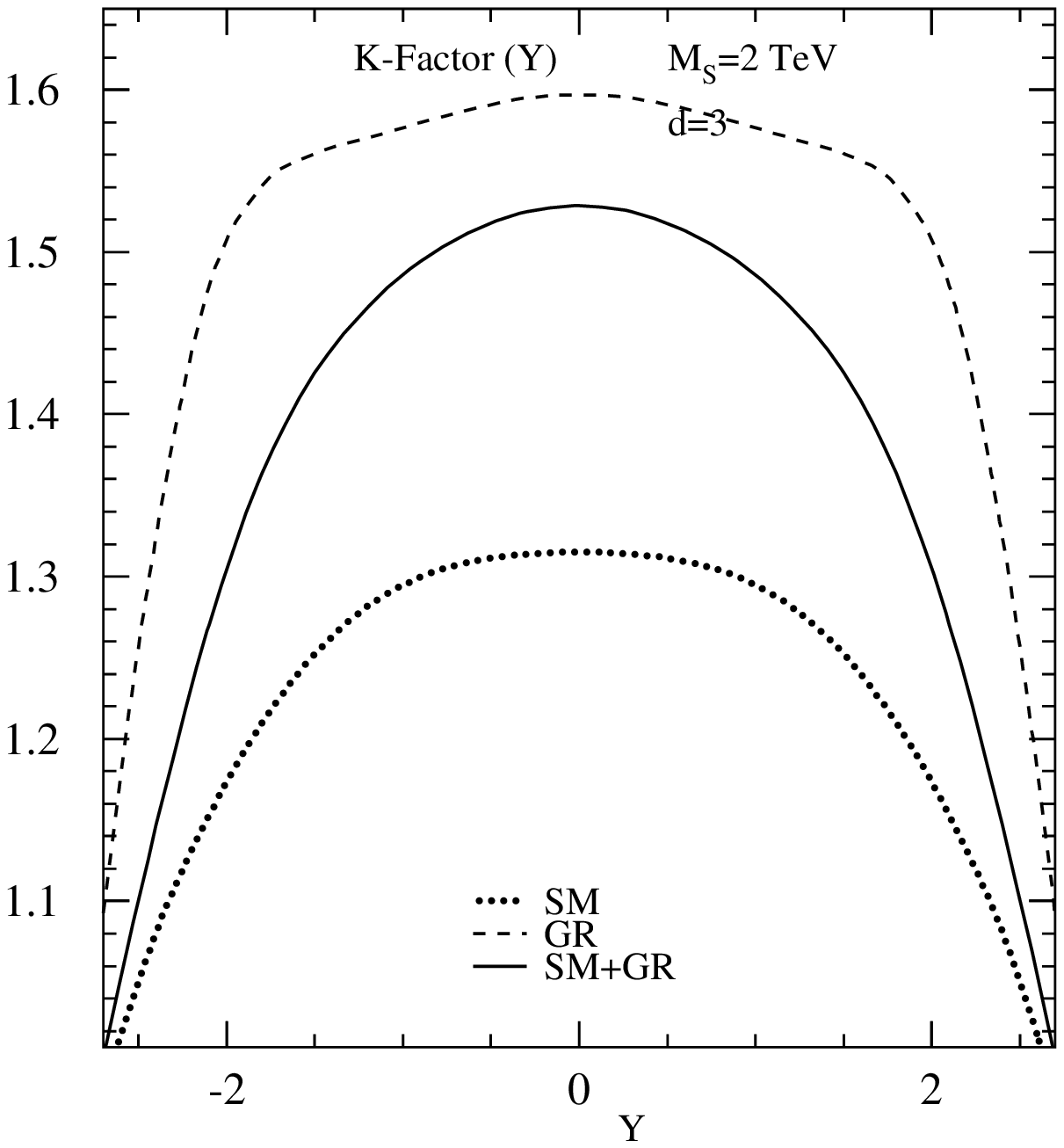,width=15cm,height=16cm,angle=0}}
\vspace{5mm}
\centerline{\bf Fig.~11c}
\end{figure}
                                                                                
\eject
                                                                                
\begin{figure}[htb]
\vspace{1mm}
\centerline{\epsfig{file=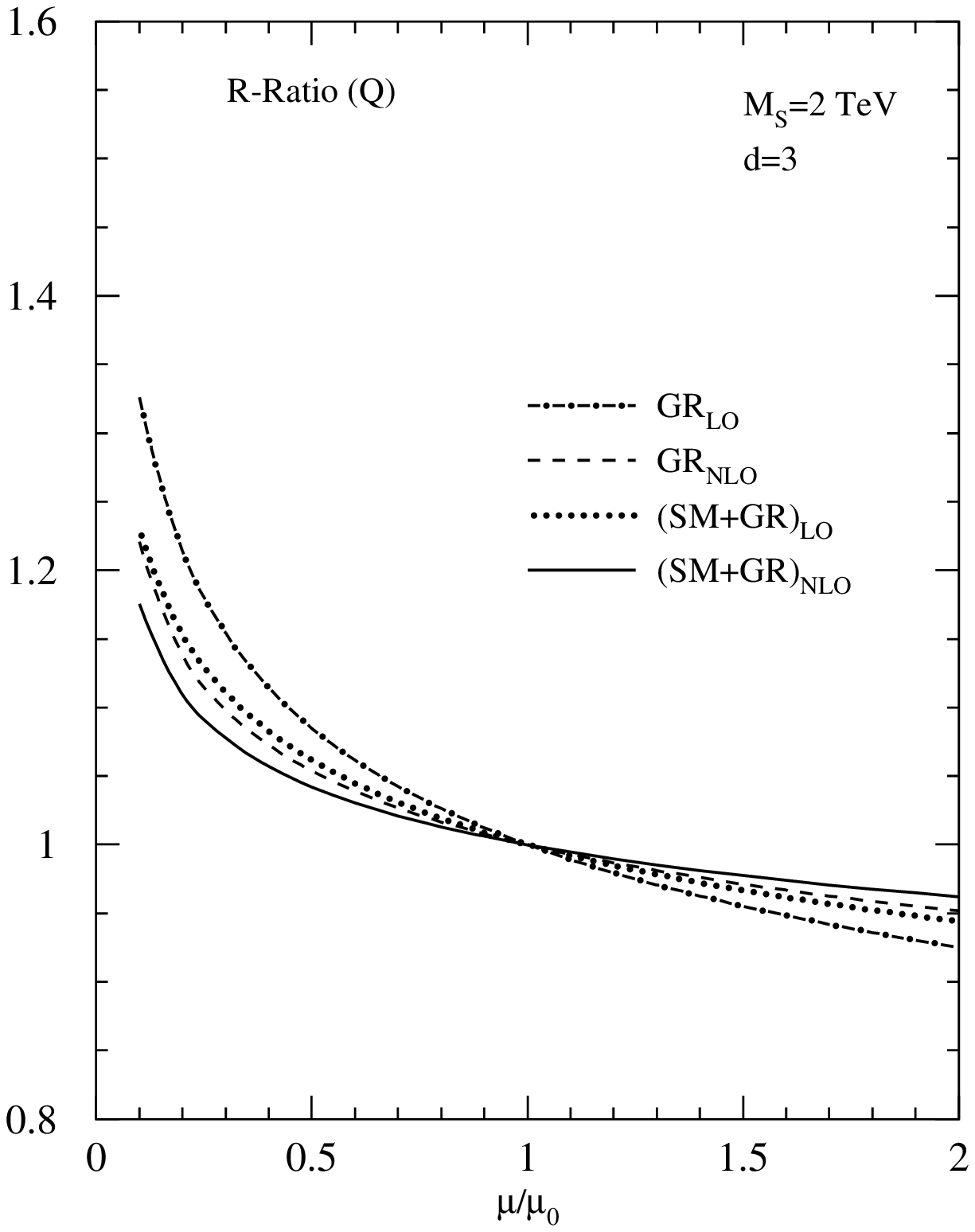,width=15cm,height=16cm,angle=0}}
\vspace{5mm}
\centerline{\bf Fig.~12a}
\label{fig12a}
\end{figure}
                                                                                
\eject
                                                                                
\begin{figure}[htb]
\vspace{1mm}
\centerline{\epsfig{file=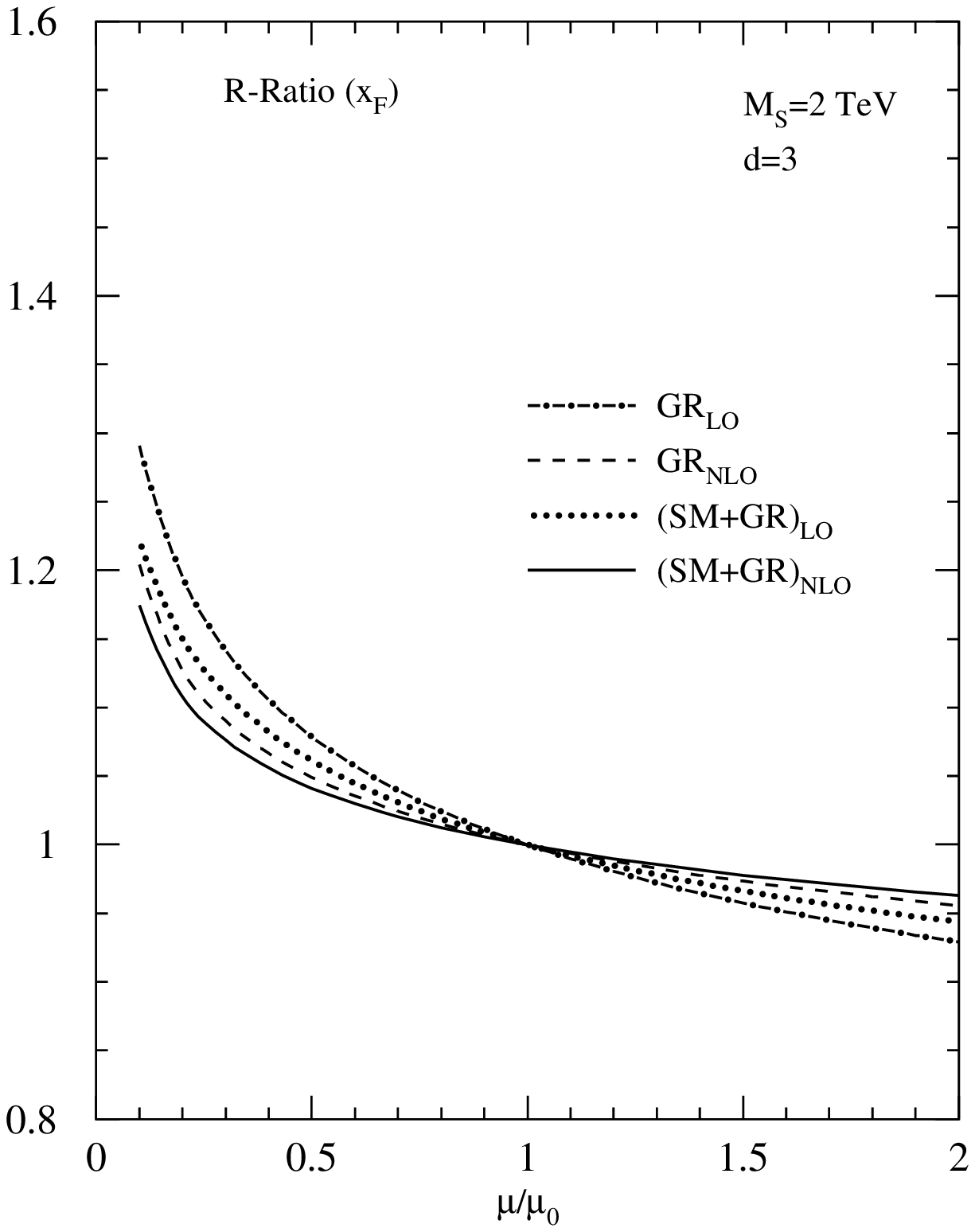,width=15cm,height=16cm,angle=0}}
\vspace{5mm}
\centerline{\bf Fig.~12b}
\end{figure}
                                                                                
\eject

\begin{figure}[htb]
\vspace{1mm}
\centerline{\epsfig{file=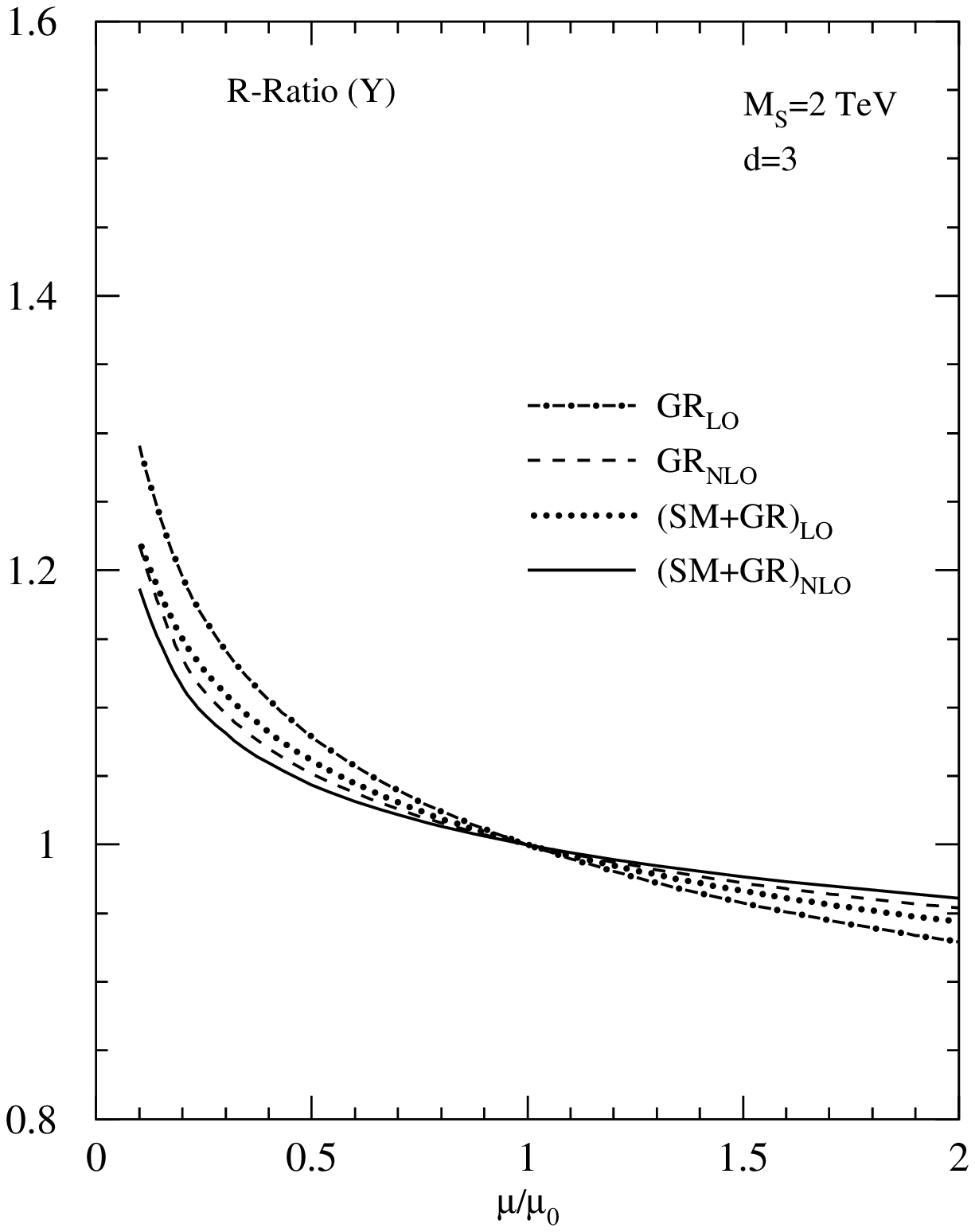,width=15cm,height=16cm,angle=0}}
\vspace{5mm}
\centerline{\bf Fig.~12c}
\end{figure}




\begin{figure}[htb]
\vspace{1mm}
\centerline{\epsfig{file=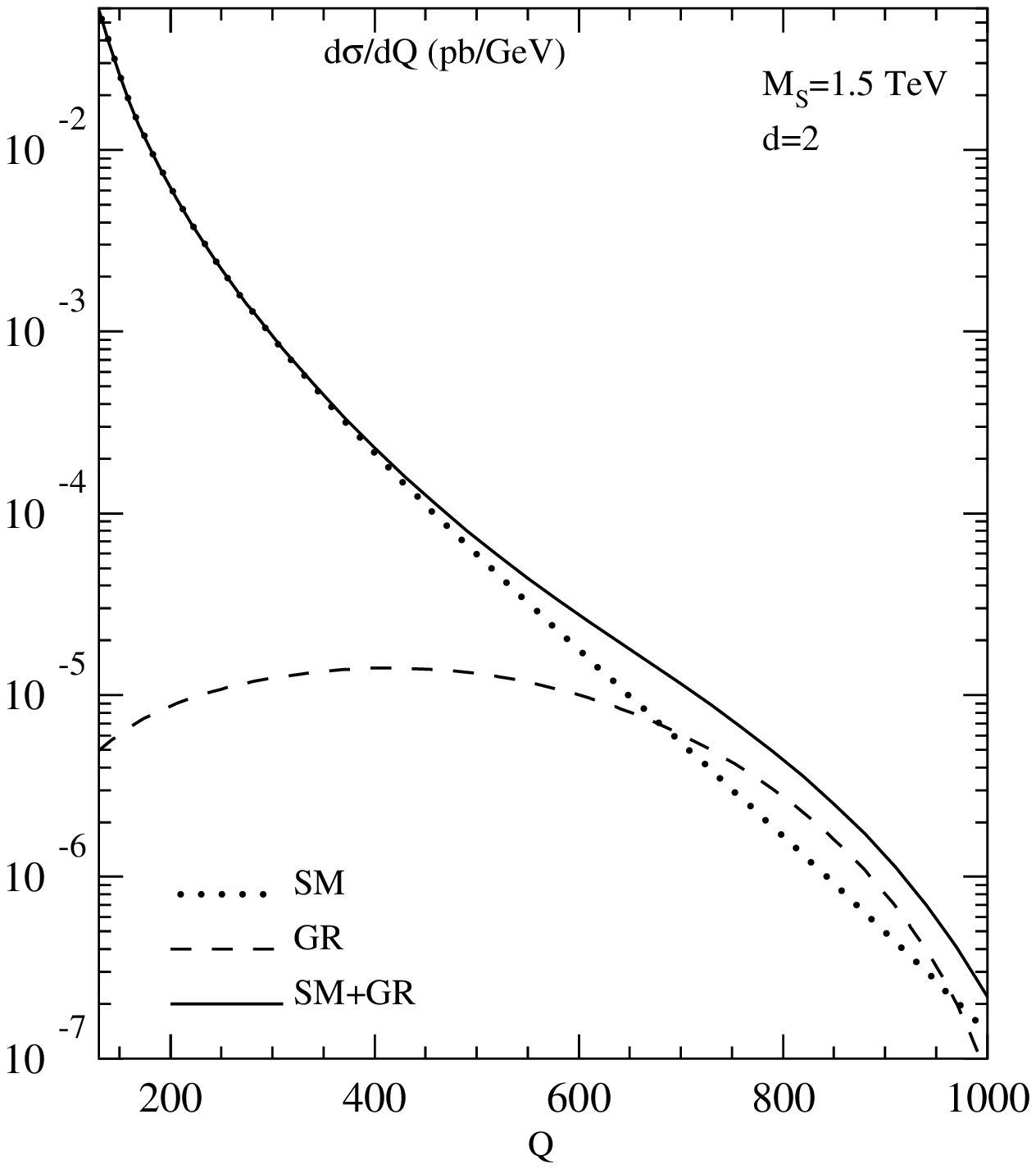,width=15cm,height=16cm,angle=0}}
\vspace{5mm}
\centerline{\bf Fig.~13a}
\label{fig13a}
\end{figure}

\eject

\begin{figure}[htb]
\vspace{1mm}
\centerline{\epsfig{file=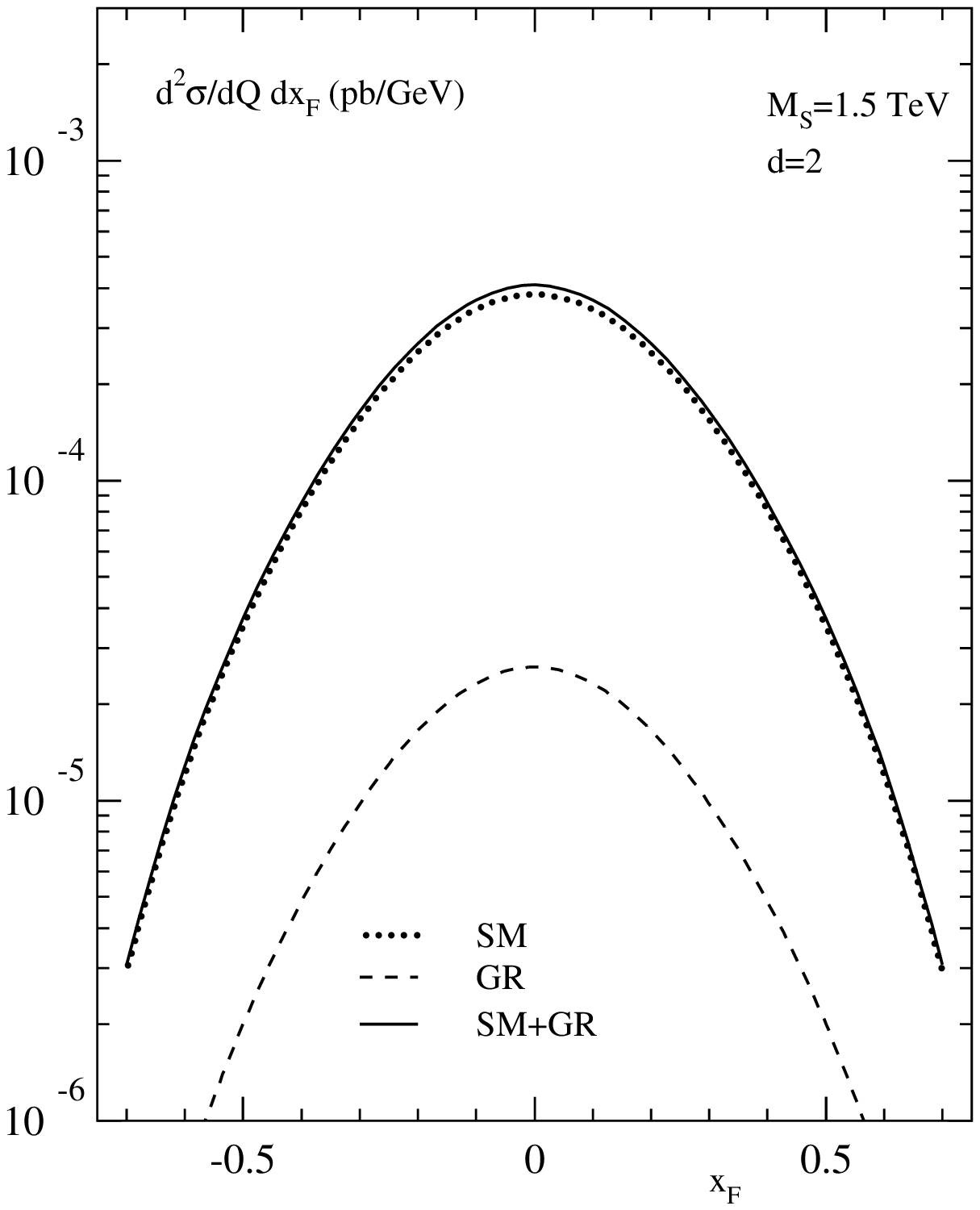,width=15cm,height=16cm,angle=0}}
\vspace{5mm}
\centerline{\bf Fig.~13b}
\end{figure}
                                                                                
\eject

\begin{figure}[htb]
\vspace{1mm}
\centerline{\epsfig{file=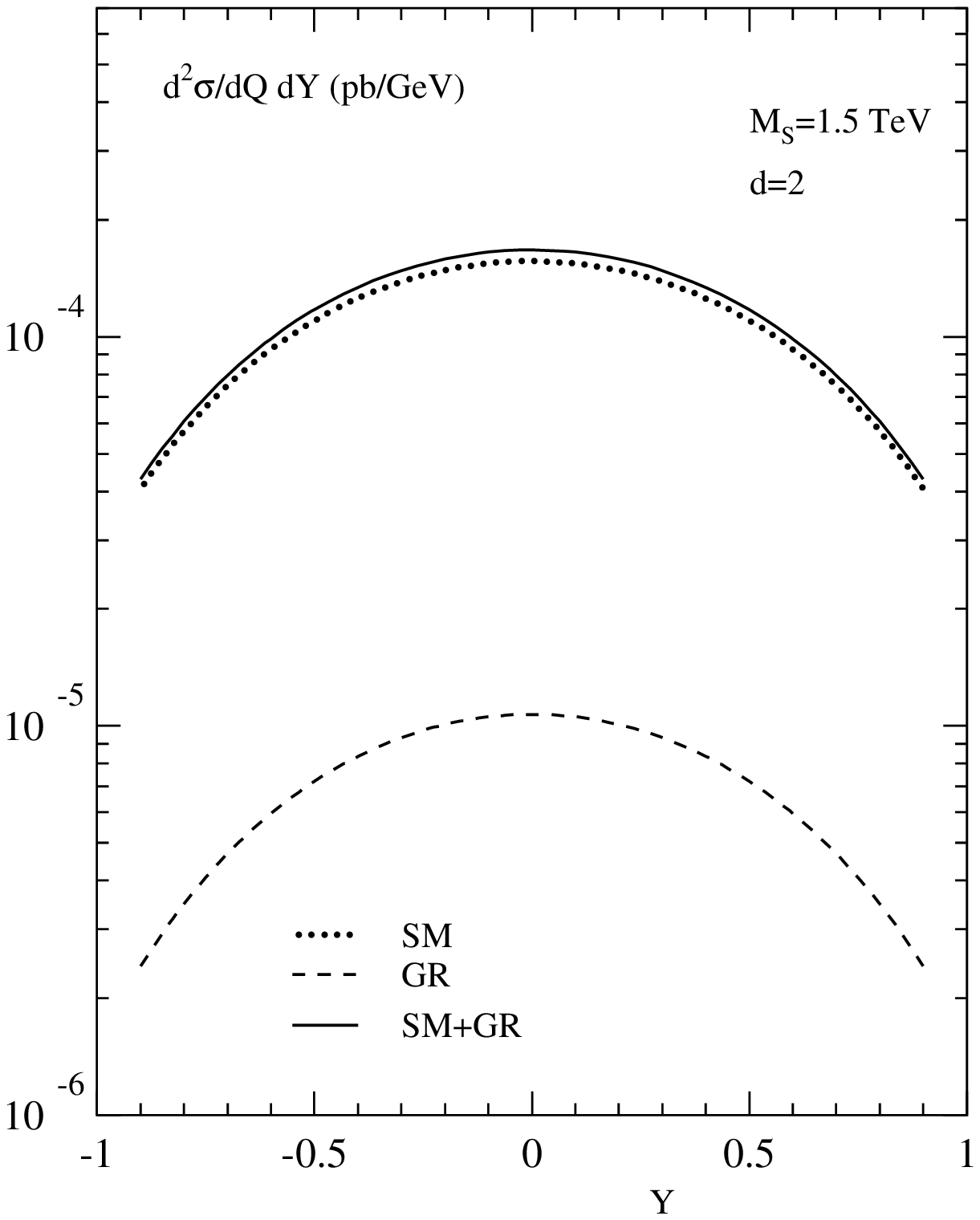,width=15cm,height=16cm,angle=0}}
\vspace{5mm}
\centerline{\bf Fig.~13c}
\label{fig13c}
\end{figure}
                                                                                
\eject

\begin{figure}[htb]
\vspace{1mm}
\centerline{\epsfig{file=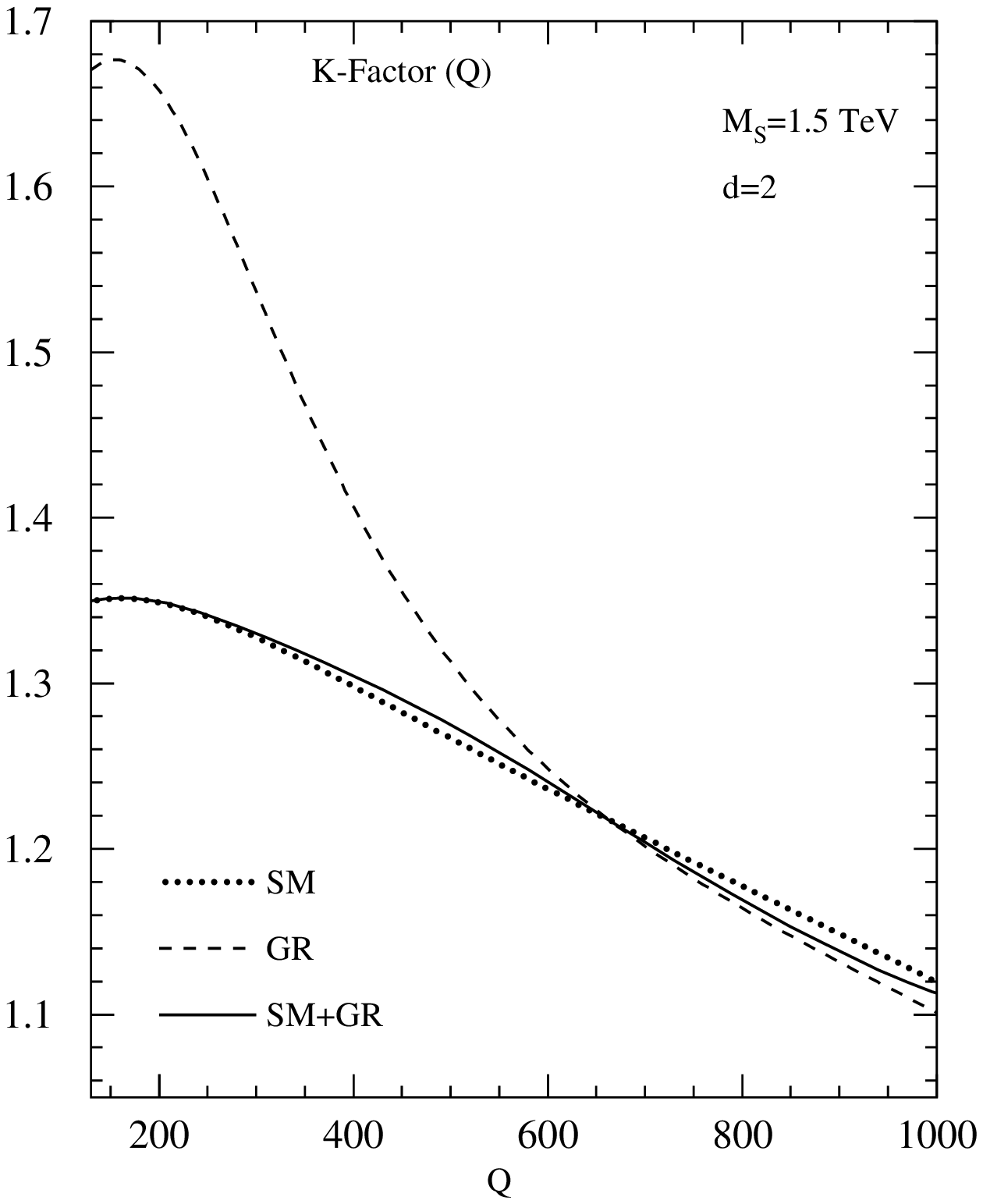,width=15cm,height=16cm,angle=0}}
\vspace{5mm}
\centerline{\bf Fig.~14a}
\label{fig14a}
\end{figure}
                                                                                
\eject

\begin{figure}[htb]
\vspace{1mm}
\centerline{\epsfig{file=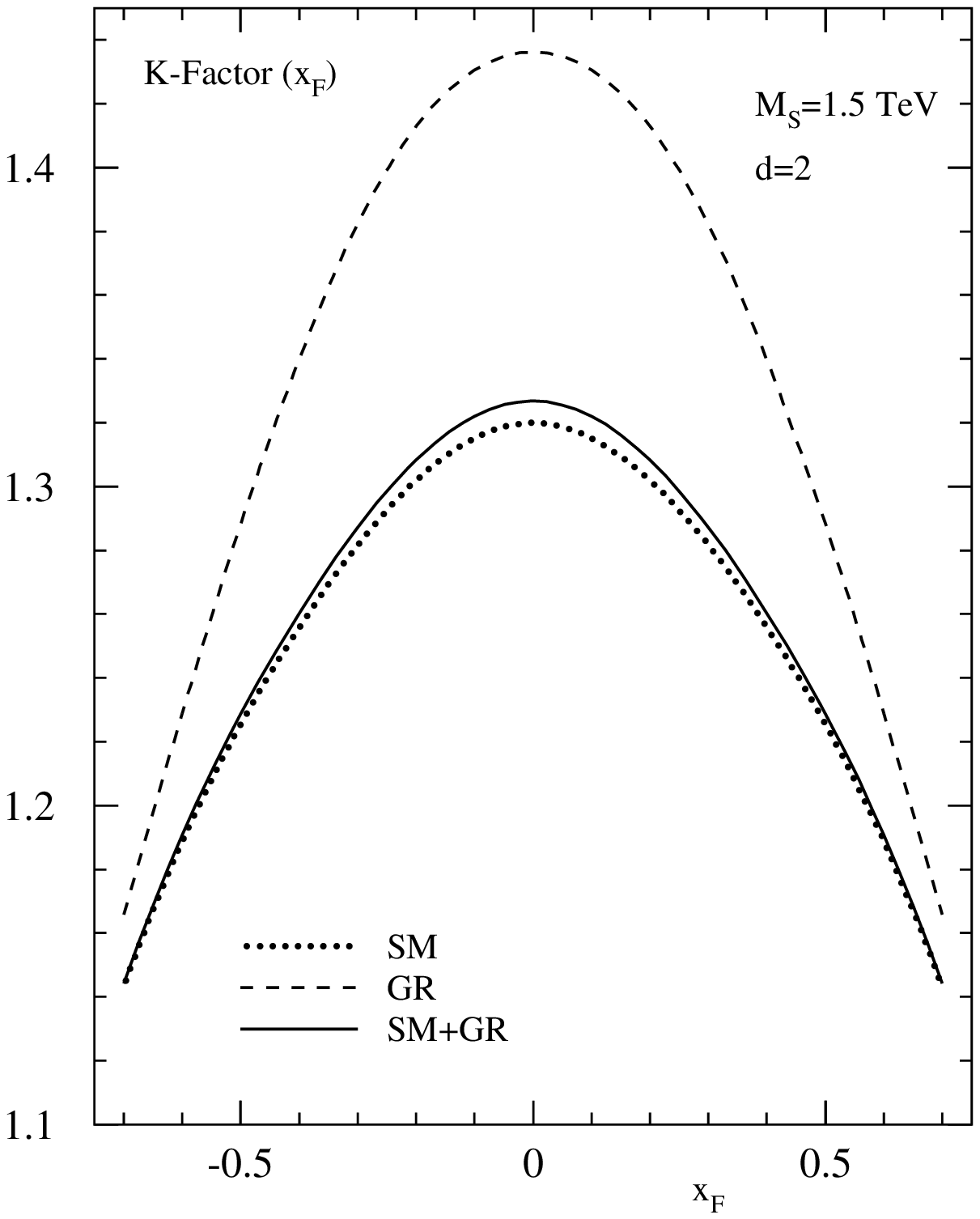,width=15cm,height=16cm,angle=0}}
\vspace{5mm}
\centerline{\bf Fig.~14b}
\label{fig14b}
\end{figure}
                                                                                
\eject

\begin{figure}[htb]
\vspace{1mm}
\centerline{\epsfig{file=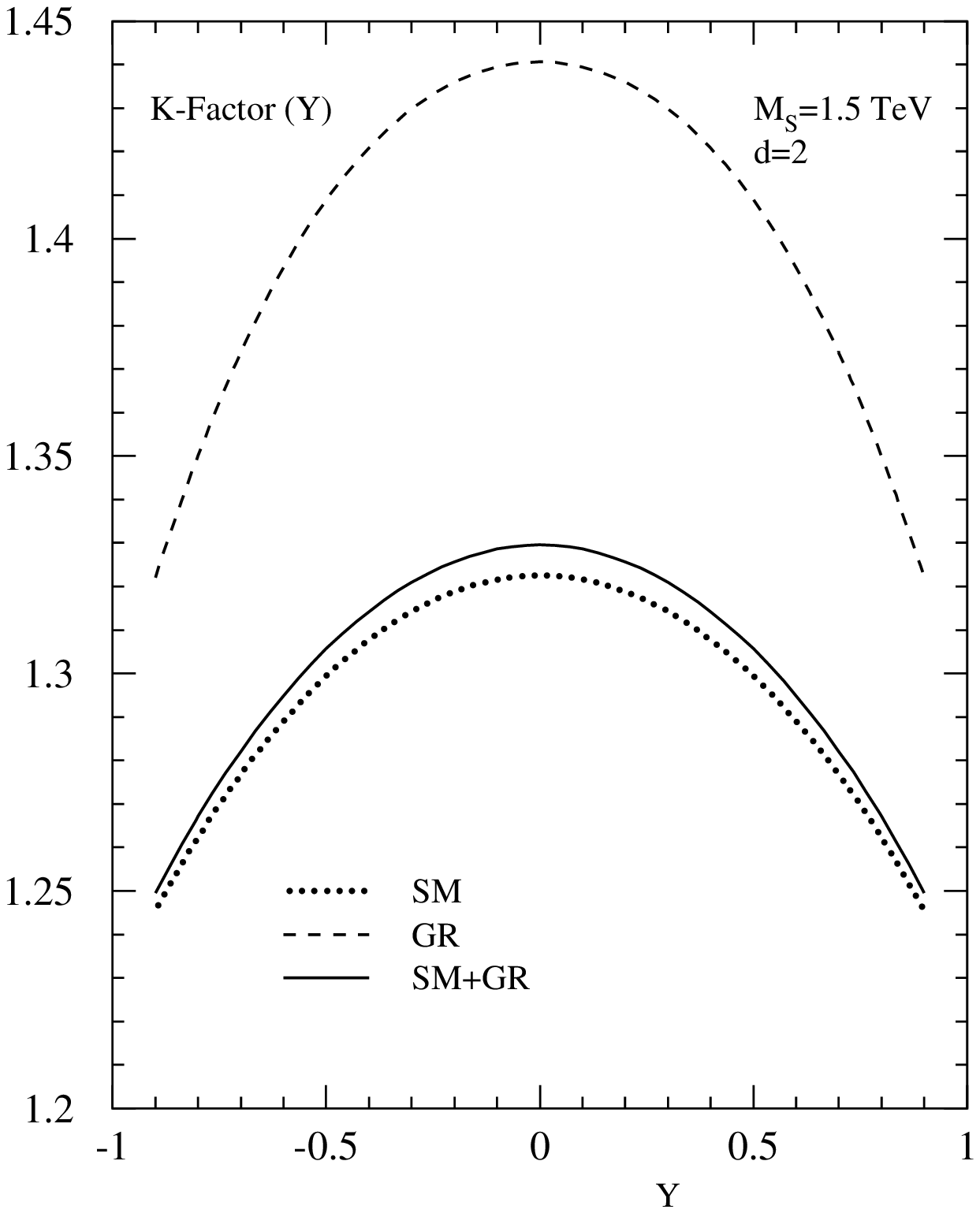,width=15cm,height=16cm,angle=0}}
\vspace{5mm}
\centerline{\bf Fig.~14c}
\label{fig14c}
\end{figure}
                                                                                
\eject

\begin{figure}[htb]
\vspace{1mm}
\centerline{\epsfig{file=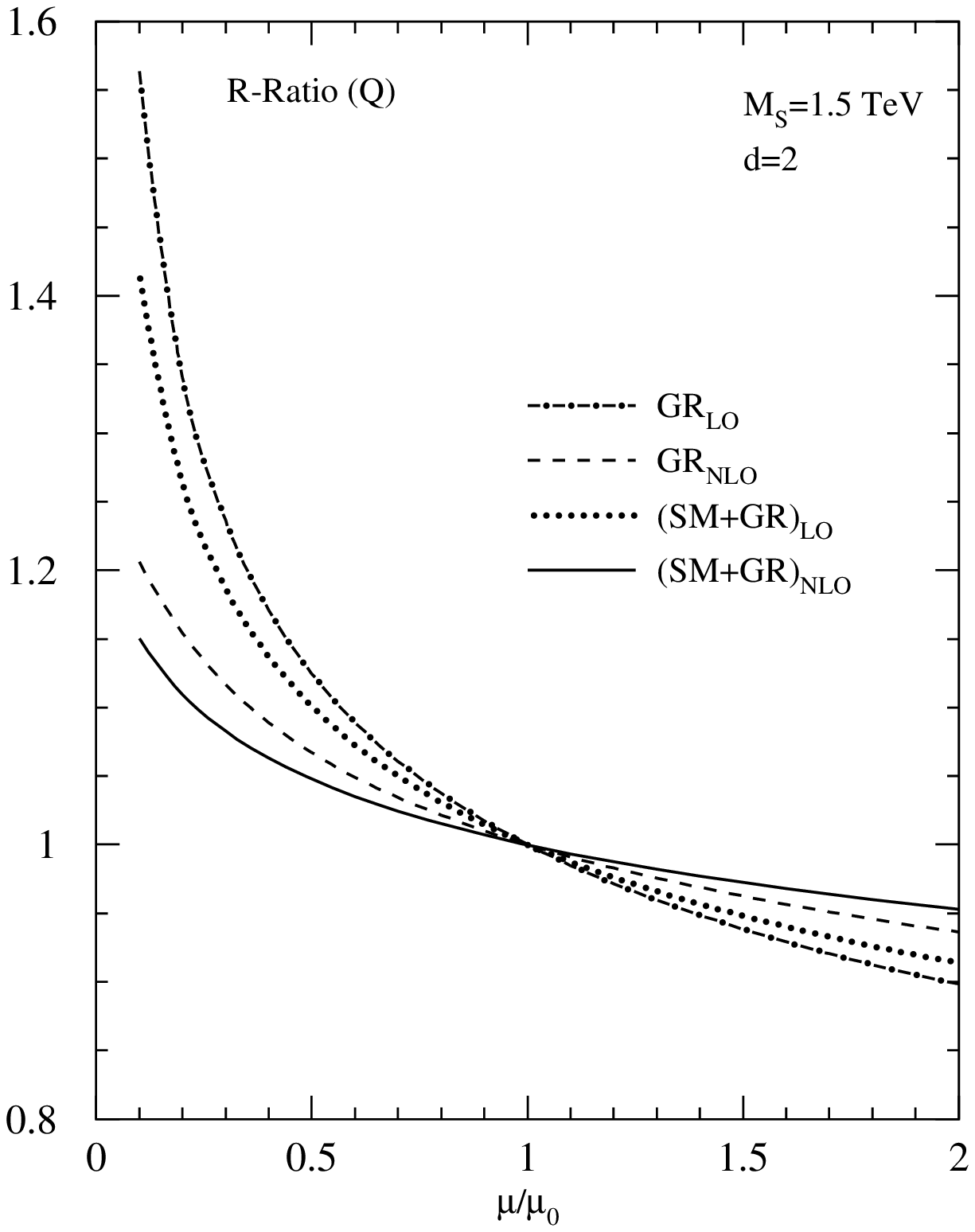,width=15cm,height=16cm,angle=0}}
\vspace{5mm}
\centerline{\bf Fig.~15a}
\label{fig15a}
\end{figure}
                                                                                
\eject

\begin{figure}[htb]
\vspace{1mm}
\centerline{\epsfig{file=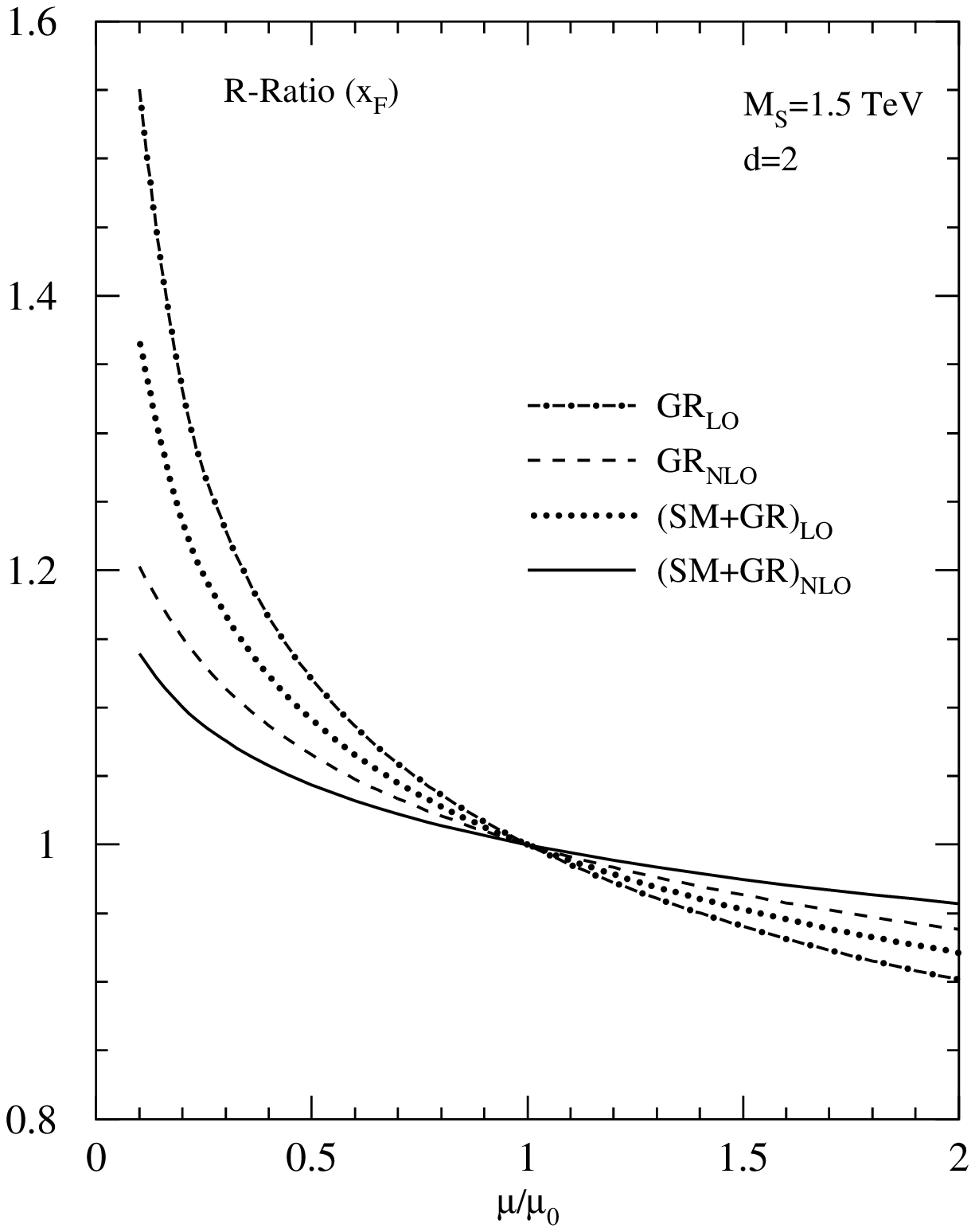,width=15cm,height=16cm,angle=0}}
\vspace{5mm}
\centerline{\bf Fig.~15b}
\label{fig15b}
\end{figure}
                                                                                
\eject

\begin{figure}[htb]
\vspace{1mm}
\centerline{\epsfig{file=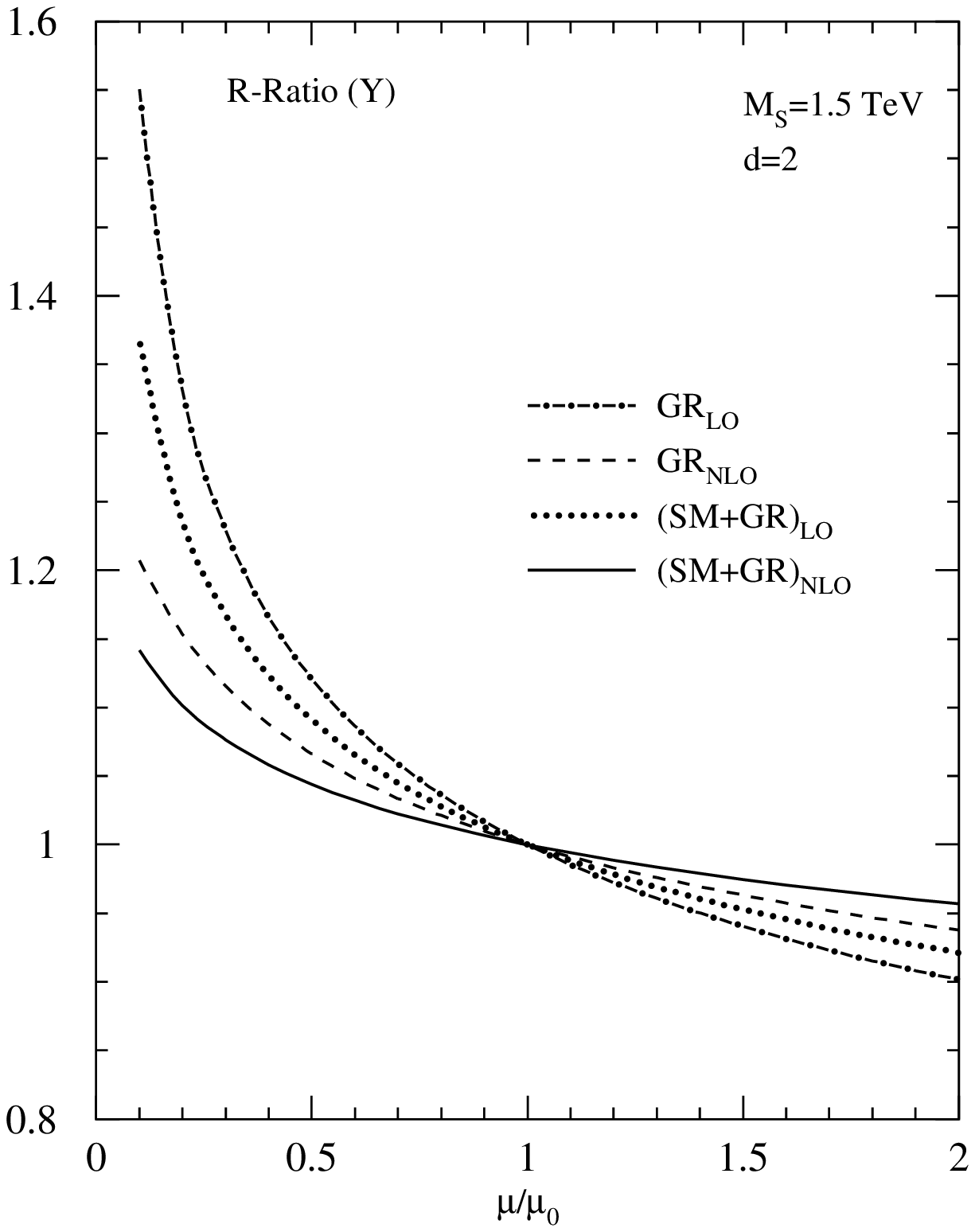,width=15cm,height=16cm,angle=0}}
\vspace{5mm}
\centerline{\bf Fig.~15c}
\label{fig15c}
\end{figure}
                                                                                
\eject

\begin{figure}[htb]
\vspace{1mm}
\centerline{\epsfig{file=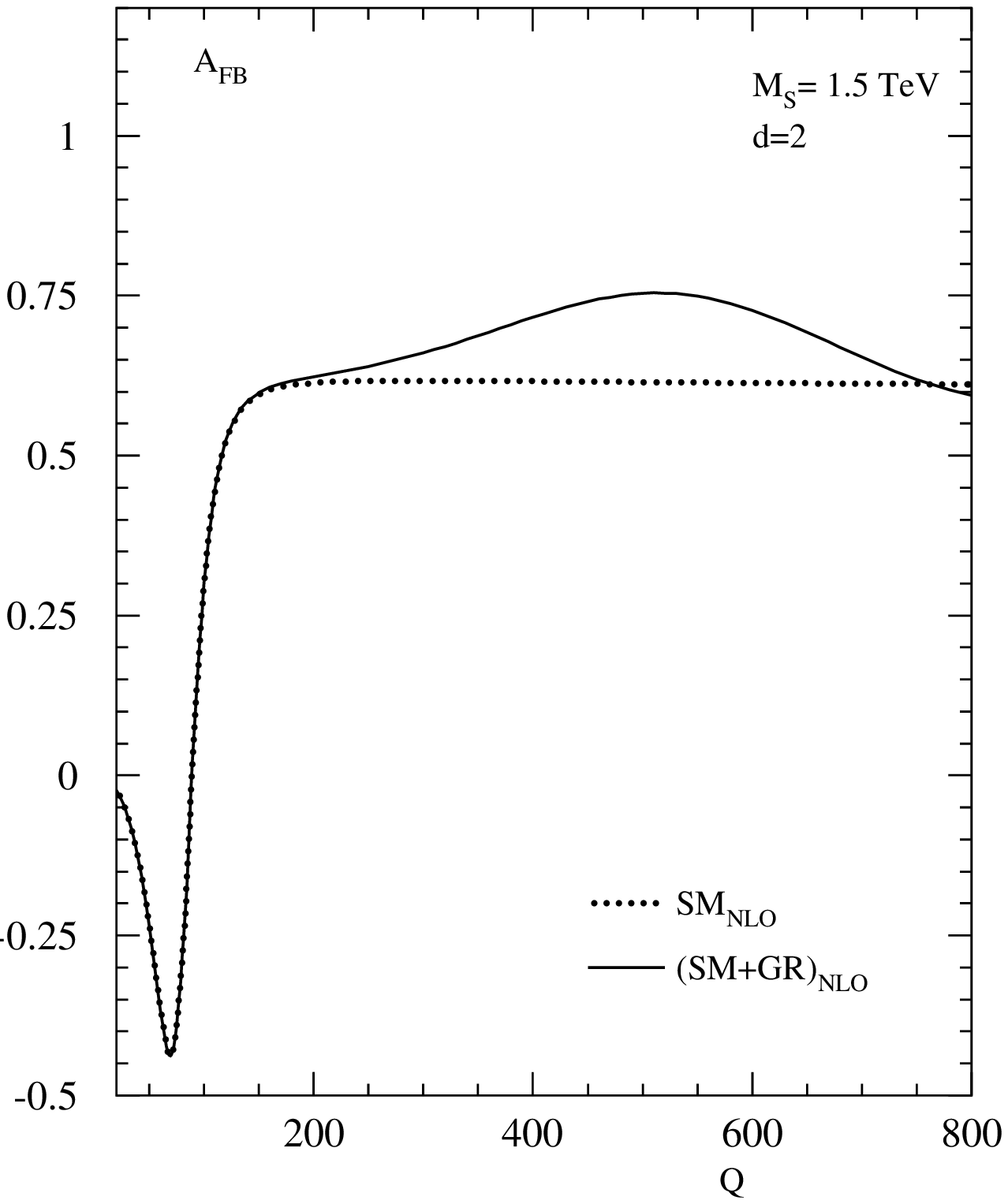,width=15cm,height=16cm,angle=0}}
\vspace{5mm}
\centerline{\bf Fig.~16}
\label{fig16}
\end{figure}
                                                                                

\end{document}